\documentclass[prb,twocolumn,superscriptaddress,floatfix,letterpaper]{revtex4}
\pdfoutput=1

\usepackage{tikz}

\def\FBox{\draw [line width=0pt, white] (0,0)--(5,0)--(5,2)--(0,2)--(0,0);}

\def\iLnkaa{\draw [dotted] (0,0) -- (0.5,1); \draw [dotted] (3,0) -- (4,0.5);}
\def\jLnkaa{\draw [dotted] (0,2) -- (0.5,1); \draw [dotted] (3,2) -- (4,1.5);}
\def\kLnkaa{\draw [dotted] (2,2) -- (1.5,1); \draw [dotted] (5,2) -- (4,1.5);}
\def\lLnkaa{\draw [dotted] (2,0) -- (1.5,1); \draw [dotted] (5,0) -- (4,0.5);}
\def\mLnkaa{\draw [dotted] (0.5,1) -- (1.5,1);}
\def\nLnkaa{\draw [dotted] (4,0.5) -- (4,1.5);}

\def\iLnkbb{\draw  (0,0) -- (0.5,1); \draw  (3,0) -- (4,0.5);}
\def\jLnkbb{\draw  (0,2) -- (0.5,1); \draw  (3,2) -- (4,1.5);}
\def\kLnkbb{\draw  (2,2) -- (1.5,1); \draw  (5,2) -- (4,1.5);}
\def\lLnkbb{\draw  (2,0) -- (1.5,1); \draw  (5,0) -- (4,0.5);}
\def\mLnkbb{\draw  (0.5,1) -- (1.5,1);}
\def\nLnkbb{\draw  (4,0.5) -- (4,1.5);}

\def\iLnkcc{\draw [red]  (0,0) -- (0.5,1); \draw [red] (3,0) -- (4,0.5);}
\def\jLnkcc{\draw [red] (0,2) -- (0.5,1); \draw [red] (3,2) -- (4,1.5);}
\def\kLnkcc{\draw [red] (2,2) -- (1.5,1); \draw [red] (5,2) -- (4,1.5);}
\def\lLnkcc{\draw [red] (2,0) -- (1.5,1); \draw [red] (5,0) -- (4,0.5);}
\def\mLnkcc{\draw [red] (0.5,1) -- (1.5,1);}
\def\nLnkcc{\draw [red] (4,0.5) -- (4,1.5);}

\def\iLnkdd{\draw [green]  (0,0) -- (0.5,1); \draw [green] (3,0) -- (4,0.5);}
\def\jLnkdd{\draw [green] (0,2) -- (0.5,1); \draw [green] (3,2) -- (4,1.5);}
\def\kLnkdd{\draw [green] (2,2) -- (1.5,1); \draw [green] (5,2) -- (4,1.5);}
\def\lLnkdd{\draw [green] (2,0) -- (1.5,1); \draw [green] (5,0) -- (4,0.5);}
\def\mLnkdd{\draw [green] (0.5,1) -- (1.5,1);}
\def\nLnkdd{\draw [green] (4,0.5) -- (4,1.5);}

\def\iLnkbc{
\draw[->]  (0,0) -- (0.25,0.5);
\draw  (0.25,0.5) -- (0.5,1);
\draw[->]  (3,0) -- (3.5,0.25);
\draw (3.5,0.25) -- (4,0.5);
}
\def\jLnkbc{
\draw[->]  (0,2) -- (0.25,1.5);
\draw  (0.25,1.5) -- (0.5,1);
\draw[->]  (3,2) -- (3.5,1.75);
\draw  (3.5,1.75) -- (4,1.5);
}
\def\kLnkbc{
\draw[->]  (2,2) -- (1.75,1.5);
\draw  (1.75,1.5) -- (1.5,1);
\draw[->]  (5,2) -- (4.5,1.75);
\draw  (4.5,1.75) -- (4,1.5);
}
\def\lLnkbc{
\draw[->]  (1.5,1) -- (1.75,0.5) ;
\draw  (1.75,0.5) -- (2,0) ;
\draw[->]  (4,0.5) -- (4.5,0.25) ;
\draw  (4.5,0.25) -- (5,0) ;
}
\def\mLnkbc{
\draw[->]  (0.5,1) -- (1,1);
\draw  (1,1) -- (1.5,1);
}
\def\nLnkbc{
\draw[->]  (4,1.5) -- (4,1);
\draw  (4,1) -- (4,0.5);
}

\def\iLnkcb{
\draw  (0,0) -- (0.25,0.5);
\draw[<-]  (0.25,0.5) -- (0.5,1);
\draw  (3,0) -- (3.5,0.25);
\draw[<-] (3.5,0.25) -- (4,0.5);
}
\def\jLnkcb{
\draw  (0,2) -- (0.25,1.5);
\draw[<-]  (0.25,1.5) -- (0.5,1);
\draw  (3,2) -- (3.5,1.75);
\draw[<-]  (3.5,1.75) -- (4,1.5);
}
\def\kLnkcb{
\draw  (2,2) -- (1.75,1.5);
\draw[<-]  (1.75,1.5) -- (1.5,1);
\draw  (5,2) -- (4.5,1.75);
\draw[<-]  (4.5,1.75) -- (4,1.5);
}
\def\lLnkcb{
\draw  (1.5,1) -- (1.75,0.5) ;
\draw[<-]  (1.75,0.5) -- (2,0) ;
\draw  (4,0.5) -- (4.5,0.25) ;
\draw[<-]  (4.5,0.25) -- (5,0) ;
}
\def\mLnkcb{
\draw  (0.5,1) -- (1,1);
\draw[<-]  (1,1) -- (1.5,1);
}
\def\nLnkcb{
\draw  (4,1.5) -- (4,1);
\draw[<-]  (4,1) -- (4,0.5);
}

\def\iLnkbd{
\draw[->]  (0,0) -- (0.25,0.5);
\draw  (0.25,0.5) -- (0.5,1);
\draw[->]  (3,0) -- (3.5,0.25);
\draw (3.5,0.25) -- (4,0.5);
}
\def\jLnkbd{
\draw[->]  (0,2) -- (0.25,1.5);
\draw  (0.25,1.5) -- (0.5,1);
\draw[->]  (3,2) -- (3.5,1.75);
\draw  (3.5,1.75) -- (4,1.5);
}
\def\kLnkbd{
\draw[->]  (2,2) -- (1.75,1.5);
\draw  (1.75,1.5) -- (1.5,1);
\draw[->]  (5,2) -- (4.5,1.75);
\draw  (4.5,1.75) -- (4,1.5);
}
\def\lLnkbd{
\draw[->]  (1.5,1) -- (1.75,0.5) ;
\draw  (1.75,0.5) -- (2,0) ;
\draw[->]  (4,0.5) -- (4.5,0.25) ;
\draw  (4.5,0.25) -- (5,0) ;
}
\def\mLnkbd{
\draw[->]  (0.5,1) -- (1,1);
\draw  (1,1) -- (1.5,1);
}
\def\nLnkbd{
\draw[->]  (4,1.5) -- (4,1);
\draw  (4,1) -- (4,0.5);
}

\def\iLnkdb{
\draw  (0,0) -- (0.25,0.5);
\draw[<-]  (0.25,0.5) -- (0.5,1);
\draw  (3,0) -- (3.5,0.25);
\draw[<-] (3.5,0.25) -- (4,0.5);
}
\def\jLnkdb{
\draw  (0,2) -- (0.25,1.5);
\draw[<-]  (0.25,1.5) -- (0.5,1);
\draw  (3,2) -- (3.5,1.75);
\draw[<-]  (3.5,1.75) -- (4,1.5);
}
\def\kLnkdb{
\draw  (2,2) -- (1.75,1.5);
\draw[<-]  (1.75,1.5) -- (1.5,1);
\draw  (5,2) -- (4.5,1.75);
\draw[<-]  (4.5,1.75) -- (4,1.5);
}
\def\lLnkdb{
\draw  (1.5,1) -- (1.75,0.5) ;
\draw[<-]  (1.75,0.5) -- (2,0) ;
\draw  (4,0.5) -- (4.5,0.25) ;
\draw[<-]  (4.5,0.25) -- (5,0) ;
}
\def\mLnkdb{
\draw  (0.5,1) -- (1,1);
\draw[<-]  (1,1) -- (1.5,1);
}
\def\nLnkdb{
\draw  (4,1.5) -- (4,1);
\draw[<-]  (4,1) -- (4,0.5);
}

\begin{document}

\title{
Modular transformations\\
 and topological orders in two dimensions
}

\author{Fangzhou Liu}
\affiliation{Perimeter Institute for Theoretical Physics, Waterloo, Ontario, N2L 2Y5 Canada} 
\affiliation{Department of Physics, Massachusetts Institute of Technology, Cambridge, Massachusetts 02139, USA}

\author{Zhenghan Wang}
\affiliation{Microsoft Station Q, CNSI Bldg. Rm 2237, University of California, Santa Barbara, CA 93106, USA}

\author{Yizhuang You}
\affiliation{Institute for Advanced Study, Tsinghua University, Beijing 10084, China}

\author{Xiao-Gang Wen}
\affiliation{Perimeter Institute for Theoretical Physics, Waterloo, Ontario, N2L 2Y5 Canada} 
\affiliation{Department of Physics, Massachusetts Institute of Technology, Cambridge, Massachusetts 02139, USA}
\affiliation{Institute for Advanced Study, Tsinghua University, Beijing,
100084, P. R. China}

%\date{Summer, 2010}
\begin{abstract}
The string-net approach by Levin and Wen and the local unitary transformation
approach by Chen, Gu and Wen provides ways to systematically label 2+1D
topological orders with gapped edge (which will be called exact topological
order).  In those approaches, different many-body wave functions for exact
topological orders are described by different fixed-point tensors.  Though
extremely powerful, the resulting fixed-point tensors are  many-to-one
description of exact topological orders.  As a result it is hard to judge if
two different fixed-point tensors describe the same quantum phase or not.  We
want to improve that approach by giving a more physical description of the
topological orders.  We find that the non-Abelian Berry's phases, $T$- and
$S$-matrices, of the topological protected degenerate ground states on a torus
give rise to a more physical description of topological orders.  
%Using the Verlinde conjecture, we can even choose the canonocal basis of the
%$T$- and $S$-matrices in some cases.  
It is conjectured that the $T$ and $S$-matrices (up to an unitary
thrasformation) form a complete and one-to-one characterization of exact
topological orders and can replace the fixed-point tensor description to give
us a more physical label for topological orders.  As a result, all the
topological properties can be obtained from the $T$- and $S$-matrices, such as
number of quasiparticle types (from the dimension of the $T$ or $S$), the
quasiparticle statistics (from the diagonal elements of $T$), the quantum
dimensions of quasiparticles, \etc.

\end{abstract}

\maketitle

{\small \setcounter{tocdepth}{2} \tableofcontents }

\section{Introduction}

One of the most important questions in condensed matter physics is the
description and the classification of different phases of matter. Landau's
symmetry-breaking theory\cite{L3726,GL5064} provides a very elegant answer to
this question: different phases are characterized by different broken
symmetries. Thus by classifying all different broken symmetries, we can get a
classification of phases.  Though extremely powerful, Landau's theory fails to
explain many new states found in experiments including the Fractional Quantum
Hall(FQH) states.\cite{TSG8259,L8395} These new states thus bring up once again
the old question of how to classify different phases of matter.

A closer look at Landau's symmetry-breaking theory reveals that the theory only
describes direct-product states up to some small local perturbations (see
\Ref{CGW1038}). Since these small perturbations can only modify the
direct-product states locally, Landau's theory can only describe states with
``short-range entanglements''. Intuitively, these states can only account for a
small fraction of all possible quantum many-body states. Thus what seems to be
missing in Landau's theory is a large class of states with ``long-range
entanglement''.  Those states were named topologically ordered states in 1989
before the concept of entanglements become popular.\cite{Wtop,WNtop}  Different
patterns of long-range-entanglements/topological-orders correspond to different
quantum phases.

But what are these patterns of long-range entanglement? What are topological
orders? It is hard to answer these questions since those patterns/orders are
new concepts that do not even have a label.  So to study
topological-order/long-range-entanglement, we first need to invent labels or
mathematical symbols for them.  The first label/symbol invented is the ground
state degeneracy $D_\text{deg}(g)$ on torus (with genus $g=1$)\cite{HR8529} and
other Riemann surfaces.\cite{Wtop,WNtop} But it was clear from the beginning
that this is not a very good label, since the same $D_\text{deg}(g)$ can
correspond to many different topological orders.  To obtain a better label, in
\Ref{Wrig,KW9327}, it was proposed to use the non-Abelian geometric
phases\cite{WZ8411} of degenerate ground states on Riemann surfaces to
characterize different topological orders.  

It was conjectured\cite{Wrig} that the non-Abelian geometric phases (the
non-Abelian part) form a complete and one-to-one description of topological
orders.  On torus,  they are described by $T$- and $S$-matrices, which generate
a projective representation of the modular transformation on the torus.  We can
view $T$- and $S$-matrices as some kind of ``non-local order parameter''.
Therefore, the concept of topological order is defined through the physical
properties of robust ground state degeneracy (called topological
degeneracy)\cite{Wtop,WNtop} and robust geomatric phases induced by the modular
transformation of the degenerate ground states,\cite{Wrig,KW9327} just like the
concept superconducting order was introduced through physical properties of
zero resistance and Meissner effect.  This led to the establishment of the
concept of topological order (and long-range entanglements).  It is thus very
desirable to develop a comprehensive theory of topological orders based on $T$-
and $S$-matrices.

However, the current comprehensive theory of 2+1D \emph{exact topological
orders} (\ie the topological orders that can have a gapped edge)\cite{KW14} is
not based on $T$- and $S$-matrices, but rather, based on the string-net
approach of \Ref{LWstrnet} and local unitary transformation approach of
\Ref{CGW1038}. They provide a systematic description of 2+1D exact topological
orders by systematically labeling the corresponding ``long-range entangled''
many-body wave functions through a set of fixed-point tensors.  These
fixed-point tensors form a mathematical structure called tensor category
theory.  Since the many-body wave functions described by the fixed-point
tensors are explicitly known, we can construct exactly soluble Hamiltonians to
realize each topological order described by a fixed-point tensor.

Although the fixed-point tensors approach (\ie the string-net or the
local-unitary-transformation approach) is closely related to the many-body wave
function, it is known that different fixed-point tensors can correspond to same
quantum phase.  To be more precise, the different fixed-point tensors actually
describe different unitary fusion categories (UFC), and two different UFC's
will describe the same 2+1D topological order if they have the same Drinfeld
center.\cite{KW14}

So to better identify the quantum phases described by the fixed-point tensors,
and in an attempt to develop a theory of topological orders based on  $T$- and
$S$-matrices, in this paper, we will calculate the $T$- and $S$-matrices from
the fixed-point tensors.  This allows us to identify various topological orders
in the the many-body wave functions described by the fixed-point tensors.  In
particular, we can calculate the number of quasiparticle types, the
quasiparticle statistics and quantum dimensions, \etc.

\section{Local unitary transformation approach and modular transformations}
\label{wvrg}

In this section, we will review the local-unitary transformation approach and
associated fixed-point tensor description of many-body wave functions of
topological ordered states.  The key in the local unitary transformation
approach is a new definition of equivalence class on many-body ground states.
Two gapped many-body ground states were defined as belonging to the same
equivalence class (\ie the same ``phase'') if and only if they are connected by
a local unitary (LU) transformation defined as follows:
\begin{align}
\label{LUdef}
 |\Phi(1)\> \sim |\Phi(0)\> \text{\ iff\ }
 |\Phi(1)\> =  \cT[e^{-i\int_0^1 dg\, \t H(g)}] |\Phi(0)\>
\end{align}
where $\cT$ is the path-ordering operator and $\t H(g)=\sum_{\v i} O_{\v i}(g)$
is a sum of local Hermitian operators.  It can be proved that the above
definition is equivalent to the standard definition through ``adiabatic
evolution'' (\ie two states belong to the same phase if and only if they can be
connected through a gapped adiabatic evolution).\cite{CGW1038} The new
definition of equivalence class is advantageous because it provides a very
operational way to determine if two states belong to the same phase, thus
provides a very natural way for renormalization.  Under this definition,
``long-range entangled'' states are those that are not in the same phase as
direct-product states.  Different ``long-range entanglement'' are then called
``topological orders''.

Following the new definition of equivalence class, a wave function
renormalization scheme can be introduced.\cite{LWstrnet,V0705,CGW1038} It
contains two basic moves called ``F-move'' and ``P-move'', both of which are
the generalized local unitary (gLU) transformations.  Under this
renormalization scheme, gapped ground states will only flow within the same
equivalence class, \ie renormalization does not change its topological order.
Thus a fixed point in this renormalization scheme can be used to represent a
whole equivalence class.  Following the convention of \Ref{CGW1038}, we
represent the two renormalization moves at fixed point by tensors:
\begin{align}
\Phi_\text{fix}
\bpm \includegraphics[scale=.40]{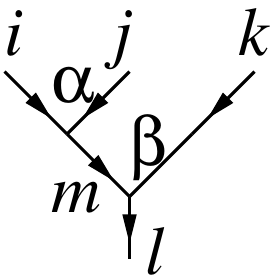} \epm   \simeq
\sum_{n=0}^N
\sum_{\chi=1}^{N_{kjn^*}}
\sum_{\del=1}^{N_{nil^*}}
 F^{ijm,\al\bt}_{kln,\chi\del}
\Phi_\text{fix}
\bpm \includegraphics[scale=.40]{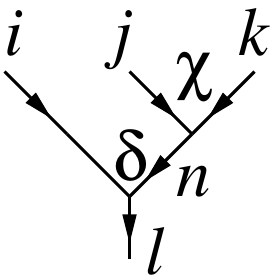} \epm, \nonumber
\end{align}
\begin{align}
\label{PhiP}
\Phi_\text{fix} \bpm \includegraphics[scale=.40]{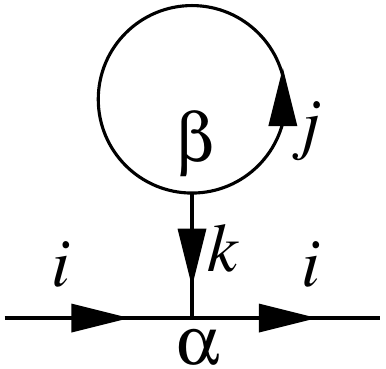} \epm
\simeq  P_i^{kj,\al\bt}
\Phi_\text{fix} \bpm \includegraphics[scale=.40]{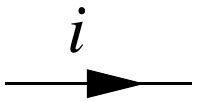} \epm ,
\end{align}
also, the simplest fixed-point wave function was represented by:
\begin{align}
\Phi_\text{fix} \bpm \includegraphics[scale=.40]{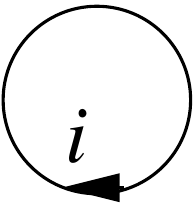} \epm
=A^i=A^{i^*}.
\end{align}

The above-mentioned fixed point tensors can be obtained through solving a set
of self-consistent equations.  Since the fixed-point state will not change
under renormalization, we can apply renormalization moves in arbitrary orders.
This arbitrariness gives us a set of self-consistent conditions that fixed
point tensors must satisfy.  The self-consistent conditions are summarized
below:\cite{CGW1038}
\begin{align}
N_{ikj}=N_{kji}=N_{i^*j^*k^*} \geq 0 ,\ \ \ \  \sum_{j,k} N_{ii^*k} N_{jk^*j^*} \geq 1 ,\nonumber
\end{align}
\begin{align}
\sum_{m=0}^N N_{jim^*} N_{kml^*}
=\sum_{n=0}^N N_{kjn^*} N_{l^*ni} ,\nonumber
\end{align}
\begin{align}
 (F^{ijm,\al\bt}_{kln,\chi\del})^*
=
 F^{jkn,\chi\del}_{l^*i^*m^*,\bt\al} ,\nonumber
\end{align}
\begin{align}
\sum_{n,\chi,\del}
F^{ijm',\al'\bt'}_{kln,\chi\del}
(F^{ijm,\al\bt}_{kln,\chi\del})^*
=\del_{m\al\bt,m'\al'\bt'} ,\nonumber
\end{align}
\begin{align}
&\ \ \ \
\sum_{t}
\sum_{\eta=1}^{N_{kjt^*}}
\sum_{\vphi=1}^{N_{tin^*}}
\sum_{\ka=1}^{N_{lts^*}}
F^{ijm,\al\bt}_{knt,\eta\vphi}
F^{itn,\vphi\chi}_{lps,\ka\ga}
F^{jkt,\eta\ka}_{lsq,\del\phi}
\nonumber\\
& = \e^{\imth \th_F}
\sum_{\eps=1}^{N_{qmp^*}}
F^{mkn,\bt\chi}_{lpq,\del\eps}
F^{ijm,\al\eps}_{qps,\phi\ga} ,\nonumber
\end{align}
\begin{align}
\e^{\imth\th_{P1}} P_i^{kj,\al\bt}=
\sum_{m,\la,\ga,l,\nu,\mu}
F^{jj^*k,\bt\al}_{i^*i^*m^*,\la\ga}
F^{i^*mj,\la\ga}_{m^*i^*l,\nu\mu}
P_{i^*}^{lm,\mu\nu} ,\nonumber
\end{align}
\begin{align}
&
\e^{\imth\th_{P2}} P^{jp,\al\eta}_i \del_{im} \del_{\bt\del}
=
\sum_{\chi=1}^{N_{kjk^*}} F^{ijm,\al\bt}_{klk,\chi\del} P^{jp,\chi\eta}_{k^*}
\nonumber\\
&\text{for all } k,i,l \text{ satisfying }
N_{kil^*}>0 ,\nonumber
\end{align}
\begin{align}
\Phi_\text{fix} \bpm \includegraphics[scale=.40]{oi} \epm
\equiv A^i= A^{i^*}\neq 0,\ \ \ \ \sum_i A^i (A^i)^*=1 ,\nonumber
\end{align}
\begin{align}
P^{mj,\ga\la}_i A^i
=\e^{\imth \th_{A1}} P^{m^*i^*,\la\ga}_{j^*} A^j ,\nonumber
\end{align}
\begin{align}
 \Phi^\th_{ikj,\al\bt}
=\e^{\imth \th'}
\sum_{m,\la,\ga}
F^{ijk^*,\al\bt}_{j^*im,\la\ga}
P^{mj,\ga\la}_i A^i ,\nonumber
\end{align}
\begin{align}
&  \Phi^\th_{ikj,\al\bt}
= \e^{\imth \th_{A2}} \Phi^\th_{kji,\al\bt},
%=  \Phi^\th_{jik,\al\bt},
\nonumber\\
& \Phi^\th_{ikj,\al\bt}=0 \text{ if } N_{ikj}=0 ,\nonumber
\end{align}
\begin{align}
\label{Phinonz}
\det[ \Phi^\th_{kji,\al\bt} ]
\neq 0 .
\end{align}
Conditions (\ref{Phinonz}) form a set of non-linear equations with variables
$N_{ijk}$, $F^{ijm,\al\bt}_{kln,\ga\la}$, $P_i^{kj,\al\bt}$, $A^i$ and
$(\th_F,\th_{P1},\th_{P2},\th_{A1},\th_{A2})$, which are tensor labels for
fixed points.  Following the previous discussion, the solutions $(N_{ijk},
F^{ijm,\al\bt}_{kln,\ga\la}, P_i^{kj,\al\bt}, A^i)$ will then give us a
labeling of different topological orders.

After obtaining such a tensor labeling for the ground state wave function, the
next natural questions to ask are: Are they in one-to-one correspondence with
different topological orders?  How can we physically understand these tensors?
Are there any physical quantities that we can at least numerically measure?

In fact, the non-Abelian Berry's phases $T$ and $S$ (up to unitary
transformations) provide an answer to the above questions.  Not only can $T$
and $S$ give a one-to-one label for different topological orders, but they can
also provide a link between the fixed-point ground states and their
corresponding quasi-particle excitations.  We believe that the description
given by the $T$- and $S$-matrices is complete, meaning that they can
completely characterize different non-chiral topological orders.  Further
calculations of various topological properties thus can all be
obtained from $T$- and $S$-matrices. 

Now we want to briefly introduce the concept of
non-Abelian-Berry's-phase description of topological order which was first
introduced in \Ref{Wrig}. The more
detailed description will be given in the next section of the paper.  The
non-Abelian Berry's phases are obtained from two types of transformations named
T- and S-transformation, both defined on a torus.  The ``T-transformation'' can
be defined by twisting a fixed-point graph along an axis (let us say the
$x$-direction) by $360^{\circ}$; whereas the ``S-transformation'' can be
defined by rotating a fixed-point graph by $90^{\circ}$.  Essentially they are
all discrete deformations of a fixed-point graph on a torus.  As hinted from
the one-to-one correspondence between the ground state degeneracy on a torus
and the number of different types of quasi-particles, the information of the
quasi-particles is always encoded in the geometry of the torus. Here again,
through the non-Abelian Berry's phases of the ground states, we can ``decode''
the information about their quasi-particles.  So $T$- and $S$-matrices give a
better description of topological orders.  In this paper, we will label
different topological orders by their resulting $T$ and $S$-matrices.

Also mathematically, the relationship between $T$- and $S$-matrix labels and
fixed-point-tensor labels can be understood within tensor category theory.  On
one hand, the $T$- and $S$-matrices describe a set quasiparticle excitations,
which may have fractional and/or non-Abelian statistics and are described by a
modular tensor category $\cT$.  On the other hand, the fixed-point tensors
correspond to the many-body wave functions of exact topological orders, which
are described by unitary fusion category $\cF$.  Thus to calculate the $T$- and
$S$-matrices from the fixed-point tensors is to calculate the quasiparticle
excitations from the many-body wave functions, which corresponds to calculating
the modular tensor category $\cT$ from a unitary fusion category $\cF$.
Mathematically, calculating $\cT$ from $\cF$ corresponds to taking the
"Drinfeld center" of the unitary fusion category $\cF$.  Our calculation of
$T$- and $S$-matrices from the fixed-point tensors corresponds exactly to
taking the "Drinfeld center" of the unitary fusion category $\cF$.

This picture agrees perfectly with our calculations: we perform modular
transformations on several fixed-point states (\ref{Z2state}, \ref{Fibonacci},
\ref{Pfaffian}, \ref{Z3state}, \ref{Z4state}), all having the structure of
unitary tensor category; our resulting $T$- and $S$-matrices for the
quasi-particle excitations are all of the structure of modular tensor category.
This result further strengthened our belief that tensor category theory is the
mathematical structure behind topological order.

In this paper, we present many fixed-point solutions to \eqn{Phinonz},
representing many exact topological orders.  We then apply ``modular
transformations'' to many of the obtained results (\ref{Z2state},
\ref{Fibonacci}, \ref{Pfaffian}, \ref{Z3state}, \ref{Z4state}) to get their $T$
and $S$-matrices, describing their corresponding quasi-particles. 

Note that there's a special solution, \ie the ``chiral'' case \ref{Chiral}.  In
that case, the labeling fixed-point tensors form the so-called ``multi-fusion
category'' and its fusion rule breaks the ``chiral'' symmetry. Although the
resulting fixed-point wave function has a trivial topological order without any
symmetry, we anticipate that the ``chiral'' case can be highly non-trivial
after introducing certain symmetries.

\section{List of different topological orders}

In this section, we present a list of all fixed-point solutions obtained so far
from solving \eqn{Phinonz}.  The solutions consist of fixed-point gLU
transformations $(N_{ijk},F^{ijm,\al\bt}_{kln,\ga\la},P_i^{kj,\al\bt})$ and
fixed-point wave functions $A^i$.  For most of the cases (\ref{Z2state},
\ref{Fibonacci}, \ref{Pfaffian}, \ref{Z3state}, \ref{Z4state}), we also present
their corresponding $T$ and $S$-matrices from modular transformations.

\subsection{$N=1$ loop states -- the $\mathbb{Z}_2$ states}
\label{Z2state}

Let us first consider a system where there are only two
states $|0\>$ and $|1\>$ on each link of the graph.  We
choose $i^*=i$ and the simplest fusion rule that satisfies
\eqn{Phinonz} 
(we call the $N$ tensors satisfying \eqn{Phinonz} as ``fusion rule''):
\begin{align}
\label{Z2fusion}
& N_{000}= N_{110}= N_{101}= N_{011}=1,
\nonumber\\
&
 \text{other } N_{ijk}=0.
\end{align}
Since $N_{ijk}\leq 1$, there are no states on the vertices.
So the indices $\al,\bt,...$ labeling the states on a vertex
can be suppressed.

The above fusion rule corresponds to the fusion rule of the
$N=1$ loop states discussed in \Ref{LWstrnet}, thus the
name $N=1$ loop states. Note further that the three edge
labels of $N$ also form a $\mathbb{Z}_2$ group: for
example, $N_{110}=1$ represents the group action
$1 \bigotimes 1 = 0$ (Here our group action is addition).
Thus we call the states obtained in this section the
$\mathbb{Z}_2$ states.

Due to relation \eqn{Phinonz},
different components of tensor $F^{ijm}_{kln}$ are not
independent.  There are only four independent potentially non-zero
components, denoted as $f_0$,...,$f_3$:
\begin{align}
\label{Floop}
F^{000}_{000}
\bmm\begin{tikzpicture}[scale=0.27]
\FBox \iLnkaa \jLnkaa \kLnkaa \lLnkaa \mLnkaa \nLnkaa
\end{tikzpicture}\emm
&=f_{0}
\nonumber\\
F^{000}_{111}
\bmm\begin{tikzpicture}[scale=0.26]
\FBox \iLnkaa \jLnkaa \kLnkbb \lLnkbb \mLnkaa \nLnkbb
\end{tikzpicture}\emm
&=(F^{011}_{100}
\bmm\begin{tikzpicture}[scale=0.26]
\FBox \iLnkaa \jLnkbb \kLnkbb \lLnkaa \mLnkbb \nLnkaa
\end{tikzpicture}\emm
)^*=(F^{101}_{010}
\bmm\begin{tikzpicture}[scale=0.26]
\FBox \iLnkbb \jLnkaa \kLnkaa \lLnkbb \mLnkbb \nLnkaa
\end{tikzpicture}\emm
)^*
\nonumber\\
&=F^{110}_{001}
\bmm\begin{tikzpicture}[scale=0.26]
\FBox \iLnkbb \jLnkbb \kLnkaa \lLnkaa \mLnkaa \nLnkbb
\end{tikzpicture}\emm
=f_{1}
\nonumber\\
F^{011}_{011}
\bmm\begin{tikzpicture}[scale=0.26]
\FBox \iLnkaa \jLnkbb \kLnkaa \lLnkbb \mLnkbb \nLnkbb
\end{tikzpicture}\emm
&=(F^{101}_{101}
\bmm\begin{tikzpicture}[scale=0.26]
\FBox \iLnkbb \jLnkaa \kLnkbb \lLnkaa \mLnkbb \nLnkbb
\end{tikzpicture}\emm
)^*=f_{2}
\nonumber\\
F^{110}_{110}
\bmm\begin{tikzpicture}[scale=0.26]
\FBox \iLnkbb \jLnkbb \kLnkbb \lLnkbb \mLnkaa \nLnkaa
\end{tikzpicture}\emm
&=f_{3}
\end{align}
We note that $F^{ijm}_{kln}$ in \eqn{PhiP} relates wave
functions on two graphs. In the above we have drawn the
two related graphs right after the $F$ tensor. The first
graph following F corresponds to the graph on the left-hand
side of \eqn{PhiP} and the second one corresponds to the
graph on the right-hand side of \eqn{PhiP}.  The doted
line corresponds to the $|0\>$-state on the link and the
solid line corresponds to the $|1\>$-state on the link.
There are four potentially non-zero components
in $P^{kj}_i$, which are denoted by
$p_0$,...,$p_3$:
\begin{align}
P^{00}_{0}=p_{0} ,\ \ \
P^{01}_{0}=p_{1} ,\ \ \
P^{00}_{1}=p_{2} ,\ \ \
P^{01}_{1}=p_{3}.
\end{align}

We can adjust the total phases of $p_i$ and $A^i$ to make
$p_0\geq 0$ and $A^0\geq 0$.  We can also use
a gauge transformation to make $f_1\geq 0$.

The fixed-point conditions (\ref{Phinonz})
form a set of non-linear equations on the ten
variables $f_i$, $p_i$, and $A^i$.  Many of the non-linear
equations are dependent or even equivalent.  Using a
computer algebraic system, we simplify the set of non-linear
equations.  The simplified equations are
(after making the phase choice described above)
\begin{align}
&
f_0= f_1 = f_2 = 1, \ \
f_3 = \eta,
\nonumber\\
&
p_1 =p_3=\eta p_0, \ \
p_2=p_0,
\nonumber\\
&
p_0^{2}+|p_1|^2 = 1,\ \
|p_2|^2+|p_3|^2=1,
\\
&
p_1 A^0=p_2 A^1, \ \
\eta p_3 A^1=p_1 A^0 ,\ \
|A^0|^2+|A^1|^2=1
\nonumber
\end{align}
where $\eta=\pm 1$.  The above simplified
equations can be solved exactly. We find two isolated
solutions parameterized by $\eta=\pm 1$:
\begin{align}
&
f_0=f_1 = f_2 = 1, \ \
f_3 = \eta,
\nonumber\\
&
p_0=p_2=\frac{1}{\sqrt 2}, \ \
p_1=p_3=\frac{\eta}{\sqrt 2},
\nonumber\\
&
A^0 = \frac{1}{\sqrt 2}, \ \
A^1 = \frac{\eta}{\sqrt 2} .
\end{align}
We also find
%\begin{align}
%\Phi^\th_{000} &= p_0 A^0 =\frac{1}{2}
%\nonumber\\
%\Phi^\th_{011} &=\Phi^\th_{101}=\Phi^\th_{110}
%=f_3 p_0 A^0
%=\frac{\eta}{2} ,
%\nonumber\\
%& \text{other } \Phi^\th_{ikj}=0 ,
%\end{align}
%and
\begin{align}
 \e^{\imth \th_F}=
 \e^{\imth \th_{P1}}=
 \e^{\imth \th_{P2}}=
 \e^{\imth \th_{A1}}=
 \e^{\imth \th_{A2}}=
1.
\end{align}

%The $\eta=1$ fixed-point state corresponds to the $\mathbb{Z}_2$ loop
%condensed state whose low energy effective field theory is
%the $\mathbb{Z}_2$ gauge theory,\cite{FNS0428,LWstrnet}
%or equivalently the $U(1)\times U(1)$ Chern-Simons gauge theory:
%\begin{align}
% \cL=\frac{1}{4\pi} K^{IJ} a_{I\mu}\prt_\nu
%a_{J\la}\eps^{\mu\nu\la} ,
%\end{align}
%with
%$ K=\bpm  0 & 2 \\
%         2 & 0 \\ \epm $.
%The $\eta=-1$ fixed-point state corresponds to the double-semion
%state whose low energy effective field theory is
%the $U(1)\times U(1)$ Chern-Simons gauge theory\cite{FNS0428,LWstrnet}
%\begin{align}
%\cL=\frac {1}{4\pi}\Big(
%2 a_{1\mu}\prt_{\nu}a_{1\la}\eps^{\mu\nu\la}
%-2 a_{2\mu}\prt_{\nu}a_{2\la}\eps^{\mu\nu\la}
%\Big) ,
%\end{align}
%Or equivalently, 
%\begin{align}
% \cL=\frac{1}{4\pi} K^{IJ} a_{I\mu}\prt_\nu
%a_{J\la}\eps^{\mu\nu\la} ,
%\end{align}
%with
%$ K=\bpm  0 & 2 \\
%         2 & 2 \\ \epm $.      

The above solutions
$(F^{ijm,\al\bt}_{kln,\ga\la},P_i^{kj,\al\bt},A^i)$
with $\eta=\pm 1$
both correspond to fixed-point states.
The $\eta=1$ state is an equal weight superposition
of all the graphic states that satisfy the fusion rule,
whereas the $\eta=-1$ state is also a superposition of
all fusion-rule-satisfying graphic states,
but some with coefficient $1$ and
some others with coefficient $-1$.

\begin{figure}[tb]
\begin{center}
\includegraphics[scale=0.4]
{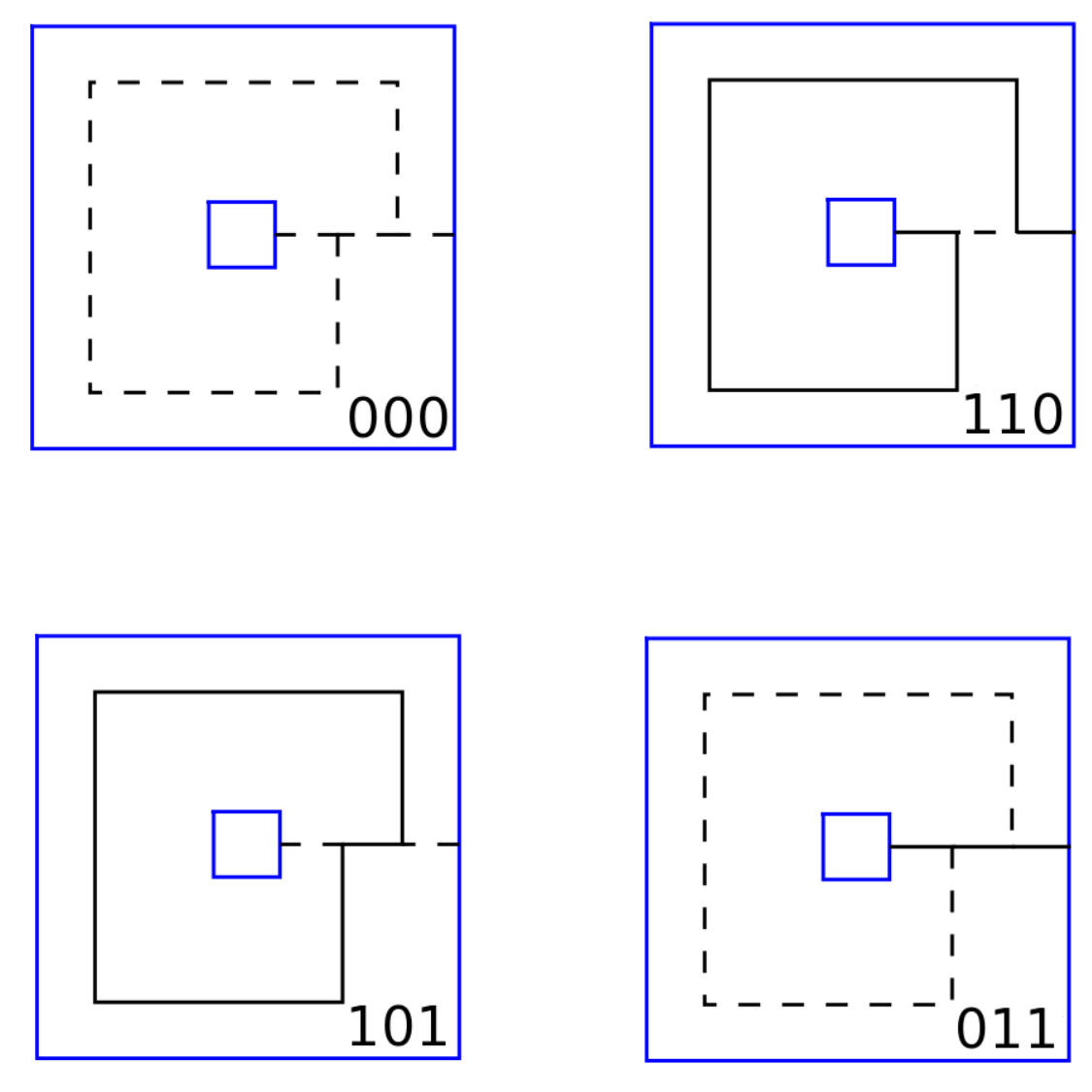}
\end{center}
\caption{
(Color online)
The 4 characteristic non-contractable graphs on a torus.
(The inner square
and the outer square are the physical boundary
of a torus.)
Any other fixed-point graphs can always be reduced
to the above four by F-moves and P-moves,
but the four types
cannot be changed into each other.
The dotted line represents the $|0\>$-state on the edges,
and
the solid line represents the $|1\>$-state
on the edges.
The three integers $ijk$ at the lower right corner
of each graph are the edge labels following the
convention of \eqn{Dtwist}.
}
\label{t4}
\end{figure}

Having obtained the abstract fixed-point solutions,
we now want to physically understand the results by 
introducing the concept of ``modular transformations''.
Modular transformations are defined on
a torus (\ie a planar graph with periodic boundary
conditions in both $x$- and $y$-directions).
Notice that after putting the above states onto a torus,
there will be four different types of non-contractable graphs 
(see Fig. \ref{t4}). They correspond to
four types of fixed-point states that
are linearly independent (meaning they
cannot be transformed into each other through
$F$ or $P$-moves) and form the
degenerate ground-state subspace
of a local Hamiltonian
(See Appendix \ref{HamTorus} for the form of the Hamiltonian).
The modular transformations can be defined
within this ground-state subspace.
We will first introduce the ``$T$-transformation'', also
known as the method of Dehn twist. 
As in \Ref{GWW1017},
we define the Dehn twist formally by requiring it to map a
fixed-point state
$
\Phi_\text{fix}^{\al\bt}(i,j,k,\al,\bt)
=\Phi_\text{fix}^{\al\bt}
\left ( \bmm \includegraphics[scale=.35]{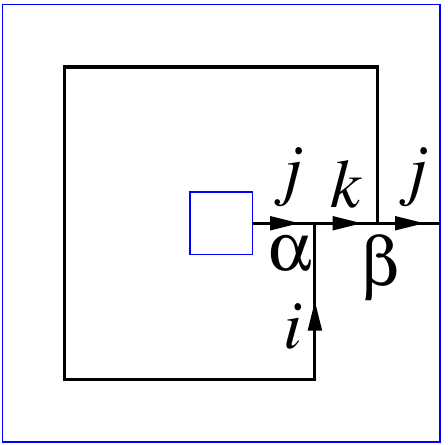} \emm \right )
$
on a torus
to another fixed-point state defined on a different graph
$
\t\Phi_\text{fix}^{\al\bt}(i,j,k,\al,\bt)
=\t\Phi_\text{fix}^{\al\bt}
\left ( \bmm \includegraphics[scale=.35]{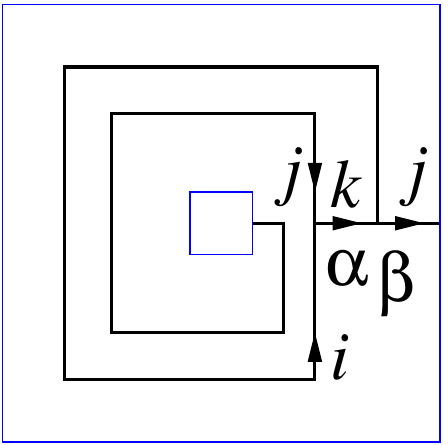} \emm \right )
$.
Then we can use an $F$-move
to deform the graph
$\bmm \includegraphics[scale=.35]{Ttrans2} \emm$
to $\bmm \includegraphics[scale=.35]{Ttrans1} \emm$,
which leads to a unitary transformation $T$
between the four fixed-point states on the torus,
called the ``$T$-transformation''.
The deformation process could be seen more clearly
from the following:
\begin{align}
\label{Dtwist}
&\ \ \
\Phi_\text{fix}^{\al\bt}
\left ( \bmm \includegraphics[scale=.40]{Ttrans1} \emm \right )
\xrightarrow{Dehn \ twist}
\Phi_\text{fix}^{\al\bt}
\left ( \bmm \includegraphics[scale=.40]{Ttrans2} \emm \right )
\nonumber\\
&=
\sum_{l,\chi\del}
F^{ijk,\al\bt}_{i^*jl^*,\chi\del}
\Phi_\text{fix}^{\chi\del}
\left ( \bmm \includegraphics[scale=.40]{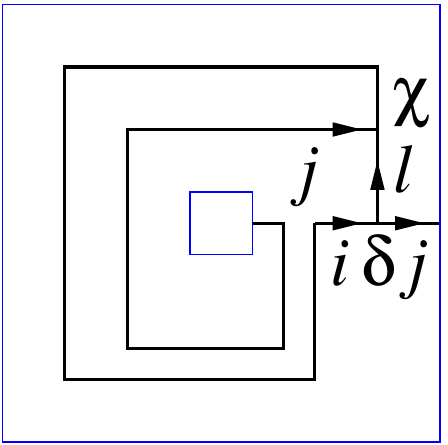} \emm \right )
\ \ (F-move)
\nonumber \\
&=
\sum_{l,\chi\del}
F^{ijk,\al\bt}_{i^*jl^*,\chi\del}
\Phi_\text{fix}^{\chi\del}
\left ( \bmm \includegraphics[scale=.40]{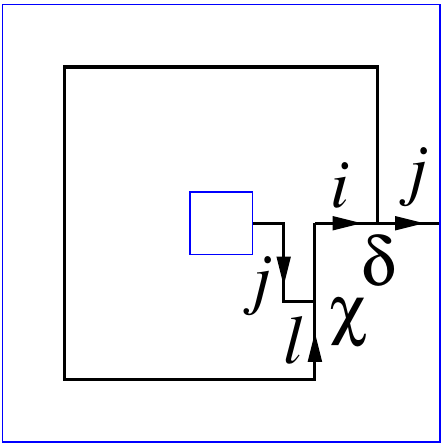} \emm \right )
\ \ (Deformation)
\nonumber \\
&=
\sum_{l}
F^{ijk}_{i^*jl^*}
\Phi_\text{fix}
\left ( \bmm \includegraphics[scale=.40]{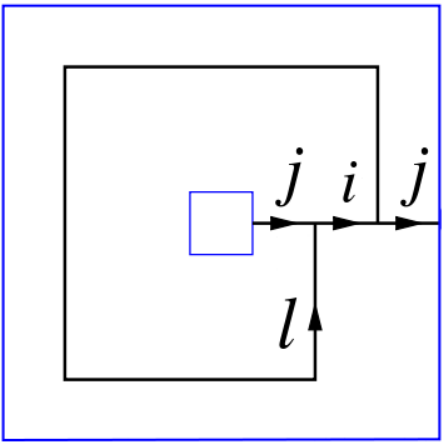} \emm
\right ) .
\end{align}
In the last step above, we have suppressed the
vertex indices $\alpha$, $\beta$, $\delta$, $\chi$
because a maximum of one state is allowed
on each vertex. 

Writing ``$T$-transformation''
in the basis of the
four non-contractable fixed point states on the
torus, we get a $4$ by $4$ unitary matrix
with elements $F^{ijk}_{i^*jl^*}$.
This matrix is called as
``$T$-matrix''.
The basis of the $4$ by $4$ matrix
is defined as follows:
\begin{align*}
\Phi_\text{fix}
\left ( \bmm \includegraphics[scale=.30]{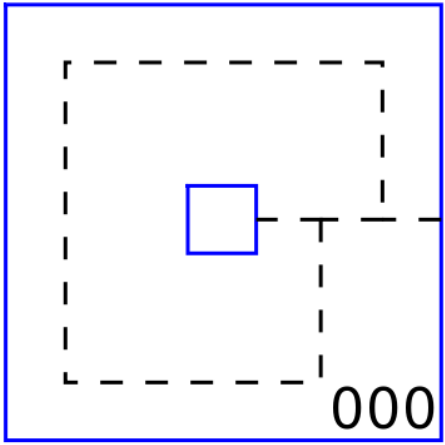} \emm \right)
=(1,0,0,0) ,
\end{align*}
\begin{align*}
\Phi_\text{fix}
\left ( \bmm \includegraphics[scale=.30]{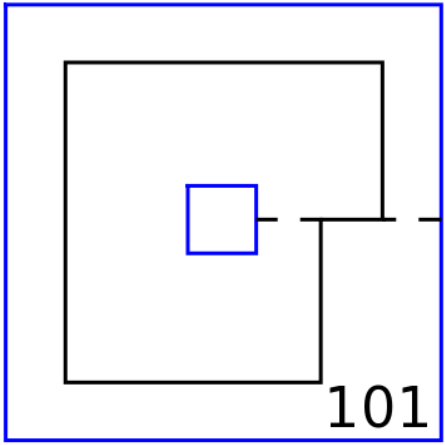} \emm \right)
=(0,1,0,0) ,
\end{align*}
\begin{align*}
\Phi_\text{fix}
\left ( \bmm \includegraphics[scale=.30]{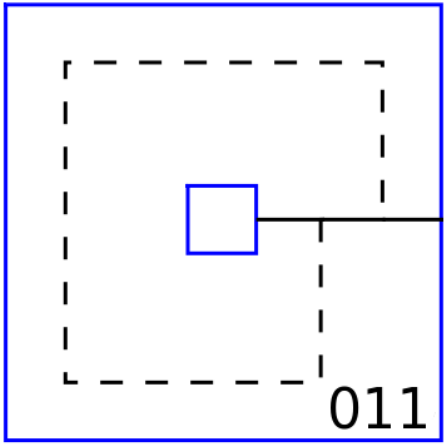} \emm \right)
=(0,0,1,0) ,
\end{align*}
\begin{align*}
\Phi_\text{fix}
\left ( \bmm \includegraphics[scale=.30]{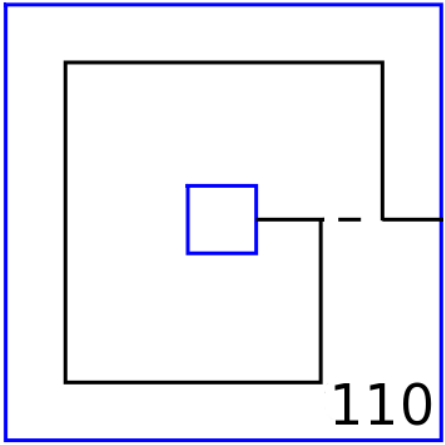} \emm \right)
=(0,0,0,1) .
\end{align*}

We now apply
Dehn twist to all the four non-contractable
states (recall Fig. \ref{t4}) to get  the $T$-matrix: 
\begin{align*}
\Phi_\text{fix}
\left ( \bmm \includegraphics[scale=.30]{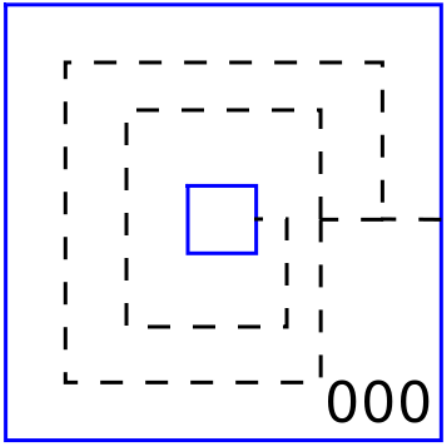} \emm \right)
=F^{000}_{000}
\Phi_\text{fix}
\left ( \bmm \includegraphics[scale=.30]{t000} \emm \right)
\end{align*}
\begin{align*}
\Phi_\text{fix}
\left ( \bmm \includegraphics[scale=.30]{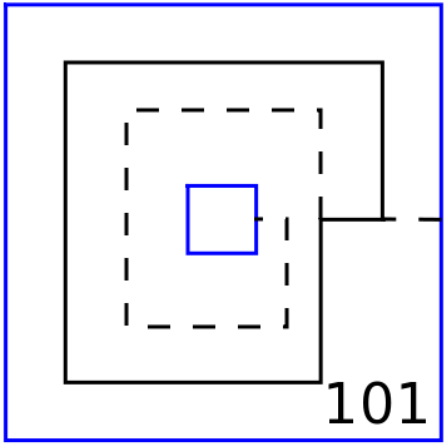} \emm \right)
=F^{101}_{101}
\Phi_\text{fix}
\left ( \bmm \includegraphics[scale=.30]{t101} \emm \right)
\end{align*}
\begin{align*}
\Phi_\text{fix}
\left ( \bmm \includegraphics[scale=.30]{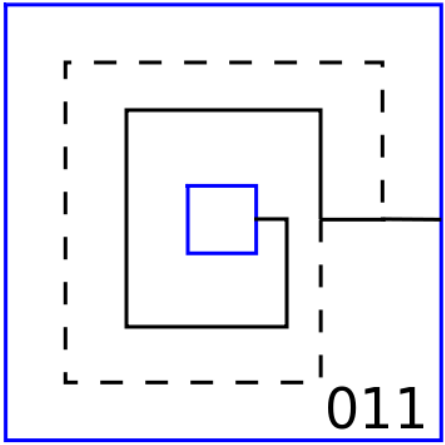} \emm \right)
=F^{011}_{011}
\Phi_\text{fix}
\left ( \bmm \includegraphics[scale=.30]{t110} \emm \right)
\end{align*}
\begin{align*}
\Phi_\text{fix}
\left ( \bmm \includegraphics[scale=.30]{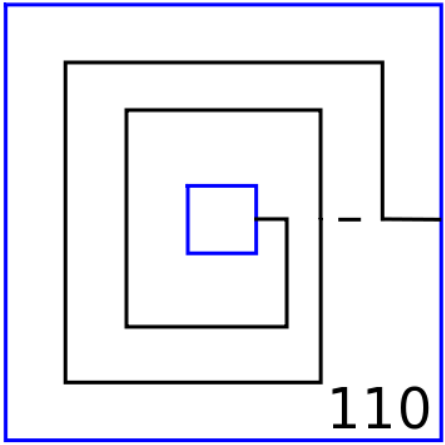} \emm \right)
=F^{110}_{110}
\Phi_\text{fix}
\left ( \bmm \includegraphics[scale=.30]{t011} \emm \right)
\end{align*}
We thus obtain the 4 by 4 $T$-matrix for the $\mathbb{Z}_2$ states
as follows:
\begin{align}
 T=
\bpm
F^{000}_{000}& 0& 0& 0& \\
0& F^{101}_{101}& 0& 0& \\
0& 0& 0& F^{110}_{110}& \\
0& 0& F^{011}_{011}& 0& \\
\epm .
\end{align}

To complete the story, we will also introduce the 
``$S$-transformation''. 
Unlike $T$-transformation which is generated by Dehn
twist on the torus,
$S$-transformation is generated by $90^{\circ}$ rotations. We define
the $90^{\circ}$ rotation
formally by mapping a
fixed-point state
$
\Phi_\text{fix}^{\al\bt}(i,j,k,\al,\bt)
=\Phi_\text{fix}^{\al\bt}
\left ( \bmm \includegraphics[scale=.35]{Ttrans1} \emm \right )
$
on a torus
to its counter-clockwise $90^{\circ}$ rotation state on a different graph
$
\t\Phi_\text{fix}^{\al\bt}(i,j,k,\al,\bt)
=\t\Phi_\text{fix}^{\al\bt}
\left ( \bmm \includegraphics[scale=.35]{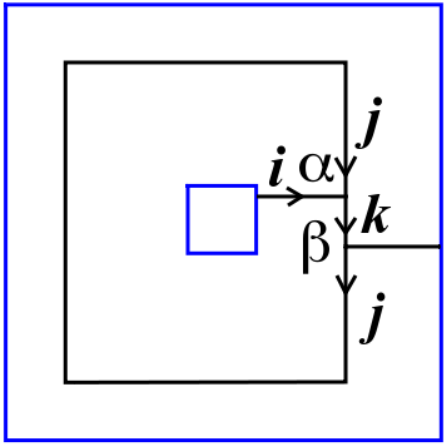} \emm \right )
$.
Then we can again use an $F$-move to deform the graph
$\bmm \includegraphics[scale=.35]{Strans2} \emm$
back to $\bmm \includegraphics[scale=.35]{Ttrans1} \emm$,
which leads to a unitary transformation $S$
between the four fixed-point states on torus.
This deformation process can be seen
more clearly through the following graph:
\begin{align}
\label{90rotate}
&\ \ \
\Phi_\text{fix}^{\al\bt}
\left ( \bmm \includegraphics[scale=.40]{Ttrans1} \emm \right )
\xrightarrow{90^{\circ} \ rotation}
\Phi_\text{fix}^{\al\bt}
\left ( \bmm \includegraphics[scale=.40]{Strans2} \emm \right )
\nonumber\\
&=
\sum_{l,\chi\del}
F^{ijk,\al\bt}_{i^*jl^*,\chi\del}
\Phi_\text{fix}^{\chi\del}
\left ( \bmm \includegraphics[scale=.40]{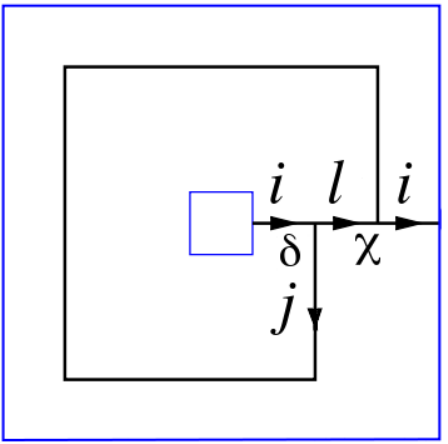} \emm \right )
\ \ (F-move)
\nonumber \\
&=
\sum_{l}
F^{ijk}_{i^*jl^*}
\Phi_\text{fix}
\left ( \bmm \includegraphics[scale=.40]{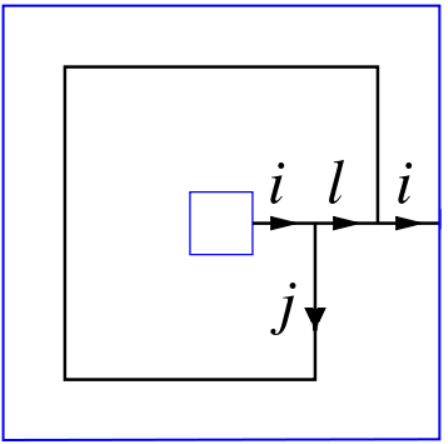} \emm
\right ) .
\end{align}
Again, we have suppressed the
vertex indices $\alpha$, $\beta$, $\delta$, $\chi$
in the last step above. 

Writing ``$S$-transformation''
in the same basis as
``$T$-transformation'',
we get a $4$ by $4$ unitary matrix
with elements $F^{ijk}_{i^*jl^*}$.
This matrix is called as the
``$S$-matrix''. We now apply
$90^{\circ}$ rotations to all the four non-contractable
states (recall Fig. \ref{t4}) to get the $S$-matrix: 
\begin{align*}
\Phi_\text{fix}
\left ( \bmm \includegraphics[scale=.30]{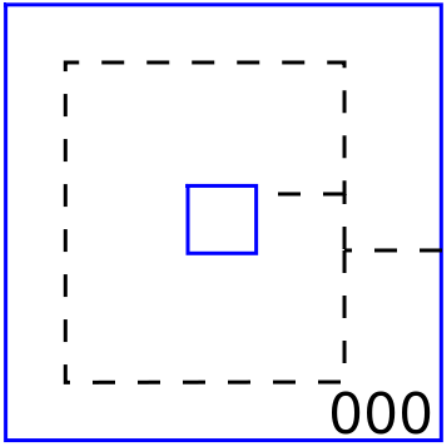} \emm \right)
=F^{000}_{000}
\Phi_\text{fix}
\left ( \bmm \includegraphics[scale=.30]{t000} \emm \right)
\end{align*}
\begin{align*}
\Phi_\text{fix}
\left ( \bmm \includegraphics[scale=.30]{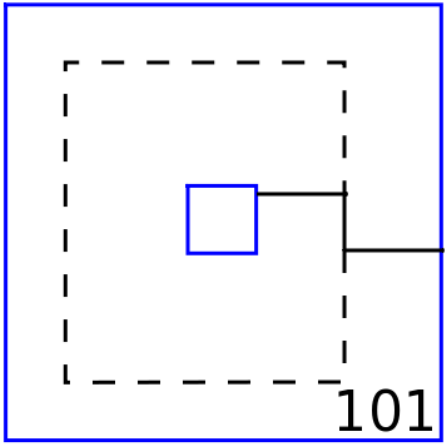} \emm \right)
=F^{101}_{101}
\Phi_\text{fix}
\left ( \bmm \includegraphics[scale=.30]{t011} \emm \right)
\end{align*}
\begin{align*}
\Phi_\text{fix}
\left ( \bmm \includegraphics[scale=.30]{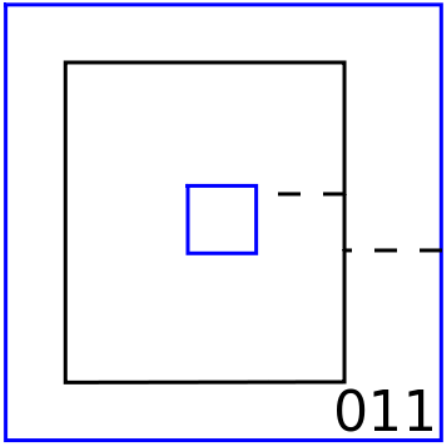} \emm \right)
=F^{011}_{011}
\Phi_\text{fix}
\left ( \bmm \includegraphics[scale=.30]{t101} \emm \right)
\end{align*}
\begin{align*}
\Phi_\text{fix}
\left ( \bmm \includegraphics[scale=.30]{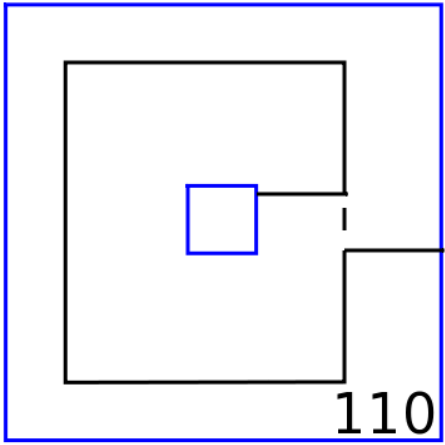} \emm \right)
=F^{110}_{110}
\Phi_\text{fix}
\left ( \bmm \includegraphics[scale=.30]{t110} \emm \right)
\end{align*}
We can thus obtain the 4 by 4 $S$-matrix
as follows:
\begin{align}
 S=
\bpm
F^{000}_{000}& 0& 0& 0& \\
0& 0& F^{011}_{011}& 0& \\
0& F^{101}_{101}& 0& 0& \\
0& 0& 0& F^{110}_{110}& \\
\epm .
\end{align}
 The $T$ and $S$-matrix obtained above
together give us the modular transformations.

When $\eta=1$, we have:
\begin{align}
 T=
\bpm
1& 0& 0& 0& \\
0& 1& 0& 0& \\
0& 0& 0& 1& \\
0& 0& 1& 0& \\
\epm
\end{align}
\begin{align}
 S=
\bpm
1& 0& 0& 0& \\
0& 0& 1& 0& \\
0& 1& 0& 0& \\
0& 0& 0& 1& \\
\epm
\end{align}
We can choose a new basis
to make the $T$ and $S$-matrices
be in a more standard form.
In this new basis, we require that
the $T$-matrix to be diagonalized
and at the same time,
$S$-matrix to satisfy the
following requirements:\cite{Wang10}
\begin{align}
\label{Sconditions}
1.&\ \ N_{ij}^k=\sum_{n} \frac{S_{n,i}S_{n,j}S^*_{n,k}}{S_{n,1}} \ \ 
\text{is a non-negative integer} , \nonumber\\
2.&\ \ S_{i,j}=S_{j,i} , \nonumber\\
3.&\ \ S_{1,i} > 0 ,\nonumber \\
4.&\ \ S^2=C,\ \ \ C^2=1,\ \ \ C|i\rangle =|i^*\rangle, \\
5.&\ \  
e^{\text{i} \sum_j A_{ij}\theta_j}
=e^{\text{i} \frac43 \sum_j A_{ij}\theta_i},\ \ \ A_{ij}=2N^j_{ii^*} N^i_{ij}
+N^j_{ii}N^i_{ji^*},
\nonumber 
\end{align}
where $C$ generates a permutation of the basis,
$C|i\rangle =|i^*\rangle$, that satisfy $i^{* *}=i$,
and $e^{\text{i}\theta_i}$ is the $i^\text{th}$ eigenvalue of $T$.
It is conjectured that the $T$ and 
$S$-matrices satisfying the above requirements
are unique (up to permutations of the basis),
and thus the basis of $T$ and $S$-matrices
are completely fixed through the above
requirements (up to permutations.)
Numerical simulations have been done in
several cases in the later sections of the paper,
all agreeing with the conjecture.

In the above, $S_{i,j}$ represents the $(i,j)$-th
matrix element of the $S$-matrix, and $S^*_{i,j}$
is its complex conjugate.
In Modular Tensor Category theory,
these are precisely the conditions
satisfied by the ``modular matrix'' $S$
and the ``diagonal twist'' matrix $T$;
thus the name for our $T$ and $S$-matrix.
The first condition above is the 
so-called Verlinde Formula; it gives the
fusion rule for the quasi-particle excitations.
(Note that this is different from the fusion rule
\ref{Z2fusion}, which is the fusion rule for fixed-point ground states.) 

Now, diagonalize the $T$-matrix and make the $S$-matrix
satisfy \ref{Sconditions}, we have:
\begin{align}
 T=
\bpm
1& 0& 0& 0& \\
0& 1& 0& 0& \\
0& 0& 1& 0& \\
0& 0& 0& -1& \\
\epm
\end{align}
\begin{align}
 S=
\frac{1}{2}
\bpm
1& 1& 1& 1& \\
1& 1& -1& -1& \\
1& -1& 1& -1& \\
1& -1& -1& 1& \\
\epm
\end{align}
Numerical simulations show that the above
$T$ and $S$-matrices are unique (up to
permutation of basis.)
The above $T$ and $S$-matrices match exactly with
the corresponding matrices in the ``Toric Code modular
tensor category'' in \Ref{RSW0777}.
From the dimension of the $T$ and $S$-matrices,
we can see that there are
four different types of quasi-particles.
From calculating \eqn{Sconditions}, we can get the
fusion rules between the quasi-particle excitations.
The quasi-particle statistical angles are given by the
eigenvalues of $T$-matrix.\cite{Wrig,FNS0428}
The quantum dimensions of these four quasi-particles
are given by the first row (column) of the $S$-matrix.

Note that the $ K=\bpm  0 & 2 \\
         2 & 0 \\ \epm $ $U(1)\times U(1)$
Chern-Simons (CS) theory also has 4 types
of quasiparticles with statistical angles
\begin{align*}
(\e^{\imth
\th_i})=(1,1,1,-1) .
\end{align*}
Therefore, the $\eta=1$ $\mathbb{Z}_2$ state can be described
by the above CS theory.

When $\eta=-1$, we have:
\begin{align}
 T=
\bpm
1& 0& 0& 0& \\
0& 1& 0& 0& \\
0& 0& 0& 1& \\
0& 0& -1& 0& \\
\epm
\end{align}
\begin{align}
 S=
\bpm
1& 0& 0& 0& \\
0& 0& 1& 0& \\
0& 1& 0& 0& \\
0& 0& 0& -1& \\
\epm
\end{align}
Again, we can diagonalize
the $T$-matrix and make the $S$-matrix
satisfy \ref{Sconditions} under
a proper basis; numerical simulations show that such
a basis is unique (up to permutations). In such a basis,
we have:
\begin{align}
 T=
\bpm
1& 0& 0& 0& \\
0& -i& 0& 0& \\
0& 0& i& 0& \\
0& 0& 0& 1& \\
\epm
\end{align}
\begin{align}
 S=
\frac{1}{2}
\bpm
1& 1& 1& 1& \\
1& -1& 1& -1& \\
1& 1& -1& -1& \\
1& -1& -1& 1& \\
\epm
\end{align}
In this particular case, the above
$T$ and $S$ matrices can be further reduced
to the following forms:
\begin{align}
 T=
\bpm
1& 0& \\
0& i& \\
\epm
\otimes
\bpm
1& 0& \\
0& -i& \\
\epm
\end{align}
\begin{align}
 S=
\frac{1}{\sqrt2}
\bpm
1& 1& \\
1& -1& \\
\epm
\otimes
\frac{1}{\sqrt2}
\bpm
1& 1& \\
1& -1& \\
\epm
\end{align}
which shows the ``doubled'' structure
of the quasi-particles.
The above $T$ and $S$-matrices match
exactly with the ``doubled Semion'' modular tensor
category in \Ref{RSW0777}. Again,
we can see that there are
four different types of quasi-particles
from the dimension of the $T$ and $S$-matrices.
From calculating \eqn{Sconditions}, we can get the
fusion rules between the quasi-particle excitations.
The quasi-particle statistics are given by the eigenvalues
of $T$-matrix, whereas the quantum dimensions
of different quasi-particles are given
by the elements on the first row of $S$-matrix.

Notice that the eigenvalues
of $T$-matrix again
match with
the $ K=\bpm  0 & 2 \\
         2 & 2 \\ \epm $ CS theory
(which is equivalent
to the $ K=\bpm  2 & 0 \\
         0 & -2 \\ \epm $ CS theory), which has 4 types
of quasiparticles with statistical angles
\begin{align*}
(\e^{\imth
\th_i})=(1,1,i,-i) .
\end{align*}
Also, the ``doubled'' structure hints that the effective
theory can be ``decoupled'', which is exactly what we
get here: $ K=\bpm  2 & 0 \\
         0 & -2 \\ \epm $ can indeed be decoupled.
Therefore, the $\eta=-1$ $\mathbb{Z}_2$ state can be described
by the  $ K=\bpm  0 & 2 \\
         2 & 2 \\ \epm $ CS theory.

\subsection{$N=1$ string-net state -- the ``Fibonacci'' state}
\label{Fibonacci}

To obtain another class of simple solutions, we modify the fusion
rule to
\begin{align}
& N_{000}= N_{110}= N_{101}= N_{011}= N_{111} =1,
\nonumber\\
& \text{other } N_{ijk}=0.
\end{align}
while keeping everything the same.
The above $N_{ijk}$ also
satisfies \eqn{Phinonz}.

The new fusion rule corresponds to the fusion rule for the
$N=1$ string-net state discussed in \Ref{LWstrnet}, thus 
the name $N=1$ string-net state. Notice the additional
fusion rule can be written as
$1 \bigotimes 1 = 0 \bigoplus 1$ which looks like Fibonacci's
golden rule, we also refer to the state as the ``Fibonacci State".

Again, due to the relation \eqn{Phinonz}, the
different components of the tensor $F^{ijm}_{kln}$ are not
independent.  Now there are seven independent potentially non-zero
components which are denoted as $f_0$,...,$f_6$:
\begin{align}
F^{000}_{000}
\bmm\begin{tikzpicture}[scale=0.26]
\FBox \iLnkaa \jLnkaa \kLnkaa \lLnkaa \mLnkaa \nLnkaa
\end{tikzpicture}\emm
&=f_{0}
\nonumber\\
F^{000}_{111}
\bmm\begin{tikzpicture}[scale=0.26]
\FBox \iLnkaa \jLnkaa \kLnkbb \lLnkbb \mLnkaa \nLnkbb
\end{tikzpicture}\emm
&=(F^{011}_{100}
\bmm\begin{tikzpicture}[scale=0.26]
\FBox \iLnkaa \jLnkbb \kLnkbb \lLnkaa \mLnkbb \nLnkaa
\end{tikzpicture}\emm
)^*=(F^{101}_{010}
\bmm\begin{tikzpicture}[scale=0.26]
\FBox \iLnkbb \jLnkaa \kLnkaa \lLnkbb \mLnkbb \nLnkaa
\end{tikzpicture}\emm
)^*
\nonumber\\
&=F^{110}_{001}
\bmm\begin{tikzpicture}[scale=0.26]
\FBox \iLnkbb \jLnkbb \kLnkaa \lLnkaa \mLnkaa \nLnkbb
\end{tikzpicture}\emm
=f_{1}
\nonumber\\
F^{011}_{011}
\bmm\begin{tikzpicture}[scale=0.26]
\FBox \iLnkaa \jLnkbb \kLnkaa \lLnkbb \mLnkbb \nLnkbb
\end{tikzpicture}\emm
&=(F^{101}_{101}
\bmm\begin{tikzpicture}[scale=0.26]
\FBox \iLnkbb \jLnkaa \kLnkbb \lLnkaa \mLnkbb \nLnkbb
\end{tikzpicture}\emm
)^*=f_{2}
\nonumber\\
F^{011}_{111}
\bmm\begin{tikzpicture}[scale=0.26]
\FBox \iLnkaa \jLnkbb \kLnkbb \lLnkbb \mLnkbb \nLnkbb
\end{tikzpicture}\emm
&=(F^{101}_{111}
\bmm\begin{tikzpicture}[scale=0.26]
\FBox \iLnkbb \jLnkaa \kLnkbb \lLnkbb \mLnkbb \nLnkbb
\end{tikzpicture}\emm
)^*=F^{111}_{011}
\bmm\begin{tikzpicture}[scale=0.26]
\FBox \iLnkbb \jLnkbb \kLnkaa \lLnkbb \mLnkbb \nLnkbb
\end{tikzpicture}\emm
\nonumber\\
&=(F^{111}_{101}
\bmm\begin{tikzpicture}[scale=0.26]
\FBox \iLnkbb \jLnkbb \kLnkbb \lLnkaa \mLnkbb \nLnkbb
\end{tikzpicture}\emm
)^*=f_{3}
\nonumber\\
F^{110}_{110}
\bmm\begin{tikzpicture}[scale=0.26]
\FBox \iLnkbb \jLnkbb \kLnkbb \lLnkbb \mLnkaa \nLnkaa
\end{tikzpicture}\emm
&=f_{4}
\nonumber\\
F^{110}_{111}
\bmm\begin{tikzpicture}[scale=0.26]
\FBox \iLnkbb \jLnkbb \kLnkbb \lLnkbb \mLnkaa \nLnkbb
\end{tikzpicture}\emm
&=(F^{111}_{110}
\bmm\begin{tikzpicture}[scale=0.26]
\FBox \iLnkbb \jLnkbb \kLnkbb \lLnkbb \mLnkbb \nLnkaa
\end{tikzpicture}\emm
)^*=f_{5}
\nonumber\\
F^{111}_{111}
\bmm\begin{tikzpicture}[scale=0.26]
\FBox \iLnkbb \jLnkbb \kLnkbb \lLnkbb \mLnkbb \nLnkbb
\end{tikzpicture}\emm
&=f_{6}
\end{align}
Note that $F$'s described by $f_1$ and $f_5$ are the only
$F$'s that change the number of $|1\>$-links and the number
of $|1\>|1\>|1\>$-vertices.  So we can use the local unitary
transformation $\e^{\imth (\th \hat M_1+\phi \hat M_{111})}$
to make $f_1$ and $f_5$ to be positive real numbers. (Here
$\hat M_1$ is the total number of $|1\>$-links and  $\hat
M_{111}$ is the total number of $|1\>|1\>|1\>$-vertices.) We
also use the freedom of adjusting the total sign of
$F^{ijm}_{kln}$ to make Re$(f_0)\geq 0$.

There are five potentially non-zero components
in $P^{kj}_i$, which are denoted by
$p_0$,...,$p_4$:
\begin{align}
P^{00}_{0} &=p_{0} ,
&
P^{01}_{0} &=p_{1} ,
&
P^{00}_{1} &=p_{2} ,
\nonumber\\
P^{01}_{1} &=p_{3} ,
&
P^{11}_{1} &=p_{4} .
\end{align}
We use the freedom of adjusting the total phase of
$P^{kj}_i$ to make $p_0$ to be a positive number.
We can also use the freedom of adjusting the total phase of
$A^i$ to make $A^0$ to be a positive number.

The fixed-point conditions (\eqn{Phinonz}) form a set of non-linear
equations on the variables $f_i$, $p_i$, and $A^i$, which can
be simplified.  The simplified equations have the following
form
\begin{align}
& f_0= f_1= f_2= f_3=1,\ \
f_4=f_5^2=-f_6>0,
\nonumber\\
&
p_1^2 f_4^2+p_1^2=1,\ \
p_0=f_4 p_1, \ \
p_2=p_0, \ \
p_3=p_1, \ \
p_4=0
\nonumber\\
&
A^0=f_4 A^1 ,\ \
(A^0)^2+(A^1)^2=1, \ \
f_4^2 + f_4 -1=0.
\end{align}
Let $\ga$ be the positive solution
of $\ga^2+\ga=1$:
$\ga=\frac{\sqrt{5}-1}{2}$.
We see that $f_5=\sqrt{\ga}$. The above can be written as
\begin{align}
& f_0= f_1= f_2= f_3=1,\ \ \
 f_4=-f_6=\ga, \ \ \
f_5=\sqrt{\ga},
\nonumber\\
&
p_0=p_2=\frac{\ga}{\sqrt{\ga^2+1}} , \ \ \
p_1 =p_3= \frac{1}{\sqrt{\ga^2+1}}, \ \ \
p_4=0,
\nonumber\\
&
 A^0=\frac{\ga}{\sqrt{\ga^2+1}},\ \ \ \
 A^1=\frac{1}{\sqrt{\ga^2+1}} .
\end{align}
We also find
%\begin{align}
%\Phi^\th_{000} &= p_0 A^0 =\frac{\ga^2}{(\ga^2+1)^2}
%\nonumber\\
%\Phi^\th_{011} &=\Phi^\th_{101}=\Phi^\th_{110}
%= p_1 A^0
%= \frac{\ga}{(\ga^2+1)^2},
%\nonumber\\
%\Phi^\th_{111} &= f_5 p_1 A^0 =\frac{\ga^{3/2}}{(\ga^2+1)^2} ,
%\nonumber\\
%& \text{other } \Phi^\th_{ikj}=0,
%\end{align}
%and
\begin{align}
 \e^{\imth \th_F}=
 \e^{\imth \th_{P1}}=
 \e^{\imth \th_{P2}}=
 \e^{\imth \th_{A1}}=
 \e^{\imth \th_{A2}}=
1.
\end{align}
The fixed-point state above 
corresponds to the $N=1$ string-net
condensed state\cite{LWstrnet} .

Following the logic from last section, we will
now apply the modular transformations.
Let us first consider
Dehn twist on a torus. Note that
we now formally 
have 5 possible non-contractable
graphs on a torus rather than 4, due to the additional
fusion rule. (Recall Fig. \ref{t4}; the additional
graph would be a graph with
all solid lines.)
Doing the Dehn twist for all 5 graphic
states using
\eqref{Dtwist}, we have:
\begin{align*}
\Phi_\text{fix}
\left ( \bmm \includegraphics[scale=.30]{tt000} \emm \right)
=F^{000}_{000}
\Phi_\text{fix}
\left ( \bmm \includegraphics[scale=.30]{t000} \emm \right)
\end{align*}
\begin{align*}
\Phi_\text{fix}
\left ( \bmm \includegraphics[scale=.30]{tt110} \emm \right)
&=F^{110}_{110}
\Phi_\text{fix}
\left ( \bmm \includegraphics[scale=.30]{t011} \emm \right)
\\
&+F^{110}_{111}
\Phi_\text{fix}
\left ( \bmm \includegraphics[scale=.30]{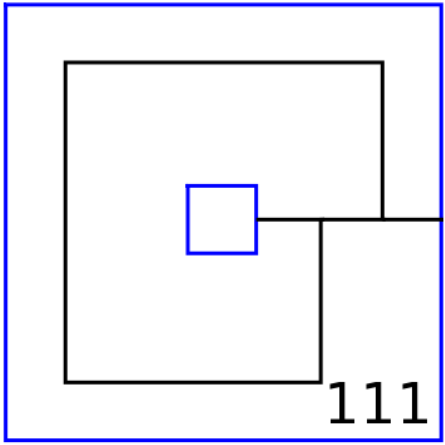} \emm \right)
\end{align*}
\begin{align*}
\Phi_\text{fix}
\left ( \bmm \includegraphics[scale=.30]{tt101} \emm \right)
=F^{101}_{101}
\Phi_\text{fix}
\left ( \bmm \includegraphics[scale=.30]{t101} \emm \right)
\end{align*}
\begin{align*}
\Phi_\text{fix}
\left ( \bmm \includegraphics[scale=.30]{tt011} \emm \right)
=F^{011}_{011}
\Phi_\text{fix}
\left ( \bmm \includegraphics[scale=.30]{t110} \emm \right)
\end{align*}
\begin{align*}
\Phi_\text{fix}
\left ( \bmm \includegraphics[scale=.30]{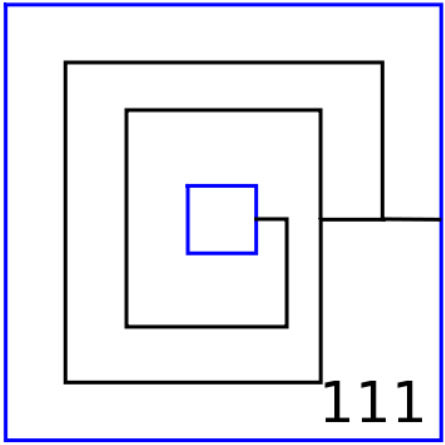} \emm \right)
&=F^{111}_{110}
\Phi_\text{fix}
\left ( \bmm \includegraphics[scale=.30]{t011} \emm \right)
\\
&+F^{111}_{111}
\Phi_\text{fix}
\left ( \bmm \includegraphics[scale=.30]{t111} \emm \right)
\end{align*}
We thus obtain the 5 by 5 $T$-matrix for the Fibonacci state
as follows:
\begin{align}
 T=
\bpm
F^{000}_{000}& 0& 0& 0& 0& \\
0& 0& 0& F^{011}_{011}& 0&\\
0& 0& F^{101}_{101}& 0& 0&\\
0& F^{110}_{110}& 0& 0& F^{111}_{110}&\\
0& F^{110}_{111}& 0& 0& F^{111}_{111}&
\epm
\end{align}

Similarly,we can also get the $S$-matrix
by applying $90^{\circ}$ rotations to all
five non-contractable states on a torus:
\begin{align*}
\Phi_\text{fix}
\left ( \bmm \includegraphics[scale=.30]{ss000} \emm \right)
=F^{000}_{000}
\Phi_\text{fix}
\left ( \bmm \includegraphics[scale=.30]{t000} \emm \right)
\end{align*}
\begin{align*}
\Phi_\text{fix}
\left ( \bmm \includegraphics[scale=.30]{ss110} \emm \right)
&=F^{110}_{110}
\Phi_\text{fix}
\left ( \bmm \includegraphics[scale=.30]{t110} \emm \right)
\\
&+F^{110}_{111}
\Phi_\text{fix}
\left ( \bmm \includegraphics[scale=.30]{t111} \emm \right)
\end{align*}
\begin{align*}
\Phi_\text{fix}
\left ( \bmm \includegraphics[scale=.30]{ss101} \emm \right)
=F^{101}_{101}
\Phi_\text{fix}
\left ( \bmm \includegraphics[scale=.30]{t011} \emm \right)
\end{align*}
\begin{align*}
\Phi_\text{fix}
\left ( \bmm \includegraphics[scale=.30]{ss011} \emm \right)
=F^{011}_{011}
\Phi_\text{fix}
\left ( \bmm \includegraphics[scale=.30]{t101} \emm \right)
\end{align*}
\begin{align*}
\Phi_\text{fix}
\left ( \bmm \includegraphics[scale=.30]{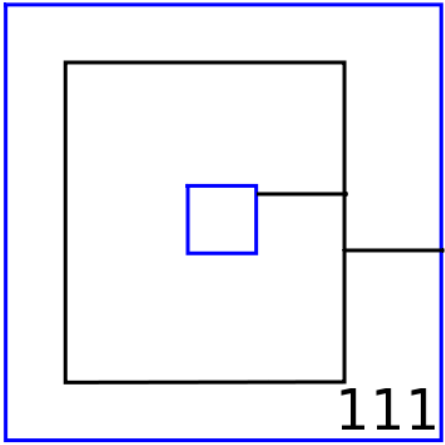} \emm \right)
&=F^{111}_{110}
\Phi_\text{fix}
\left ( \bmm \includegraphics[scale=.30]{t110} \emm \right)
\\
&+F^{111}_{111}
\Phi_\text{fix}
\left ( \bmm \includegraphics[scale=.30]{t111} \emm \right)
\end{align*}
From which we can get the 5 by 5 $S$-matrix for the Fibonacci state:
\begin{align}
 S=
\bpm
F^{000}_{000}& 0& 0& 0& 0& \\
0& F^{110}_{110}& 0& 0& F^{111}_{110}&\\
0& 0& 0& F^{011}_{011}& 0&\\
0& 0& F^{101}_{101}& 0& 0&\\
0& F^{110}_{111}& 0& 0& F^{111}_{111}&
\epm
\end{align}
Now we can substitute in the values of
the $F$ tensors, and thus obtain:
\begin{align}
 T=
\bpm
1& 0& 0& 0& 0& \\
0& 0& 0& 1& 0&\\
0& 0& 1& 0& 0&\\
0& \ga& 0& 0& \sqrt{\ga}&\\
0& \sqrt{\ga}& 0& 0& -\ga&
\epm
\end{align}
\begin{align}
 S=
\bpm
1& 0& 0& 0& 0& \\
0& \ga& 0& 0& \sqrt{\ga}&\\
0& 0& 0& 1& 0&\\
0& 0& 1& 0& 0&\\
0& \sqrt{\ga}& 0& 0& -\ga&
\epm
\end{align}
where $\ga=\frac{\sqrt{5}-1}{2}$.

There is, however, a small complication here compared to the previous
$\mathbb{Z}_2$ case. Although there are now five
possible non-contractable graphs on a torus, they are no longer
linearly independent (meaning they
can be transformed into each other
through $F$ or $P$-moves) and
are not all fixed-point graphs.
Thus they don't all correspond to ground states
of a local Hamiltonian. As can be checked using
the Hamiltonian construction in Appendix \ref{HamTorus}, 
the ground state subspace in this case is only $4$-fold degenerate.
We thus need to restrict our modular
transformations to be within this ground state subspace.
In order to get the $T$ and $S$-matrix within the ground states, we need to 
project the above obtained matrices onto its $4\times4$ ground-state subspace.
Such a process can be easily done by diagonalizing the Hamiltonian
and projecting the $T$ and $S$-matrices only to the ground-state subspace;
see Appendix \ref{HamTorus} for details. 

After the projection, we can again diagonalize
the $T$-matrix and make the $S$-matrix
satisfy \ref{Sconditions} under
a proper basis; numerical simulations again show that such
a basis is unique (up to permutations).
We thus have the final form of $T$ and $S$:
\begin{align}
 T=
\bpm
1& 0& 0& 0&\\
0& e^{i\frac{4\pi}{5}}& 0& 0&\\
0& 0& e^{-i\frac{4\pi}{5}}& 0&\\
0& 0& 0& 1&
\epm
\end{align}
\begin{align}
 S=
 \frac{5-\sqrt{5}}{10}
\bpm
1&  \frac{1+\sqrt5}{2}& \frac{1+\sqrt5}{2}& \frac{3+\sqrt5}{2} \\
\frac{1+\sqrt5}{2}&  -1& \frac{3+\sqrt5}{2}& \frac{-1-\sqrt5}{2} \\
\frac{1+\sqrt5}{2}& \frac{3+\sqrt5}{2}& -1& \frac{-1-\sqrt5}{2}\\
\frac{3+\sqrt5}{2}& \frac{-1-\sqrt5}{2}& \frac{-1-\sqrt5}{2}& 1
\epm
\end{align}
As can be easily seen, the above
$S$-matrix is real
and symmetric.
Also in this particular case, the above
$T$ and $S$ matrices can be further reduced
to the following forms:
\begin{align}
 T=
\bpm
1& 0& \\
0& e^{-i\frac{4\pi}{5}}& \\
\epm
\otimes
\bpm
1& 0& \\
0& e^{i\frac{4\pi}{5}}& \\
\epm
\end{align}
\begin{align}
 S=
\frac{5-\sqrt{5}}{10}
\bpm
1& \frac{1+\sqrt5}{2}& \\
\frac{1+\sqrt5}{2}& -1& \\
\epm
\otimes
\bpm
1& \frac{1+\sqrt5}{2}& \\
\frac{1+\sqrt5}{2}& -1& \\
\epm
\end{align}
which shows the ``doubled'' structure
of the quasi-particles.
The above $T$ and $S$-matrices match exactly
with the doubled ``Fibonacci" modular tensor category
in \Ref{RSW0777}. As before, 
we can tell there are four different types of
quasi-particles from the dimension of
$T$ and $S$; we can also get the
fusion rules between quasi-particle excitations
from calculating \eqn{Sconditions}.
The quasi-particle statistical angles are given by
the eigenvalues of $T$, whereas
the quantum dimensions of different quasi-particles
are given by the first row (column) of the $S$-matrix.

Following the last section, here we want to also
comment
on the eigenvalues
of $T$-matrix.
These eigenvalues
correspond exactly to the quasiparticle statistical angles
of the doubled $SO(3)$ Chern-Simons gauge
theory:
\begin{align*}
(\e^{\imth
\th_i})=(1,1,e^{i\frac{4\pi}{5}},e^{-i\frac{4\pi}{5}}) .
\end{align*}
Thus the ``Fibonacci'' state can be described by
the doubled $SO(3)$ CS theory.

\subsection{An $N=2$ string-net state -- the ``Pfaffian'' state}
\label{Pfaffian}

Here we will give a more complicated example of a non-orientable
string-net state. We choose $N=2$, $0^*=0$, $1^*=1$, $2^*=2$, and
\begin{align}
N_{000}&=
N_{011}=
N_{110}=
N_{101}=
N_{022}=
N_{202}=
N_{220}
\nonumber\\
&=
N_{112}=
N_{121}=
N_{211}=1.
\end{align}
The above $N_{ijk}$
satisfies \eqn{Phinonz}.

Similar to the $\mathbb{Z}_2$ case, the three edge states
of $N$ form a structure called a ``tensor category''
with the following fusion rules between its three elements:
$1 \bigotimes \sigma = \sigma$,
$1 \bigotimes \psi = \psi$,
$\psi \bigotimes \psi =1$,
$\sigma \bigotimes \sigma =1 \bigoplus \psi$.
In the above fusion rules, ``1'' represents state 0
on the edge, ``$\sigma$'' represents state 1
and ``$\psi$'' represents state 2. (For example,
$N_{112}$ corresponds to
$\sigma \bigotimes \sigma = \psi$.)
Since the state also corresponds to the ``Pfaffian"
state in the Ising model, we have the name --
``Pfaffian state".

Due to relation \eqn{Phinonz},
different components of the tensor $F^{ijm}_{kln}$ are not
independent.  There are fourteen independent potentially non-zero
components which are denoted as $f_0$,...,$f_{13}$:
\begin{align}
F^{000}_{000}
\bmm\begin{tikzpicture}[scale=0.26]
\FBox \iLnkaa \jLnkaa \kLnkaa \lLnkaa \mLnkaa \nLnkaa 
\end{tikzpicture}\emm
&=f_{0}
\nonumber\\
F^{000}_{111}
\bmm\begin{tikzpicture}[scale=0.26]
\FBox \iLnkaa \jLnkaa \kLnkbb \lLnkbb \mLnkaa \nLnkbb 
\end{tikzpicture}\emm
&=(F^{011}_{100}
\bmm\begin{tikzpicture}[scale=0.26]
\FBox \iLnkaa \jLnkbb \kLnkbb \lLnkaa \mLnkbb \nLnkaa 
\end{tikzpicture}\emm
)^*=(F^{101}_{010}
\bmm\begin{tikzpicture}[scale=0.26]
\FBox \iLnkbb \jLnkaa \kLnkaa \lLnkbb \mLnkbb \nLnkaa 
\end{tikzpicture}\emm
)^*
\nonumber\\
&=F^{110}_{001}
\bmm\begin{tikzpicture}[scale=0.26]
\FBox \iLnkbb \jLnkbb \kLnkaa \lLnkaa \mLnkaa \nLnkbb 
\end{tikzpicture}\emm
=f_{1}
\nonumber\\
F^{000}_{222}
\bmm\begin{tikzpicture}[scale=0.26]
\FBox \iLnkaa \jLnkaa \kLnkcc \lLnkcc \mLnkaa \nLnkcc 
\end{tikzpicture}\emm
&=(F^{022}_{200}
\bmm\begin{tikzpicture}[scale=0.26]
\FBox \iLnkaa \jLnkcc \kLnkcc \lLnkaa \mLnkcc \nLnkaa 
\end{tikzpicture}\emm
)^*=(F^{202}_{020}
\bmm\begin{tikzpicture}[scale=0.26]
\FBox \iLnkcc \jLnkaa \kLnkaa \lLnkcc \mLnkcc \nLnkaa 
\end{tikzpicture}\emm
)^*
\nonumber\\
&=F^{220}_{002}
\bmm\begin{tikzpicture}[scale=0.26]
\FBox \iLnkcc \jLnkcc \kLnkaa \lLnkaa \mLnkaa \nLnkcc 
\end{tikzpicture}\emm
=f_{2}
\nonumber\\
F^{011}_{011}
\bmm\begin{tikzpicture}[scale=0.26]
\FBox \iLnkaa \jLnkbb \kLnkaa \lLnkbb \mLnkbb \nLnkbb 
\end{tikzpicture}\emm
&=(F^{101}_{101}
\bmm\begin{tikzpicture}[scale=0.26]
\FBox \iLnkbb \jLnkaa \kLnkbb \lLnkaa \mLnkbb \nLnkbb 
\end{tikzpicture}\emm
)^*=f_{3}
\nonumber\\
F^{011}_{122}
\bmm\begin{tikzpicture}[scale=0.26]
\FBox \iLnkaa \jLnkbb \kLnkbb \lLnkcc \mLnkbb \nLnkcc 
\end{tikzpicture}\emm
&=(F^{112}_{201}
\bmm\begin{tikzpicture}[scale=0.26]
\FBox \iLnkbb \jLnkbb \kLnkcc \lLnkaa \mLnkcc \nLnkbb 
\end{tikzpicture}\emm
)^*=F^{121}_{012}
\bmm\begin{tikzpicture}[scale=0.26]
\FBox \iLnkbb \jLnkcc \kLnkaa \lLnkbb \mLnkbb \nLnkcc 
\end{tikzpicture}\emm
\nonumber\\
&=(F^{202}_{111}
\bmm\begin{tikzpicture}[scale=0.26]
\FBox \iLnkcc \jLnkaa \kLnkbb \lLnkbb \mLnkcc \nLnkbb 
\end{tikzpicture}\emm
)^*=f_{4}
\nonumber
\end{align}

\begin{align}
F^{011}_{211}
\bmm\begin{tikzpicture}[scale=0.26]
\FBox \iLnkaa \jLnkbb \kLnkcc \lLnkbb \mLnkbb \nLnkbb 
\end{tikzpicture}\emm
&=(F^{101}_{121}
\bmm\begin{tikzpicture}[scale=0.26]
\FBox \iLnkbb \jLnkaa \kLnkbb \lLnkcc \mLnkbb \nLnkbb 
\end{tikzpicture}\emm
)^*=(F^{121}_{101}
\bmm\begin{tikzpicture}[scale=0.26]
\FBox \iLnkbb \jLnkcc \kLnkbb \lLnkaa \mLnkbb \nLnkbb 
\end{tikzpicture}\emm
)^*
\nonumber\\
&=F^{211}_{011}
\bmm\begin{tikzpicture}[scale=0.26]
\FBox \iLnkcc \jLnkbb \kLnkaa \lLnkbb \mLnkbb \nLnkbb 
\end{tikzpicture}\emm
=f_{5}
\nonumber\\
F^{022}_{022}
\bmm\begin{tikzpicture}[scale=0.26]
\FBox \iLnkaa \jLnkcc \kLnkaa \lLnkcc \mLnkcc \nLnkcc 
\end{tikzpicture}\emm
&=(F^{202}_{202}
\bmm\begin{tikzpicture}[scale=0.26]
\FBox \iLnkcc \jLnkaa \kLnkcc \lLnkaa \mLnkcc \nLnkcc 
\end{tikzpicture}\emm
)^*=f_{6}
\nonumber\\
F^{022}_{111}
\bmm\begin{tikzpicture}[scale=0.26]
\FBox \iLnkaa \jLnkcc \kLnkbb \lLnkbb \mLnkcc \nLnkbb 
\end{tikzpicture}\emm
&=(F^{101}_{212}
\bmm\begin{tikzpicture}[scale=0.26]
\FBox \iLnkbb \jLnkaa \kLnkcc \lLnkbb \mLnkbb \nLnkcc 
\end{tikzpicture}\emm
)^*=F^{112}_{021}
\bmm\begin{tikzpicture}[scale=0.26]
\FBox \iLnkbb \jLnkbb \kLnkaa \lLnkcc \mLnkcc \nLnkbb 
\end{tikzpicture}\emm
\nonumber\\
&=(F^{211}_{102}
\bmm\begin{tikzpicture}[scale=0.26]
\FBox \iLnkcc \jLnkbb \kLnkbb \lLnkaa \mLnkbb \nLnkcc 
\end{tikzpicture}\emm
)^*=f_{7}
\nonumber\\
F^{110}_{110}
\bmm\begin{tikzpicture}[scale=0.26]
\FBox \iLnkbb \jLnkbb \kLnkbb \lLnkbb \mLnkaa \nLnkaa 
\end{tikzpicture}\emm
&=f_{8}
\nonumber\\
F^{110}_{112}
\bmm\begin{tikzpicture}[scale=0.26]
\FBox \iLnkbb \jLnkbb \kLnkbb \lLnkbb \mLnkaa \nLnkcc 
\end{tikzpicture}\emm
&=(F^{112}_{110}
\bmm\begin{tikzpicture}[scale=0.26]
\FBox \iLnkbb \jLnkbb \kLnkbb \lLnkbb \mLnkcc \nLnkaa 
\end{tikzpicture}\emm
)^*=f_{9}
\nonumber
\end{align}

\begin{align}
F^{112}_{112}
\bmm\begin{tikzpicture}[scale=0.26]
\FBox \iLnkbb \jLnkbb \kLnkbb \lLnkbb \mLnkcc \nLnkcc 
\end{tikzpicture}\emm
&=f_{10}
\nonumber\\
F^{110}_{221}
\bmm\begin{tikzpicture}[scale=0.26]
\FBox \iLnkbb \jLnkbb \kLnkcc \lLnkcc \mLnkaa \nLnkbb 
\end{tikzpicture}\emm
&=(F^{121}_{210}
\bmm\begin{tikzpicture}[scale=0.26]
\FBox \iLnkbb \jLnkcc \kLnkcc \lLnkbb \mLnkbb \nLnkaa 
\end{tikzpicture}\emm
)^*=(F^{211}_{120}
\bmm\begin{tikzpicture}[scale=0.26]
\FBox \iLnkcc \jLnkbb \kLnkbb \lLnkcc \mLnkbb \nLnkaa 
\end{tikzpicture}\emm
)^*
\nonumber\\
&=F^{220}_{111}
\bmm\begin{tikzpicture}[scale=0.26]
\FBox \iLnkcc \jLnkcc \kLnkbb \lLnkbb \mLnkaa \nLnkbb 
\end{tikzpicture}\emm
=f_{11}
\nonumber\\
F^{121}_{121}
\bmm\begin{tikzpicture}[scale=0.26]
\FBox \iLnkbb \jLnkcc \kLnkbb \lLnkcc \mLnkbb \nLnkbb 
\end{tikzpicture}\emm
&=(F^{211}_{211}
\bmm\begin{tikzpicture}[scale=0.26]
\FBox \iLnkcc \jLnkbb \kLnkcc \lLnkbb \mLnkbb \nLnkbb 
\end{tikzpicture}\emm
)^*=f_{12}
\nonumber\\
F^{220}_{220}
\bmm\begin{tikzpicture}[scale=0.26]
\FBox \iLnkcc \jLnkcc \kLnkcc \lLnkcc \mLnkaa \nLnkaa 
\end{tikzpicture}\emm
&=f_{13}
\end{align}
There are ten potentially non-zero components
in $P^{kj}_i$, which are denoted by
$p_0$,...,$p_9$:
\begin{align}
&
P^{00}_{0}=p_{0},\ \
P^{01}_{0}=p_{1},\ \
P^{02}_{0}=p_{2},\ \
P^{00}_{1}=p_{3},\ \
P^{01}_{1}=p_{4},
\nonumber\\
&
P^{21}_{1}=p_{5},\ \
P^{02}_{1}=p_{6},\ \
P^{00}_{2}=p_{7},\ \
P^{01}_{2}=p_{8},\ \
P^{02}_{2}=p_{9}.
\end{align}
Using the ``gauge fixing'' discussed in section
\Ref{CGW1038}, we can consistently fix the phases
of $f_1$, $f_2$, $f_4$, $f_7$, $f_9$, $f_{11}$,
$p_0$ and $A^0$ to make them positive. (Note that
this `gauge fixing' is non-trivial, since these
phases are not completely independent;
for e.g. the phases of $f_4$, $f_7$ are complex
conjugates of each other.)

The fixed-point conditions (\eqn{Phinonz}) form a set of non-linear
equations on the variables $f_i$, $p_i$, and $A^i$, which can
be solved exactly. After applying the ``gauge fixing''
discussed above, we find
only one solution
\begin{align}
&
f_0=f_1=...=f_7=f_{11}=f_{13}=1,\ \
\nonumber\\
&
f_8=f_9=-f_{10}=\frac{1}{\sqrt{2}},\ \ f_{12}=-1,
\nonumber\\
&
p_1=p_4=p_8=\frac{1}{\sqrt 2},\ \ p_5=0,
\nonumber\\
&
p_0=p_2=p_3=p_6=p_7=p_9=\frac{1}{2},
\nonumber\\
&
A^0=A^2=\frac{1}{2},\ \ A^1=\frac{1}{\sqrt 2}.
\end{align}
We also find
\begin{align}
 \e^{\imth \th_F}=
 \e^{\imth \th_{P1}}=
 \e^{\imth \th_{P2}}=
 \e^{\imth \th_{A1}}=
 \e^{\imth \th_{A2}}=
1.
\end{align}

Now we are ready to apply the Modular transformations. First we
try to get the $T$-matrix by applying
the Dehn twists. In the ``Pfaffian''
case we have altogether $10$ non-contractable
graphs on a torus as can be seen
in below.
Applying Dehn twists to all of them using
\eqref{Dtwist}, we have:
\begin{align*}
\Phi_\text{fix}
\left ( \bmm \includegraphics[scale=.30]{tt000} \emm \right)
=F^{000}_{000}
\Phi_\text{fix}
\left ( \bmm \includegraphics[scale=.30]{t000} \emm \right)
\end{align*}
\begin{align*}
\Phi_\text{fix}
\left ( \bmm \includegraphics[scale=.30]{tt110} \emm \right)
&=F^{110}_{110}
\Phi_\text{fix}
\left ( \bmm \includegraphics[scale=.30]{t011} \emm \right)
\\
&+F^{110}_{112}
\Phi_\text{fix}
\left ( \bmm \includegraphics[scale=.30]{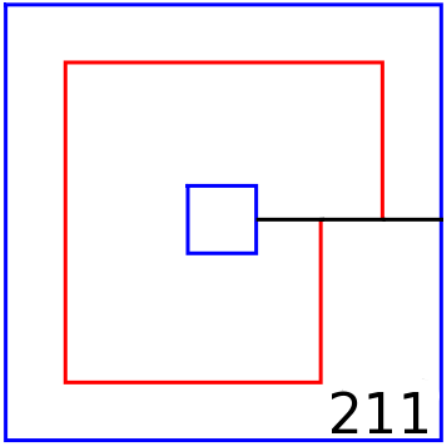} \emm \right)
\end{align*}
\begin{align*}
\Phi_\text{fix}
\left ( \bmm \includegraphics[scale=.30]{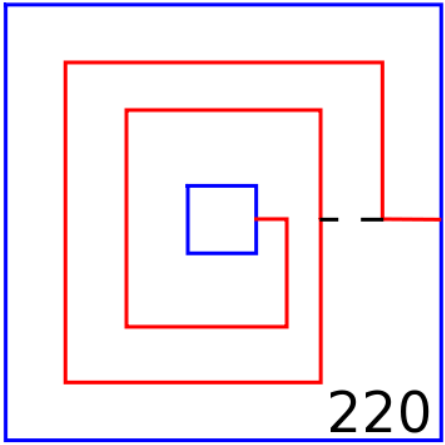} \emm \right)
=F^{220}_{220}
\Phi_\text{fix}
\left ( \bmm \includegraphics[scale=.30]{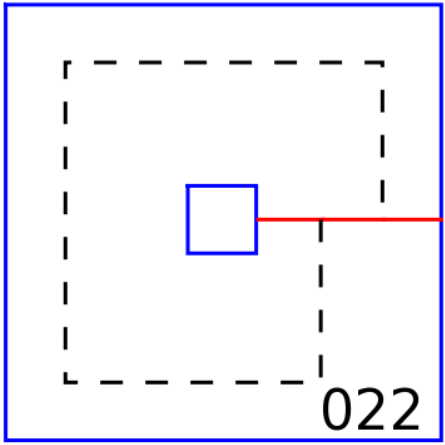} \emm \right)
\end{align*}
\begin{align*}
\Phi_\text{fix}
\left ( \bmm \includegraphics[scale=.30]{tt101} \emm \right)
=F^{101}_{101}
\Phi_\text{fix}
\left ( \bmm \includegraphics[scale=.30]{t101} \emm \right)
\end{align*}
\begin{align*}
\Phi_\text{fix}
\left ( \bmm \includegraphics[scale=.30]{tt011} \emm \right)
=F^{011}_{011}
\Phi_\text{fix}
\left ( \bmm \includegraphics[scale=.30]{t110} \emm \right)
\end{align*}
\begin{align*}
\Phi_\text{fix}
\left ( \bmm \includegraphics[scale=.30]{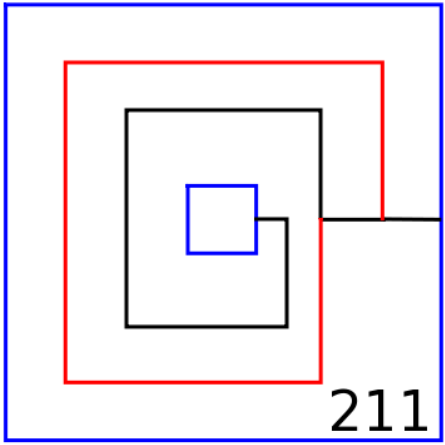} \emm \right)
=F^{211}_{211}
\Phi_\text{fix}
\left ( \bmm \includegraphics[scale=.30]{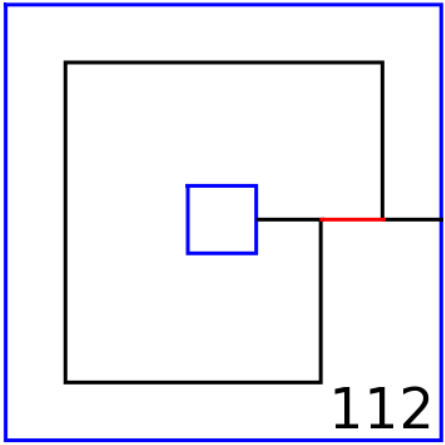} \emm \right)
\end{align*}
\begin{align*}
\Phi_\text{fix}
\left ( \bmm \includegraphics[scale=.30]{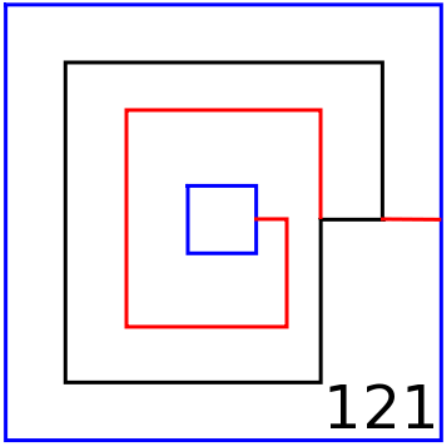} \emm \right)
=F^{121}_{121}
\Phi_\text{fix}
\left ( \bmm \includegraphics[scale=.30]{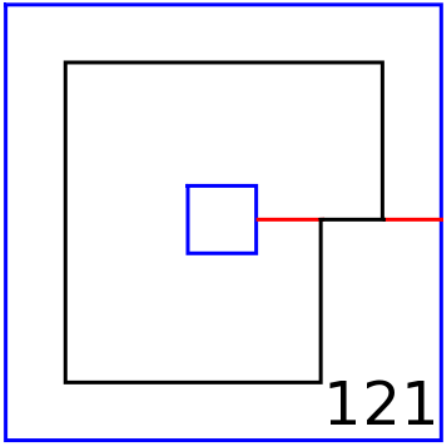} \emm \right)
\end{align*}
\begin{align*}
\Phi_\text{fix}
\left ( \bmm \includegraphics[scale=.30]{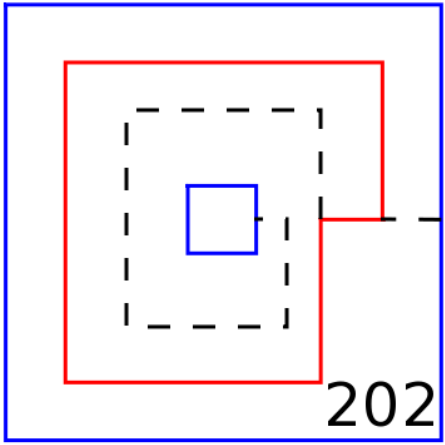} \emm \right)
=F^{202}_{202}
\Phi_\text{fix}
\left ( \bmm \includegraphics[scale=.30]{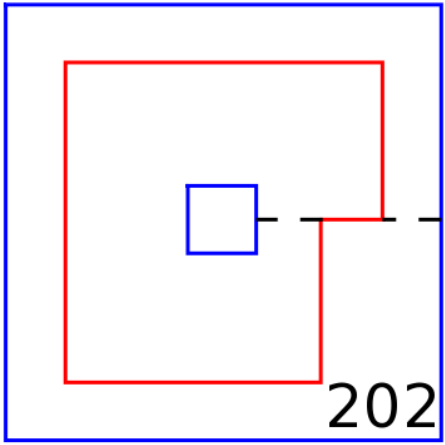} \emm \right)
\end{align*}
\begin{align*}
\Phi_\text{fix}
\left ( \bmm \includegraphics[scale=.30]{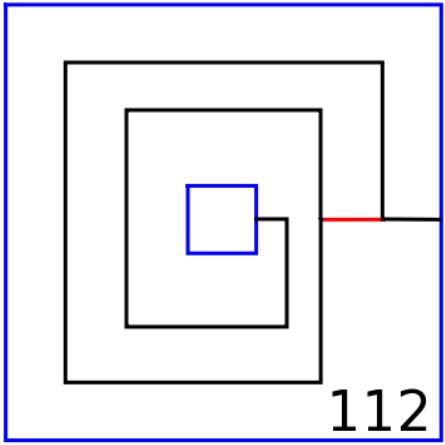} \emm \right)
&=F^{112}_{112}
\Phi_\text{fix}
\left ( \bmm \includegraphics[scale=.30]{t211} \emm \right)
\\
&+F^{112}_{110}
\Phi_\text{fix}
\left ( \bmm \includegraphics[scale=.30]{t011} \emm \right)
\end{align*}
\begin{align*}
\Phi_\text{fix}
\left ( \bmm \includegraphics[scale=.30]{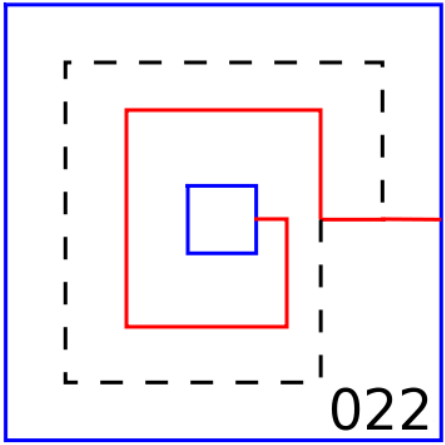} \emm \right)
=F^{022}_{022}
\Phi_\text{fix}
\left ( \bmm \includegraphics[scale=.30]{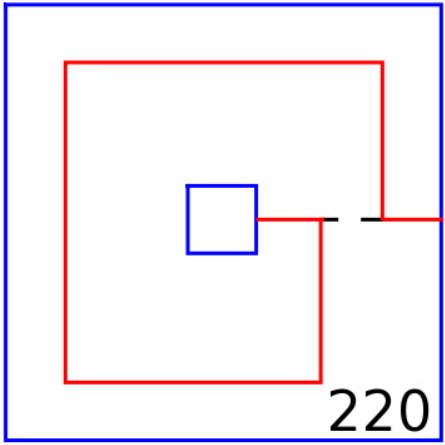} \emm \right)
\end{align*}
We thus obtain the $10$ by $10$ $T$-matrix for the Pfaffian
state:
\begin{align}
T=
\bpm
1& 0& 0& 0& 0& 0& 0& 0& 0& 0\\
0& 0& 0& 0& 1& 0& 0& 0& 0& 0\\
0& 0& 0& 0& 0& 0& 0& 0& 0& 1\\
0& 0& 0& 1& 0& 0& 0& 0& 0& 0\\
0& \frac{1}{\sqrt{2}}& 0& 0& 0& 0& 0& 0& \frac{1}{\sqrt{2}}& 0\\
0& \frac{1}{\sqrt{2}}& 0& 0& 0& 0& 0& 0& -\frac{1}{\sqrt{2}}& 0\\
0& 0& 0& 0& 0& 0& -1& 0& 0& 0\\
0& 0& 0& 0& 0& 0& 0& 1& 0& 0\\
0& 0& 0& 0& 0& -1& 0& 0& 0& 0\\
0& 0& 1& 0& 0& 0& 0& 0& 0& 0
\epm
\end{align}
Similarly, we can apply the $90^{\circ}$ rotations and obtain the
$S$-matrix. Without going into to much details, we give here
the final result (under the same basis):
\begin{align}
S=
\bpm
1& 0& 0& 0& 0& 0& 0& 0& 0& 0\\
0& \frac{1}{\sqrt{2}}& 0& 0& 0& 0& 0& 0& \frac{1}{\sqrt{2}}& 0\\
0& 0& 1& 0& 0& 0& 0& 0& 0& 0\\
0& 0& 0& 0& 1& 0& 0& 0& 0& 0\\
0& 0& 0& 1& 0& 0& 0& 0& 0& 0\\
0& 0& 0& 0& 0& 0& -1& 0& 0& 0\\
0& 0& 0& 0& 0& -1& 0& 0& 0& 0\\
0& 0& 0& 0& 0& 0& 0& 0& 0& 1\\
0& \frac{1}{\sqrt{2}}& 0& 0& 0& 0& 0& 0& -\frac{1}{\sqrt{2}}& 0\\
0& 0& 0& 0& 0& 0& 0& 1& 0& 0
\epm.
\end{align}

As in the ``Fibonacci'' case, 
the above $10$ non-contractable graphs
are not all fixed-point graphs and do not all
correspond to ground states of a local Hamiltonian.
Using the Hamiltonian construction in
Appendix \ref{HamTorus},
it's easy to check that the true ground state subspace
is actually $9$-fold;
we can thus project the above
matrices onto its ground state
subspace. By carefully choosing the basis,
we can again make the resulting
$9 \times 9$ $T$-matrix diagonalized and
the $9 \times 9$
$S$-matrix satisfy \ref{Sconditions}
at the same time, as shown below. 
Numerical simulations again show that such
a basis is unique (up to permutations.)
\begin{align}
T=
\bpm
1& 0& 0& 0& 0& 0& 0& 0& 0\\
0& e^{-{i\frac{\pi}{8}}}& 0& 0& 0& 0& 0& 0& 0\\
0& 0& -1& 0& 0& 0& 0& 0& 0\\
0& 0& 0& e^{{i\frac{\pi}{8}}}& 0& 0& 0& 0& 0\\
0& 0& 0& 0& 1& 0& 0& 0& 0\\
0& 0& 0& 0& 0& e^{-{i\frac{7\pi}{8}}}& 0& 0& 0\\
0& 0& 0& 0& 0& 0& -1& 0& 0\\
0& 0& 0& 0& 0& 0& 0& e^{{i\frac{7\pi}{8}}}& 0\\
0& 0& 0& 0& 0& 0& 0& 0& 1
\epm
\end{align}
\small
\setlength{\arraycolsep}{2pt}
\begin{align}
S=\frac{1}{4}
\bpm
1& \sqrt{2}& 1& \sqrt{2}& 2& \sqrt{2}& 1& \sqrt{2}& 1\\
\sqrt{2}& 0& -\sqrt{2}& 2& 0& -2& \sqrt{2}& 0& -\sqrt{2}\\
1& -\sqrt{2}& 1& \sqrt{2}& -2& \sqrt{2}& 1& -\sqrt{2}& 1\\
\sqrt{2}& 2& \sqrt{2}& 0& 0& 0& -\sqrt{2}& -2& -\sqrt{2}\\
2& 0& -2& 0& 0& 0& -2& 0& 2\\
\sqrt{2}& -2& \sqrt{2}& 0& 0& 0& -\sqrt{2}& 2& -\sqrt{2}\\
1& \sqrt{2}& 1& -\sqrt{2}& -2& -\sqrt{2}& 1& \sqrt{2}& 1\\
\sqrt{2}& 0& -\sqrt{2}& -2& 0& 2& \sqrt{2}& 0& -\sqrt{2}\\
1& -\sqrt{2}& 1& -\sqrt{2}& 2& -\sqrt{2}& 1& -\sqrt{2}& 1
\epm.
\end{align}
\setlength{\arraycolsep}{2pt}
\normalsize
The above $S$-matrix is again real and symmetric.
Also in this particular case, the above
$T$ and $S$ matrices can be further reduced
to the following forms:
\begin{align}
 T=
\bpm
1& 0& 0&\\
0& e^{i\frac{\pi}{8}}& 0&\\
0& 0& -1&\\
\epm
\otimes
\bpm
1& 0& 0&\\
0& e^{-i\frac{\pi}{8}}& 0&\\
0& 0& -1&\\
\epm
\end{align}
\begin{align}
 S=
\frac{1}{2}
\bpm
1& \sqrt2& 1& \\
\sqrt2& 0& -\sqrt2& \\
1& -\sqrt2& 1& \\
\epm
\otimes
\frac{1}{2}
\bpm
1& \sqrt2& 1& \\
\sqrt2& 0& -\sqrt2& \\
1& -\sqrt2& 1& \\
\epm
\end{align}
which shows the ``doubled'' structure
of the quasi-particles.
The above $T$ and $S$-matrices match exactly
with the doubled ``Ising'' modular tensor category
in \Ref{RSW0777}. As before,
we can tell that there are 9 types of different 
quasi-particles from the dimension of $T$ and $S$
and through calculating
the Verlinde Formula \ref{Sconditions}, we can get 
the fusion rule between them. The quasi-particle
statistics can be obtained from the eigenvalues of $T$-matrix
and the quantum dimensions of these quasiparticles
are given by the elements
on the fist row of $S$-matrix.

\subsection{Another $N=2$ string-net state -- the $S_3$ state}

By adding an additional fusion rule on top of the previous result, we can get
yet another example of a non-orientable string-net state. Here we still 
choose $N=2$, $0^*=0$, $1^*=1$, $2^*=2$, and
\begin{align}
N_{000}&=
N_{011}=
N_{110}=
N_{101}=
N_{022}=
N_{202}=
N_{220}
\nonumber\\
&=
N_{111}=
N_{112}=
N_{121}=
N_{211}=1.
\end{align}
The above $N_{ijk}$
satisfies \eqn{Phinonz}.

These fusion rules correspond to the $N=2$ string-net
model with non-orientable strings in \Ref{LWstrnet}.
Following that paper, we will also
call this state the $S_3$ state. As can be seen in the
following, after gauge fixing, the fixed-point
solutions completely agree with the ``local rules'' obtained in
\Ref{LWstrnet}, thus we re-obtain another string-net
state from local unitary transformation point of view.
(Note that edge labels 1 and 2 are reversed in
\Ref{LWstrnet} compared to this paper.)

Due to relation \eqn{Phinonz},
different components of the tensor $F^{ijm}_{kln}$ are not
independent.  There are fourteen independent potentially non-zero
components which are denoted as $f_0$,...,$f_{18}$:
\begin{align}
F^{000}_{000}
\bmm\begin{tikzpicture}[scale=0.26]
\FBox \iLnkaa \jLnkaa \kLnkaa \lLnkaa \mLnkaa \nLnkaa 
\end{tikzpicture}\emm
&=f_{0}
\nonumber\\
F^{000}_{111}
\bmm\begin{tikzpicture}[scale=0.26]
\FBox \iLnkaa \jLnkaa \kLnkbb \lLnkbb \mLnkaa \nLnkbb 
\end{tikzpicture}\emm
&=(F^{011}_{100}
\bmm\begin{tikzpicture}[scale=0.26]
\FBox \iLnkaa \jLnkbb \kLnkbb \lLnkaa \mLnkbb \nLnkaa 
\end{tikzpicture}\emm
)^*=(F^{101}_{010}
\bmm\begin{tikzpicture}[scale=0.26]
\FBox \iLnkbb \jLnkaa \kLnkaa \lLnkbb \mLnkbb \nLnkaa 
\end{tikzpicture}\emm
)^*
\nonumber\\
&=F^{110}_{001}
\bmm\begin{tikzpicture}[scale=0.26]
\FBox \iLnkbb \jLnkbb \kLnkaa \lLnkaa \mLnkaa \nLnkbb 
\end{tikzpicture}\emm
=f_{1}
\nonumber\\
F^{000}_{222}
\bmm\begin{tikzpicture}[scale=0.26]
\FBox \iLnkaa \jLnkaa \kLnkcc \lLnkcc \mLnkaa \nLnkcc 
\end{tikzpicture}\emm
&=(F^{022}_{200}
\bmm\begin{tikzpicture}[scale=0.26]
\FBox \iLnkaa \jLnkcc \kLnkcc \lLnkaa \mLnkcc \nLnkaa 
\end{tikzpicture}\emm
)^*=(F^{202}_{020}
\bmm\begin{tikzpicture}[scale=0.26]
\FBox \iLnkcc \jLnkaa \kLnkaa \lLnkcc \mLnkcc \nLnkaa 
\end{tikzpicture}\emm
)^*
\nonumber\\
&=F^{220}_{002}
\bmm\begin{tikzpicture}[scale=0.26]
\FBox \iLnkcc \jLnkcc \kLnkaa \lLnkaa \mLnkaa \nLnkcc 
\end{tikzpicture}\emm
=f_{2}
\nonumber\\
F^{011}_{011}
\bmm\begin{tikzpicture}[scale=0.26]
\FBox \iLnkaa \jLnkbb \kLnkaa \lLnkbb \mLnkbb \nLnkbb 
\end{tikzpicture}\emm
&=(F^{101}_{101}
\bmm\begin{tikzpicture}[scale=0.26]
\FBox \iLnkbb \jLnkaa \kLnkbb \lLnkaa \mLnkbb \nLnkbb 
\end{tikzpicture}\emm
)^*=f_{3}
\nonumber\\
F^{011}_{111}
\bmm\begin{tikzpicture}[scale=0.26]
\FBox \iLnkaa \jLnkbb \kLnkbb \lLnkbb \mLnkbb \nLnkbb 
\end{tikzpicture}\emm
&=(F^{101}_{111}
\bmm\begin{tikzpicture}[scale=0.26]
\FBox \iLnkbb \jLnkaa \kLnkbb \lLnkbb \mLnkbb \nLnkbb 
\end{tikzpicture}\emm
)^*=F^{111}_{011}
\bmm\begin{tikzpicture}[scale=0.26]
\FBox \iLnkbb \jLnkbb \kLnkaa \lLnkbb \mLnkbb \nLnkbb 
\end{tikzpicture}\emm
\nonumber\\
&=(F^{111}_{101}
\bmm\begin{tikzpicture}[scale=0.26]
\FBox \iLnkbb \jLnkbb \kLnkbb \lLnkaa \mLnkbb \nLnkbb 
\end{tikzpicture}\emm
)^*=f_{4}
\nonumber\\
F^{011}_{122}
\bmm\begin{tikzpicture}[scale=0.26]
\FBox \iLnkaa \jLnkbb \kLnkbb \lLnkcc \mLnkbb \nLnkcc 
\end{tikzpicture}\emm
&=(F^{112}_{201}
\bmm\begin{tikzpicture}[scale=0.26]
\FBox \iLnkbb \jLnkbb \kLnkcc \lLnkaa \mLnkcc \nLnkbb 
\end{tikzpicture}\emm
)^*=F^{121}_{012}
\bmm\begin{tikzpicture}[scale=0.26]
\FBox \iLnkbb \jLnkcc \kLnkaa \lLnkbb \mLnkbb \nLnkcc 
\end{tikzpicture}\emm
\nonumber\\
&=(F^{202}_{111}
\bmm\begin{tikzpicture}[scale=0.26]
\FBox \iLnkcc \jLnkaa \kLnkbb \lLnkbb \mLnkcc \nLnkbb 
\end{tikzpicture}\emm
)^*=f_{5}
\nonumber
\end{align}

\begin{align}
F^{011}_{211}
\bmm\begin{tikzpicture}[scale=0.26]
\FBox \iLnkaa \jLnkbb \kLnkcc \lLnkbb \mLnkbb \nLnkbb 
\end{tikzpicture}\emm
&=(F^{101}_{121}
\bmm\begin{tikzpicture}[scale=0.26]
\FBox \iLnkbb \jLnkaa \kLnkbb \lLnkcc \mLnkbb \nLnkbb 
\end{tikzpicture}\emm
)^*=(F^{121}_{101}
\bmm\begin{tikzpicture}[scale=0.26]
\FBox \iLnkbb \jLnkcc \kLnkbb \lLnkaa \mLnkbb \nLnkbb 
\end{tikzpicture}\emm
)^*
\nonumber\\
&=F^{211}_{011}
\bmm\begin{tikzpicture}[scale=0.26]
\FBox \iLnkcc \jLnkbb \kLnkaa \lLnkbb \mLnkbb \nLnkbb 
\end{tikzpicture}\emm
=f_{6}
\nonumber\\
F^{022}_{022}
\bmm\begin{tikzpicture}[scale=0.26]
\FBox \iLnkaa \jLnkcc \kLnkaa \lLnkcc \mLnkcc \nLnkcc 
\end{tikzpicture}\emm
&=(F^{202}_{202}
\bmm\begin{tikzpicture}[scale=0.26]
\FBox \iLnkcc \jLnkaa \kLnkcc \lLnkaa \mLnkcc \nLnkcc 
\end{tikzpicture}\emm
)^*=f_{7}
\nonumber\\
F^{022}_{111}
\bmm\begin{tikzpicture}[scale=0.26]
\FBox \iLnkaa \jLnkcc \kLnkbb \lLnkbb \mLnkcc \nLnkbb 
\end{tikzpicture}\emm
&=(F^{101}_{212}
\bmm\begin{tikzpicture}[scale=0.26]
\FBox \iLnkbb \jLnkaa \kLnkcc \lLnkbb \mLnkbb \nLnkcc 
\end{tikzpicture}\emm
)^*=F^{112}_{021}
\bmm\begin{tikzpicture}[scale=0.26]
\FBox \iLnkbb \jLnkbb \kLnkaa \lLnkcc \mLnkcc \nLnkbb 
\end{tikzpicture}\emm
\nonumber\\
&=(F^{211}_{102}
\bmm\begin{tikzpicture}[scale=0.26]
\FBox \iLnkcc \jLnkbb \kLnkbb \lLnkaa \mLnkbb \nLnkcc 
\end{tikzpicture}\emm
)^*=f_{8}
\nonumber\\
F^{110}_{110}
\bmm\begin{tikzpicture}[scale=0.26]
\FBox \iLnkbb \jLnkbb \kLnkbb \lLnkbb \mLnkaa \nLnkaa 
\end{tikzpicture}\emm
&=f_{9}
\nonumber\\
F^{110}_{111}
\bmm\begin{tikzpicture}[scale=0.26]
\FBox \iLnkbb \jLnkbb \kLnkbb \lLnkbb \mLnkaa \nLnkbb 
\end{tikzpicture}\emm
&=(F^{111}_{110}
\bmm\begin{tikzpicture}[scale=0.26]
\FBox \iLnkbb \jLnkbb \kLnkbb \lLnkbb \mLnkbb \nLnkaa 
\end{tikzpicture}\emm
)^*=f_{10}
\nonumber\\
F^{110}_{112}
\bmm\begin{tikzpicture}[scale=0.26]
\FBox \iLnkbb \jLnkbb \kLnkbb \lLnkbb \mLnkaa \nLnkcc 
\end{tikzpicture}\emm
&=(F^{112}_{110}
\bmm\begin{tikzpicture}[scale=0.26]
\FBox \iLnkbb \jLnkbb \kLnkbb \lLnkbb \mLnkcc \nLnkaa 
\end{tikzpicture}\emm
)^*=f_{11}
\nonumber
\end{align}

\begin{align}
F^{111}_{111}
\bmm\begin{tikzpicture}[scale=0.26]
\FBox \iLnkbb \jLnkbb \kLnkbb \lLnkbb \mLnkbb \nLnkbb 
\end{tikzpicture}\emm
&=f_{12}
\nonumber\\
F^{111}_{112}
\bmm\begin{tikzpicture}[scale=0.26]
\FBox \iLnkbb \jLnkbb \kLnkbb \lLnkbb \mLnkbb \nLnkcc 
\end{tikzpicture}\emm
&=(F^{112}_{111}
\bmm\begin{tikzpicture}[scale=0.26]
\FBox \iLnkbb \jLnkbb \kLnkbb \lLnkbb \mLnkcc \nLnkbb 
\end{tikzpicture}\emm
)^*=f_{13}
\nonumber\\
F^{112}_{112}
\bmm\begin{tikzpicture}[scale=0.26]
\FBox \iLnkbb \jLnkbb \kLnkbb \lLnkbb \mLnkcc \nLnkcc 
\end{tikzpicture}\emm
&=f_{14}
\nonumber\\
F^{111}_{121}
\bmm\begin{tikzpicture}[scale=0.26]
\FBox \iLnkbb \jLnkbb \kLnkbb \lLnkcc \mLnkbb \nLnkbb 
\end{tikzpicture}\emm
&=(F^{111}_{211}
\bmm\begin{tikzpicture}[scale=0.26]
\FBox \iLnkbb \jLnkbb \kLnkcc \lLnkbb \mLnkbb \nLnkbb 
\end{tikzpicture}\emm
)^*=F^{121}_{111}
\bmm\begin{tikzpicture}[scale=0.26]
\FBox \iLnkbb \jLnkcc \kLnkbb \lLnkbb \mLnkbb \nLnkbb 
\end{tikzpicture}\emm
\nonumber\\
&=(F^{211}_{111}
\bmm\begin{tikzpicture}[scale=0.26]
\FBox \iLnkcc \jLnkbb \kLnkbb \lLnkbb \mLnkbb \nLnkbb 
\end{tikzpicture}\emm
)^*=f_{15}
\nonumber\\
F^{110}_{221}
\bmm\begin{tikzpicture}[scale=0.26]
\FBox \iLnkbb \jLnkbb \kLnkcc \lLnkcc \mLnkaa \nLnkbb 
\end{tikzpicture}\emm
&=(F^{121}_{210}
\bmm\begin{tikzpicture}[scale=0.26]
\FBox \iLnkbb \jLnkcc \kLnkcc \lLnkbb \mLnkbb \nLnkaa 
\end{tikzpicture}\emm
)^*=(F^{211}_{120}
\bmm\begin{tikzpicture}[scale=0.26]
\FBox \iLnkcc \jLnkbb \kLnkbb \lLnkcc \mLnkbb \nLnkaa 
\end{tikzpicture}\emm
)^*
\nonumber\\
&=F^{220}_{111}
\bmm\begin{tikzpicture}[scale=0.26]
\FBox \iLnkcc \jLnkcc \kLnkbb \lLnkbb \mLnkaa \nLnkbb 
\end{tikzpicture}\emm
=f_{16}
\nonumber\\
F^{121}_{121}
\bmm\begin{tikzpicture}[scale=0.26]
\FBox \iLnkbb \jLnkcc \kLnkbb \lLnkcc \mLnkbb \nLnkbb 
\end{tikzpicture}\emm
&=(F^{211}_{211}
\bmm\begin{tikzpicture}[scale=0.26]
\FBox \iLnkcc \jLnkbb \kLnkcc \lLnkbb \mLnkbb \nLnkbb 
\end{tikzpicture}\emm
)^*=f_{17}
\nonumber\\
F^{220}_{220}
\bmm\begin{tikzpicture}[scale=0.26]
\FBox \iLnkcc \jLnkcc \kLnkcc \lLnkcc \mLnkaa \nLnkaa 
\end{tikzpicture}\emm
&=f_{18}
\end{align}
There are ten potentially non-zero components
in $P^{kj}_i$, which are denoted by
$p_0$,...,$p_9$:
\begin{align}
&
P^{00}_{0}=p_{0},\ \
P^{01}_{0}=p_{1},\ \
P^{02}_{0}=p_{2},\ \
P^{00}_{1}=p_{3},\ \
P^{01}_{1}=p_{4},
\nonumber\\
&
P^{11}_{1}=p_{5},\ \
P^{21}_{1}=p_{6},\ \
P^{02}_{1}=p_{7},\ \
P^{00}_{2}=p_{8},\ \
P^{01}_{2}=p_{9},
\nonumber\\
&
P^{02}_{2}=p_{10}.
\end{align}
Using the ``gauge fixing'' discussed in section
\Ref{CGW1038}, we can fix the phases of $f_1$,
$f_2$, $f_5$, $f_8$, $f_{10}$, $f_{11}$, $f_{16}$,
$p_0$ and $A^0$ to make them positive. Again, these
phases are not completely independent, so the
``gauge fixing'' has to be done self-consistently.

The fixed-point conditions (\eqn{Phinonz}) form a set of non-linear
equations on the variables $f_i$, $p_i$, and $A^i$, which can
be solved exactly. After applying the ``gauge fixing''
discussed above, we find
only one solution
\begin{align}
&
f_0=f_1=...=f_8=f_{16}=f_{17}=f_{18}=1,\ \
\nonumber\\
&
f_9=f_{11}=f_{14}=\frac{1}{2},\ \ f_{12}=0,\ \ 
\nonumber\\
&
f_{10}=-f_{13}=\frac{1}{\sqrt{2}},\ \ f_{15}=-1,
\nonumber\\
&
p_1=p_4=p_9=\frac{2}{\sqrt 6},\ \ p_5=p_6=0,
\nonumber\\
&
p_0=p_2=p_3=p_7=p_8=p_{10}=\frac{1}{\sqrt{6}},
\nonumber\\
&
A^0=A^2=\frac{1}{\sqrt 6},\ \ A^1=\frac{2}{\sqrt 6}.
\end{align}
We also find
\begin{align}
 \e^{\imth \th_F}=
 \e^{\imth \th_{P1}}=
 \e^{\imth \th_{P2}}=
 \e^{\imth \th_{A1}}=
 \e^{\imth \th_{A2}}=
1.
\end{align}
A careful comparison with \Ref{LWstrnet} shows
that the F-tensors obtained above perfectly match
the $F$ local rules for the $N=2$ non-orientable
string-net state in \Ref{LWstrnet} (after switching the
edge labels 1 and 2). Thus the above state
corresponds to the standard lattice gauge theory
with gauge group $S_3$ -- the permutation group
of 3 objects.

\subsection{An $N=2$ string-net state -- the $\mathbb{Z}_3$ state}
\label{Z3state}

The above four examples all correspond to non-orientable
string-net states. Here we will give an example of an orientable
string-net state.  We choose $N=2$, $0^*=0$, $1^*=2$,
$2^*=1$, and
\begin{align}
N_{000}&=
N_{012}=
N_{120}=
N_{201}=
N_{021}=
N_{102}=
N_{210}
\nonumber\\
&=
N_{111}=
N_{222}=1.
\end{align}
The above $N_{ijk}$
satisfies \eqn{Phinonz}.

This state also appeared in \Ref{LWstrnet} as the other example
of $N=2$ string-net state. As in the $\mathbb{Z}_2$ case,
the three edge labels of $N$ form a $\mathbb{Z}_3$ group
after we switch the definition of positive direction in the third
label of $N$. For example, $N_{012}$ becomes $N_{011}$
after switching, which will then correspond
to group action $0 \bigotimes 1 = 1$.
we will call such a state the $\mathbb{Z}_3$ state.

Due to relation \eqn{Phinonz},
different components of the tensor $F^{ijm}_{kln}$ are not
independent.  There are eight independent potentially non-zero
components which are denoted as $f_0$,...,$f_7$:
\begin{align}
F^{000}_{000}
\bmm\begin{tikzpicture}[scale=0.26]
\FBox \iLnkaa \jLnkaa \kLnkaa \lLnkaa \mLnkaa \nLnkaa
\end{tikzpicture}\emm
&=f_{0}
\nonumber\\
F^{000}_{111}
\bmm\begin{tikzpicture}[scale=0.26]
\FBox \iLnkaa \jLnkaa \kLnkbc \lLnkbc \mLnkaa \nLnkbc
\end{tikzpicture}\emm
&=(F^{011}_{200}
\bmm\begin{tikzpicture}[scale=0.26]
\FBox \iLnkaa \jLnkbc \kLnkcb \lLnkaa \mLnkbc \nLnkaa
\end{tikzpicture}\emm
)^*=F^{120}_{002}
\bmm\begin{tikzpicture}[scale=0.26]
\FBox \iLnkbc \jLnkcb \kLnkaa \lLnkaa \mLnkaa \nLnkcb
\end{tikzpicture}\emm
\nonumber\\
&=(F^{202}_{020}
\bmm\begin{tikzpicture}[scale=0.26]
\FBox \iLnkcb \jLnkaa \kLnkaa \lLnkcb \mLnkcb \nLnkaa
\end{tikzpicture}\emm
)^*=f_{1}
\nonumber\\
F^{000}_{222}
\bmm\begin{tikzpicture}[scale=0.26]
\FBox \iLnkaa \jLnkaa \kLnkcb \lLnkcb \mLnkaa \nLnkcb
\end{tikzpicture}\emm
&=(F^{022}_{100}
\bmm\begin{tikzpicture}[scale=0.26]
\FBox \iLnkaa \jLnkcb \kLnkbc \lLnkaa \mLnkcb \nLnkaa
\end{tikzpicture}\emm
)^*=(F^{101}_{010}
\bmm\begin{tikzpicture}[scale=0.26]
\FBox \iLnkbc \jLnkaa \kLnkaa \lLnkbc \mLnkbc \nLnkaa
\end{tikzpicture}\emm
)^*\nonumber\\
&=F^{210}_{001}
\bmm\begin{tikzpicture}[scale=0.26]
\FBox \iLnkcb \jLnkbc \kLnkaa \lLnkaa \mLnkaa \nLnkbc
\end{tikzpicture}\emm
=f_{2}
\nonumber\\
F^{011}_{011}
\bmm\begin{tikzpicture}[scale=0.26]
\FBox \iLnkaa \jLnkbc \kLnkaa \lLnkbc \mLnkbc \nLnkbc
\end{tikzpicture}\emm
&=F^{022}_{022}
\bmm\begin{tikzpicture}[scale=0.26]
\FBox \iLnkaa \jLnkcb \kLnkaa \lLnkcb \mLnkcb \nLnkcb
\end{tikzpicture}\emm
=(F^{101}_{202}
\bmm\begin{tikzpicture}[scale=0.26]
\FBox \iLnkbc \jLnkaa \kLnkcb \lLnkaa \mLnkbc \nLnkcb
\end{tikzpicture}\emm
)^*\nonumber\\
&=(F^{202}_{101}
\bmm\begin{tikzpicture}[scale=0.26]
\FBox \iLnkcb \jLnkaa \kLnkbc \lLnkaa \mLnkcb \nLnkbc
\end{tikzpicture}\emm
)^*=f_{3}
\nonumber\\
F^{011}_{122}
\bmm\begin{tikzpicture}[scale=0.26]
\FBox \iLnkaa \jLnkbc \kLnkbc \lLnkcb \mLnkbc \nLnkcb
\end{tikzpicture}\emm
&=(F^{101}_{121}
\bmm\begin{tikzpicture}[scale=0.26]
\FBox \iLnkbc \jLnkaa \kLnkbc \lLnkcb \mLnkbc \nLnkbc
\end{tikzpicture}\emm
)^*=F^{112}_{021}
\bmm\begin{tikzpicture}[scale=0.26]
\FBox \iLnkbc \jLnkbc \kLnkaa \lLnkcb \mLnkcb \nLnkbc
\end{tikzpicture}\emm
\nonumber\\
&=(F^{112}_{102}
\bmm\begin{tikzpicture}[scale=0.26]
\FBox \iLnkbc \jLnkbc \kLnkbc \lLnkaa \mLnkcb \nLnkcb
\end{tikzpicture}\emm
)^*=f_{4}
\nonumber\\
F^{022}_{211}
\bmm\begin{tikzpicture}[scale=0.26]
\FBox \iLnkaa \jLnkcb \kLnkcb \lLnkbc \mLnkcb \nLnkbc
\end{tikzpicture}\emm
&=(F^{202}_{212}
\bmm\begin{tikzpicture}[scale=0.26]
\FBox \iLnkcb \jLnkaa \kLnkcb \lLnkbc \mLnkcb \nLnkcb
\end{tikzpicture}\emm
)^*=F^{221}_{012}
\bmm\begin{tikzpicture}[scale=0.26]
\FBox \iLnkcb \jLnkcb \kLnkaa \lLnkbc \mLnkbc \nLnkcb
\end{tikzpicture}\emm
\nonumber\\
&=(F^{221}_{201}
\bmm\begin{tikzpicture}[scale=0.26]
\FBox \iLnkcb \jLnkcb \kLnkcb \lLnkaa \mLnkbc \nLnkbc
\end{tikzpicture}\emm
)^*=f_{5}
\nonumber\\
F^{112}_{210}
\bmm\begin{tikzpicture}[scale=0.26]
\FBox \iLnkbc \jLnkbc \kLnkcb \lLnkbc \mLnkcb \nLnkaa
\end{tikzpicture}\emm
&=(F^{120}_{221}
\bmm\begin{tikzpicture}[scale=0.26]
\FBox \iLnkbc \jLnkcb \kLnkcb \lLnkcb \mLnkaa \nLnkbc
\end{tikzpicture}\emm
)^*=(F^{210}_{112}
\bmm\begin{tikzpicture}[scale=0.26]
\FBox \iLnkcb \jLnkbc \kLnkbc \lLnkbc \mLnkaa \nLnkcb
\end{tikzpicture}\emm
)^*
\nonumber\\
&=F^{221}_{120}
\bmm\begin{tikzpicture}[scale=0.26]
\FBox \iLnkcb \jLnkcb \kLnkbc \lLnkcb \mLnkbc \nLnkaa
\end{tikzpicture}\emm
=f_{6}
\nonumber\\
F^{120}_{110}
\bmm\begin{tikzpicture}[scale=0.26]
\FBox \iLnkbc \jLnkcb \kLnkbc \lLnkbc \mLnkaa \nLnkaa
\end{tikzpicture}\emm
&=(F^{210}_{220}
\bmm\begin{tikzpicture}[scale=0.26]
\FBox \iLnkcb \jLnkbc \kLnkcb \lLnkcb \mLnkaa \nLnkaa
\end{tikzpicture}\emm
)^*=f_{7}
\end{align}
There are nine potentially non-zero components
in $P^{kj}_i$, which are denoted by
$p_0$,...,$p_8$:
\begin{align}
&
P^{00}_{0}=p_{0},\ \
P^{01}_{0}=p_{1},\ \
P^{02}_{0}=p_{2},\ \
P^{00}_{1}=p_{3},\ \
P^{01}_{1}=p_{4},
\nonumber\\
&
P^{02}_{1}=p_{5},\ \
P^{00}_{2}=p_{6},\ \
P^{01}_{2}=p_{7},\ \
P^{02}_{2}=p_{8}.
\end{align}
Using the ``gauge fixing'' discussed in section
\Ref{CGW1038}, we can fix the phases of $f_1$, $f_2$, $f_6$,
$p_0$ and $A^0$ to make them positive.

The fixed-point conditions (\eqn{Phinonz}) form a set of non-linear
equations on the variables $f_i$, $p_i$, and $A^i$, which can
be solved exactly. After applying the ``gauge fixing''
discussed above, we find
only one solution
\begin{align}
&
f_i=1,\ \ i=0,1,...,7,
\nonumber\\
&
p_i=\frac{1}{\sqrt 3},\ \ i=0,1,...,8,
\nonumber\\
&
A^0=A^1=A^2=\frac{1}{\sqrt 3}.
\end{align}
We also find
\begin{align}
 \e^{\imth \th_F}=
 \e^{\imth \th_{P1}}=
 \e^{\imth \th_{P2}}=
 \e^{\imth \th_{A1}}=
 \e^{\imth \th_{A2}}=
1.
\end{align}
The fixed-point state corresponds to the $\mathbb{Z}_3$
string-net condensed state\cite{LWstrnet}.

The process of applying the Modular transformations are
exactly the same as in \ref{Z2state}. By applying Dehn twists
to all the $9$ non-contractable fixed-point states on a torus,
we can get the $T$-matrix. 
Note that as in \ref{Z2state},
here all the $9$ non-contractable graphs
correspond to fixed-point states and
are ground states of a
certain local Hamiltonian.
Without going into too much details, we present here the resulting
$9$ by $9$ $T$-matrix:
\begin{align}
T=
\bpm
1& 0& 0& 0& 0& 0& 0& 0& 0& \\
0& 1& 0& 0& 0& 0& 0& 0& 0& \\
0& 0& 1& 0& 0& 0& 0& 0& 0& \\
0& 0& 0& 0& 0& 1& 0& 0& 0& \\
0& 0& 0& 1& 0& 0& 0& 0& 0& \\
0& 0& 0& 0& 1& 0& 0& 0& 0& \\
0& 0& 0& 0& 0& 0& 0& 1& 0& \\
0& 0& 0& 0& 0& 0& 0& 0& 1& \\
0& 0& 0& 0& 0& 0& 1& 0& 0&
\epm
\end{align}
Similarly, $S$-matrix can be obtained by applying the
$90^{\circ}$ rotations to all the
$9$ fixed-point states.
The resulting $9$ by $9$ $S$-matrix is then:
\begin{align}
S=
\bpm
1& 0& 0& 0& 0& 0& 0& 0& 0& \\
0& 0& 0& 0& 0& 0& 1& 0& 0& \\
0& 0& 0& 1& 0& 0& 0& 0& 0& \\
0& 1& 0& 0& 0& 0& 0& 0& 0& \\
0& 0& 0& 0& 0& 0& 0& 1& 0& \\
0& 0& 0& 0& 1& 0& 0& 0& 0& \\
0& 0& 1& 0& 0& 0& 0& 0& 0& \\
0& 0& 0& 0& 0& 0& 0& 0& 1& \\
0& 0& 0& 0& 0& 1& 0& 0& 0&
\epm.
\end{align}

As before, we can choose a particular
basis in which $T$-matrix is diagonalized
and $S$-matrix satisfies
eqn. \ref{Sconditions}
in modular tensor category theory.
Without going into too much details,
we present here the resulting
$T$ and $S$-matrix under
a change of basis:
\begin{align}
T=
\bpm
1& 0& 0& 0& 0& 0& 0& 0& 0& \\
0& 1& 0& 0& 0& 0& 0& 0& 0& \\
0& 0& 1& 0& 0& 0& 0& 0& 0& \\
0& 0& 0& 1& 0& 0& 0& 0& 0& \\
0& 0& 0& 0& \xi& 0& 0& 0& 0& \\
0& 0& 0& 0& 0& \xi^*& 0& 0& 0& \\
0& 0& 0& 0& 0& 0& 1& 0& 0& \\
0& 0& 0& 0& 0& 0& 0& \xi& 0& \\
0& 0& 0& 0& 0& 0& 0& 0& \xi^*&
\epm
\end{align}
and
\begin{align}
S=\frac{1}{3}
\bpm
1& 1& 1& 1& 1& 1& 1& 1& 1& \\
1& 1& 1&  \xi^*&  \xi^*& \xi^*& \xi& \xi& \xi& \\
1& 1& 1&  \xi&  \xi& \xi& \xi^*& \xi^*& \xi^*& \\
1& \xi^*& \xi& 1&  \xi^*& \xi& 1& \xi& \xi^*& \\
1& \xi^*& \xi&  \xi^*&  \xi& 1& \xi& (\xi)^2& 1& \\
1& \xi^*& \xi&  \xi& 1& (\xi)^2& \xi^*& 1& \xi& \\
1& \xi& \xi^*& 1&  \xi& \xi^*& 1& \xi^*& \xi& \\
1& \xi& \xi^*&  \xi& (\xi)^2& 1& \xi^*& \xi& 1& \\
1& \xi& \xi^*&  \xi^*& 1& \xi& \xi& 1& (\xi)^2&
\epm
\end{align}
where $\xi= e^{{i\frac{2\pi}{3}}}$.
As before, all information of the quasi-particles can be obtained
from the above $T$ and $S$ matrices.

Note that the
corresponding eigenvalues of $T$ are:
\begin{align*}
(1,1,1,1,1,e^{{i\frac{2\pi}{3}}},e^{{i\frac{2\pi}{3}}},e^{{-i\frac{2\pi}{3}}},e^{{-i\frac{2\pi}{3}}}).
\end{align*}
which exactly correspond to the statistical
angles of the $U(1)\times U(1)$ Chern-Simons
theory\cite{LWstrnet,KLW0834}
\begin{align}
 \cL=\frac{1}{4\pi} K^{IJ} a_{I\mu}\prt_\nu
a_{J\la}\eps^{\mu\nu\la} ,
\end{align}
with
$ K=\bpm  0 & 3 \\
         3 & 0 \\ \epm $.
This is also equivalent to $\mathbb{Z}_3$ gauge theory.

%Note that we have thus obtained all the string-net
%examples in \Ref{LWstrnet} except for one --
%the $\Phi_-$ state which correspond to another
%$U(1)\times U(1)$ Chern-Simons theory\cite{LWstrnet,KLW}
%\begin{align}
%\cL=\frac {1}{4\pi}\Big(
%3 a_{1\mu}\prt_{\nu}a_{2\la}\eps^{\mu\nu\la}
%-3 a_{2\mu}\prt_{\nu}a_{1\la}\eps^{\mu\nu\la}
%\Big).
%\end{align}
%This state cannot be obtained from a bosonic system.

%We note further that all the above solutions also satisfy the
%standard pentagon identity, although we solved the weaker
%projective pentagon identity.  It is not clear if we can
%find non-trivial solutions that do not satisfy the standard
%pentagon identity.

\subsection{$N=3$ string-net states -- the $\mathbb{Z}_4$ states}
\label{Z4state}

Now we will allow four states on each edge of the graph,
namely $|0\>$, $|1\>$, $|2\>$ and $|3\>$. We will first
give an example of orientable string-net state.
We choose $N=3$, $0^*=0$, $1^*=3$,
$2^*=2$, $3^*=1$, and
\begin{align}
N_{000}&=
N_{013}=
N_{130}=
N_{301}=
N_{022}=
N_{202}=
N_{220}
\nonumber\\
&=
N_{031}=
N_{103}=
N_{310}=
N_{112}=
N_{121}=
N_{211}
\nonumber\\
&=
N_{233}=
N_{323}=
N_{332}=1.
\end{align}
The above $N_{ijk}$
satisfies \eqn{Phinonz}.

We will see that one of the solutions to be obtained corresponds
to the $\mathbb{Z}_4$ gauge theory. Again, as in the
$\mathbb{Z}_3$ case, the three edge labels of $N$ form
a $\mathbb{Z}_4$ group after switching the positive direction
of the third label of $N$: for example, $N_{121}$ becomes
$N_{123}$ which corresponds to group action
$1 \bigotimes 2 = 3$. Thus, we will call states obtained
from the above fusion rules the $\mathbb{Z}_4$ states.

Again, due to relation \eqn{Phinonz},
different components of tensor $F^{ijm}_{kln}$ are not
independent.  There are now twenty independent potentially 
non-zero components which are denoted as $f_0$,...,$f_{19}$:
\begin{align}
F^{000}_{000}
\bmm\begin{tikzpicture}[scale=0.26]
\FBox \iLnkaa \jLnkaa \kLnkaa \lLnkaa \mLnkaa \nLnkaa 
\end{tikzpicture}\emm
&=f_{0}
\nonumber\\
F^{000}_{111}
\bmm\begin{tikzpicture}[scale=0.26]
\FBox \iLnkaa \jLnkaa \kLnkbd \lLnkbd \mLnkaa \nLnkbd 
\end{tikzpicture}\emm
&=(F^{011}_{300}
\bmm\begin{tikzpicture}[scale=0.26]
\FBox \iLnkaa \jLnkbd \kLnkdb \lLnkaa \mLnkbd \nLnkaa 
\end{tikzpicture}\emm
)^*=F^{130}_{003}
\bmm\begin{tikzpicture}[scale=0.26]
\FBox \iLnkbd \jLnkdb \kLnkaa \lLnkaa \mLnkaa \nLnkdb 
\end{tikzpicture}\emm
\nonumber\\
&=(F^{303}_{030}
\bmm\begin{tikzpicture}[scale=0.26]
\FBox \iLnkdb \jLnkaa \kLnkaa \lLnkdb \mLnkdb \nLnkaa 
\end{tikzpicture}\emm
)^*=f_{1}
\nonumber\\
F^{000}_{222}
\bmm\begin{tikzpicture}[scale=0.26]
\FBox \iLnkaa \jLnkaa \kLnkcc \lLnkcc \mLnkaa \nLnkcc 
\end{tikzpicture}\emm
&=(F^{022}_{200}
\bmm\begin{tikzpicture}[scale=0.26]
\FBox \iLnkaa \jLnkcc \kLnkcc \lLnkaa \mLnkcc \nLnkaa 
\end{tikzpicture}\emm
)^*=(F^{202}_{020}
\bmm\begin{tikzpicture}[scale=0.26]
\FBox \iLnkcc \jLnkaa \kLnkaa \lLnkcc \mLnkcc \nLnkaa 
\end{tikzpicture}\emm
)^*
\nonumber\\
&=F^{220}_{002}
\bmm\begin{tikzpicture}[scale=0.26]
\FBox \iLnkcc \jLnkcc \kLnkaa \lLnkaa \mLnkaa \nLnkcc 
\end{tikzpicture}\emm
=f_{2}
\nonumber\\
F^{000}_{333}
\bmm\begin{tikzpicture}[scale=0.26]
\FBox \iLnkaa \jLnkaa \kLnkdb \lLnkdb \mLnkaa \nLnkdb 
\end{tikzpicture}\emm
&=(F^{033}_{100}
\bmm\begin{tikzpicture}[scale=0.26]
\FBox \iLnkaa \jLnkdb \kLnkbd \lLnkaa \mLnkdb \nLnkaa 
\end{tikzpicture}\emm
)^*=(F^{101}_{010}
\bmm\begin{tikzpicture}[scale=0.26]
\FBox \iLnkbd \jLnkaa \kLnkaa \lLnkbd \mLnkbd \nLnkaa 
\end{tikzpicture}\emm
)^*
\nonumber\\
&=F^{310}_{001}
\bmm\begin{tikzpicture}[scale=0.26]
\FBox \iLnkdb \jLnkbd \kLnkaa \lLnkaa \mLnkaa \nLnkbd 
\end{tikzpicture}\emm
=f_{3}
\nonumber\\
F^{011}_{011}
\bmm\begin{tikzpicture}[scale=0.26]
\FBox \iLnkaa \jLnkbd \kLnkaa \lLnkbd \mLnkbd \nLnkbd 
\end{tikzpicture}\emm
&=F^{033}_{033}
\bmm\begin{tikzpicture}[scale=0.26]
\FBox \iLnkaa \jLnkdb \kLnkaa \lLnkdb \mLnkdb \nLnkdb 
\end{tikzpicture}\emm
=(F^{101}_{303}
\bmm\begin{tikzpicture}[scale=0.26]
\FBox \iLnkbd \jLnkaa \kLnkdb \lLnkaa \mLnkbd \nLnkdb 
\end{tikzpicture}\emm
)^*
\nonumber\\
&=(F^{303}_{101}
\bmm\begin{tikzpicture}[scale=0.26]
\FBox \iLnkdb \jLnkaa \kLnkbd \lLnkaa \mLnkdb \nLnkbd 
\end{tikzpicture}\emm
)^*=f_{4}
\nonumber
\end{align}

\begin{align}
F^{011}_{122}
\bmm\begin{tikzpicture}[scale=0.26]
\FBox \iLnkaa \jLnkbd \kLnkbd \lLnkcc \mLnkbd \nLnkcc 
\end{tikzpicture}\emm
&=(F^{112}_{203}
\bmm\begin{tikzpicture}[scale=0.26]
\FBox \iLnkbd \jLnkbd \kLnkcc \lLnkaa \mLnkcc \nLnkdb 
\end{tikzpicture}\emm
)^*=F^{123}_{032}
\bmm\begin{tikzpicture}[scale=0.26]
\FBox \iLnkbd \jLnkcc \kLnkaa \lLnkdb \mLnkdb \nLnkcc 
\end{tikzpicture}\emm
\nonumber\\
&=(F^{202}_{131}
\bmm\begin{tikzpicture}[scale=0.26]
\FBox \iLnkcc \jLnkaa \kLnkbd \lLnkdb \mLnkcc \nLnkbd 
\end{tikzpicture}\emm
)^*=f_{5}
\nonumber\\
F^{011}_{233}
\bmm\begin{tikzpicture}[scale=0.26]
\FBox \iLnkaa \jLnkbd \kLnkcc \lLnkdb \mLnkbd \nLnkdb 
\end{tikzpicture}\emm
&=(F^{101}_{121}
\bmm\begin{tikzpicture}[scale=0.26]
\FBox \iLnkbd \jLnkaa \kLnkbd \lLnkcc \mLnkbd \nLnkbd 
\end{tikzpicture}\emm
)^*=(F^{123}_{103}
\bmm\begin{tikzpicture}[scale=0.26]
\FBox \iLnkbd \jLnkcc \kLnkbd \lLnkaa \mLnkdb \nLnkdb 
\end{tikzpicture}\emm
)^*
\nonumber\\
&=F^{213}_{031}
\bmm\begin{tikzpicture}[scale=0.26]
\FBox \iLnkcc \jLnkbd \kLnkaa \lLnkdb \mLnkdb \nLnkbd 
\end{tikzpicture}\emm
=f_{6}
\nonumber\\
F^{022}_{022}
\bmm\begin{tikzpicture}[scale=0.26]
\FBox \iLnkaa \jLnkcc \kLnkaa \lLnkcc \mLnkcc \nLnkcc 
\end{tikzpicture}\emm
&=(F^{202}_{202}
\bmm\begin{tikzpicture}[scale=0.26]
\FBox \iLnkcc \jLnkaa \kLnkcc \lLnkaa \mLnkcc \nLnkcc 
\end{tikzpicture}\emm
)^*=f_{7}
\nonumber\\
F^{022}_{133}
\bmm\begin{tikzpicture}[scale=0.26]
\FBox \iLnkaa \jLnkcc \kLnkbd \lLnkdb \mLnkcc \nLnkdb 
\end{tikzpicture}\emm
&=(F^{101}_{232}
\bmm\begin{tikzpicture}[scale=0.26]
\FBox \iLnkbd \jLnkaa \kLnkcc \lLnkdb \mLnkbd \nLnkcc 
\end{tikzpicture}\emm
)^*=F^{112}_{021}
\bmm\begin{tikzpicture}[scale=0.26]
\FBox \iLnkbd \jLnkbd \kLnkaa \lLnkcc \mLnkcc \nLnkbd 
\end{tikzpicture}\emm
\nonumber\\
&=(F^{213}_{102}
\bmm\begin{tikzpicture}[scale=0.26]
\FBox \iLnkcc \jLnkbd \kLnkbd \lLnkaa \mLnkdb \nLnkcc 
\end{tikzpicture}\emm
)^*=f_{8}
\nonumber\\
F^{022}_{311}
\bmm\begin{tikzpicture}[scale=0.26]
\FBox \iLnkaa \jLnkcc \kLnkdb \lLnkbd \mLnkcc \nLnkbd 
\end{tikzpicture}\emm
&=(F^{231}_{302}
\bmm\begin{tikzpicture}[scale=0.26]
\FBox \iLnkcc \jLnkdb \kLnkdb \lLnkaa \mLnkbd \nLnkcc 
\end{tikzpicture}\emm
)^*=(F^{303}_{212}
\bmm\begin{tikzpicture}[scale=0.26]
\FBox \iLnkdb \jLnkaa \kLnkcc \lLnkbd \mLnkdb \nLnkcc 
\end{tikzpicture}\emm
)^*
\nonumber\\
&=F^{332}_{023}
\bmm\begin{tikzpicture}[scale=0.26]
\FBox \iLnkdb \jLnkdb \kLnkaa \lLnkcc \mLnkcc \nLnkdb 
\end{tikzpicture}\emm
=f_{9}
\nonumber
\end{align}

\begin{align}
F^{033}_{211}
\bmm\begin{tikzpicture}[scale=0.26]
\FBox \iLnkaa \jLnkdb \kLnkcc \lLnkbd \mLnkdb \nLnkbd 
\end{tikzpicture}\emm
&=F^{231}_{013}
\bmm\begin{tikzpicture}[scale=0.26]
\FBox \iLnkcc \jLnkdb \kLnkaa \lLnkbd \mLnkbd \nLnkdb 
\end{tikzpicture}\emm
=(F^{303}_{323}
\bmm\begin{tikzpicture}[scale=0.26]
\FBox \iLnkdb \jLnkaa \kLnkdb \lLnkcc \mLnkdb \nLnkdb 
\end{tikzpicture}\emm
)^*
\nonumber\\
&=(F^{321}_{301}
\bmm\begin{tikzpicture}[scale=0.26]
\FBox \iLnkdb \jLnkcc \kLnkdb \lLnkaa \mLnkbd \nLnkbd 
\end{tikzpicture}\emm
)^*=f_{10}
\nonumber\\
F^{033}_{322}
\bmm\begin{tikzpicture}[scale=0.26]
\FBox \iLnkaa \jLnkdb \kLnkdb \lLnkcc \mLnkdb \nLnkcc 
\end{tikzpicture}\emm
&=(F^{202}_{313}
\bmm\begin{tikzpicture}[scale=0.26]
\FBox \iLnkcc \jLnkaa \kLnkdb \lLnkbd \mLnkcc \nLnkdb 
\end{tikzpicture}\emm
)^*=F^{321}_{012}
\bmm\begin{tikzpicture}[scale=0.26]
\FBox \iLnkdb \jLnkcc \kLnkaa \lLnkbd \mLnkbd \nLnkcc 
\end{tikzpicture}\emm
\nonumber\\
&=(F^{332}_{201}
\bmm\begin{tikzpicture}[scale=0.26]
\FBox \iLnkdb \jLnkdb \kLnkcc \lLnkaa \mLnkcc \nLnkbd 
\end{tikzpicture}\emm
)^*=f_{11}
\nonumber\\
F^{112}_{132}
\bmm\begin{tikzpicture}[scale=0.26]
\FBox \iLnkbd \jLnkbd \kLnkbd \lLnkdb \mLnkcc \nLnkcc 
\end{tikzpicture}\emm
&=f_{12}
\nonumber\\
F^{112}_{310}
\bmm\begin{tikzpicture}[scale=0.26]
\FBox \iLnkbd \jLnkbd \kLnkdb \lLnkbd \mLnkcc \nLnkaa 
\end{tikzpicture}\emm
&=(F^{130}_{332}
\bmm\begin{tikzpicture}[scale=0.26]
\FBox \iLnkbd \jLnkdb \kLnkdb \lLnkdb \mLnkaa \nLnkcc 
\end{tikzpicture}\emm
)^*=(F^{310}_{112}
\bmm\begin{tikzpicture}[scale=0.26]
\FBox \iLnkdb \jLnkbd \kLnkbd \lLnkbd \mLnkaa \nLnkcc 
\end{tikzpicture}\emm
)^*
\nonumber\\
&=F^{332}_{130}
\bmm\begin{tikzpicture}[scale=0.26]
\FBox \iLnkdb \jLnkdb \kLnkbd \lLnkdb \mLnkcc \nLnkaa 
\end{tikzpicture}\emm
=f_{13}
\nonumber\\
F^{123}_{210}
\bmm\begin{tikzpicture}[scale=0.26]
\FBox \iLnkbd \jLnkcc \kLnkcc \lLnkbd \mLnkdb \nLnkaa 
\end{tikzpicture}\emm
&=(F^{220}_{331}
\bmm\begin{tikzpicture}[scale=0.26]
\FBox \iLnkcc \jLnkcc \kLnkdb \lLnkdb \mLnkaa \nLnkbd 
\end{tikzpicture}\emm
)^*=F^{231}_{120}
\bmm\begin{tikzpicture}[scale=0.26]
\FBox \iLnkcc \jLnkdb \kLnkbd \lLnkcc \mLnkbd \nLnkaa 
\end{tikzpicture}\emm
\nonumber\\
&=(F^{310}_{223}
\bmm\begin{tikzpicture}[scale=0.26]
\FBox \iLnkdb \jLnkbd \kLnkcc \lLnkcc \mLnkaa \nLnkdb 
\end{tikzpicture}\emm
)^*=f_{14}
\nonumber
\end{align}

\begin{align}
F^{123}_{321}
\bmm\begin{tikzpicture}[scale=0.26]
\FBox \iLnkbd \jLnkcc \kLnkdb \lLnkcc \mLnkdb \nLnkbd 
\end{tikzpicture}\emm
&=(F^{213}_{213}
\bmm\begin{tikzpicture}[scale=0.26]
\FBox \iLnkcc \jLnkbd \kLnkcc \lLnkbd \mLnkdb \nLnkdb 
\end{tikzpicture}\emm
)^*=(F^{231}_{231}
\bmm\begin{tikzpicture}[scale=0.26]
\FBox \iLnkcc \jLnkdb \kLnkcc \lLnkdb \mLnkbd \nLnkbd 
\end{tikzpicture}\emm
)^*
\nonumber\\
&=F^{321}_{123}
\bmm\begin{tikzpicture}[scale=0.26]
\FBox \iLnkdb \jLnkcc \kLnkbd \lLnkcc \mLnkbd \nLnkdb 
\end{tikzpicture}\emm
=f_{15}
\nonumber\\
F^{130}_{110}
\bmm\begin{tikzpicture}[scale=0.26]
\FBox \iLnkbd \jLnkdb \kLnkbd \lLnkbd \mLnkaa \nLnkaa 
\end{tikzpicture}\emm
&=(F^{310}_{330}
\bmm\begin{tikzpicture}[scale=0.26]
\FBox \iLnkdb \jLnkbd \kLnkdb \lLnkdb \mLnkaa \nLnkaa 
\end{tikzpicture}\emm
)^*=f_{16}
\nonumber\\
F^{130}_{221}
\bmm\begin{tikzpicture}[scale=0.26]
\FBox \iLnkbd \jLnkdb \kLnkcc \lLnkcc \mLnkaa \nLnkbd 
\end{tikzpicture}\emm
&=(F^{213}_{320}
\bmm\begin{tikzpicture}[scale=0.26]
\FBox \iLnkcc \jLnkbd \kLnkdb \lLnkcc \mLnkdb \nLnkaa 
\end{tikzpicture}\emm
)^*=F^{220}_{113}
\bmm\begin{tikzpicture}[scale=0.26]
\FBox \iLnkcc \jLnkcc \kLnkbd \lLnkbd \mLnkaa \nLnkdb 
\end{tikzpicture}\emm
\nonumber\\
&=(F^{321}_{230}
\bmm\begin{tikzpicture}[scale=0.26]
\FBox \iLnkdb \jLnkcc \kLnkcc \lLnkdb \mLnkbd \nLnkaa 
\end{tikzpicture}\emm
)^*=f_{17}
\nonumber\\
F^{220}_{220}
\bmm\begin{tikzpicture}[scale=0.26]
\FBox \iLnkcc \jLnkcc \kLnkcc \lLnkcc \mLnkaa \nLnkaa 
\end{tikzpicture}\emm
&=f_{18}
\nonumber\\
F^{332}_{312}
\bmm\begin{tikzpicture}[scale=0.26]
\FBox \iLnkdb \jLnkdb \kLnkdb \lLnkbd \mLnkcc \nLnkcc 
\end{tikzpicture}\emm
&=f_{19}
\end{align}
There are now sixteen potentially non-zero components
in $P^{kj}_i$, which are denoted by
$p_0$,...,$p_{15}$:
\begin{align}
&
P^{00}_{0}=p_{0},\ \
P^{01}_{0}=p_{1},\ \
P^{02}_{0}=p_{2},\ \
P^{03}_{0}=p_{3},\ \
P^{00}_{1}=p_{4},
\nonumber\\
&
P^{01}_{1}=p_{5},\ \
P^{02}_{1}=p_{6},\ \
P^{03}_{1}=p_{7},\ \
P^{00}_{2}=p_{8},\ \
P^{01}_{2}=p_{9},
\nonumber\\
&
P^{02}_{2}=p_{10},\ \
P^{03}_{2}=p_{11},\ \
P^{00}_{3}=p_{12},\ \
P^{01}_{3}=p_{13},
\nonumber\\
&
P^{02}_{3}=p_{14},\ \
P^{03}_{3}=p_{15}.
\end{align}
Using the ``gauge fixing'' discussed in section
\Ref{CGW1038}, we can fix the phases of $f_1$, $f_2$, $f_3$,
$f_5$, $f_8$, $f_9$, $f_{11}$, $f_{13}$, $f_{14}$, $f_{17}$,
$p_0$ and $A^0$ to make them positive.
Again, these phases are not completely independent,
so the ``gauge fixing'' has to be done self-consistently.

The fixed-point conditions (\eqn{Phinonz}) form a set of non-linear
equations on the variables $f_i$, $p_i$, and $A^i$, which can
be solved exactly. After applying the "gauge fixing"
discussed above, we find
two solutions
\begin{align}
&
f_{12}=f_{15}=f_{19}=\eta,
\nonumber\\
&
f_i=1,\ \ i=otherwise,
\nonumber\\
&
p_i=\frac{1}{2},\ \ i=0,1,...,15,
\nonumber\\
&
A^0=A^1=A^2=A^3=\frac{1}{2}
\end{align}
where $\eta = \pm 1$.
We also find
\begin{align}
 \e^{\imth \th_F}=
 \e^{\imth \th_{P1}}=
 \e^{\imth \th_{P2}}=
 \e^{\imth \th_{A1}}=
 \e^{\imth \th_{A2}}=
1.
\end{align}

Similar to the $\mathbb{Z}_2$ and $\mathbb{Z}_3$ case,
we can obtain the $T$ and $S$-matrices of the two states above
using modular transformations. By relating the eigenvalues of
the $T$-matrices to quasi-particle statistics, we can
also find the
corresponding effective theory for these two states. 
First, by applying Dehn twists to all the $16$ fixed-point states
on a torus, we obtain the following $T$-matrix:
\begin{align}
T=
\bpm
1\; \ \ \; \ \ \; \ \ \; \ \ \; \ \ \; \ \ \; \ \ \; \ \ \; \ \ \; \ \ \; \ \ \; \ \ \; \ \ \; \ \ \; \ \ \\
\ \ \; 1\; \ \ \; \ \ \; \ \ \; \ \ \; \ \ \; \ \ \; \ \ \; \ \ \; \ \ \; \ \ \; \ \ \; \ \ \; \ \ \; \ \ \\
\ \ \; \ \ \; 1\; \ \ \; \ \ \; \ \ \; \ \ \; \ \ \; \ \ \; \ \ \; \ \ \; \ \ \; \ \ \; \ \ \; \ \ \; \ \ \\
\ \ \; \ \ \; \ \ \; 1\; \ \ \; \ \ \; \ \ \; \ \ \; \ \ \; \ \ \; \ \ \; \ \ \; \ \ \; \ \ \; \ \ \; \ \ \\
\ \ \; \ \ \; \ \ \; \ \ \; \ \ \; \ \ \; \ \ \; 1\; \ \ \; \ \ \; \ \ \; \ \ \; \ \ \; \ \ \; \ \ \; \ \ \\
\ \ \; \ \ \; \ \ \; \ \ \; \ \ \; \ \ \; \eta\; \ \ \; \ \ \; \ \ \; \ \ \; \ \ \; \ \ \; \ \ \; \ \ \; \ \ \\
\ \ \; \ \ \; \ \ \; \ \ \; \ \ \; \eta\; \ \ \; \ \ \; \ \ \; \ \ \; \ \ \; \ \ \; \ \ \; \ \ \; \ \ \; \ \ \\
\ \ \; \ \ \; \ \ \; \ \ \; 1\; \ \ \; \ \ \; \ \ \; \ \ \; \ \ \; \ \ \; \ \ \; \ \ \; \ \ \; \ \ \; \ \ \\
\ \ \; \ \ \; \ \ \; \ \ \; \ \ \; \ \ \; \ \ \; \ \ \; \ \ \; 1\; \ \ \; \ \ \; \ \ \; \ \ \; \ \ \; \ \ \\
\ \ \; \ \ \; \ \ \; \ \ \; \ \ \; \ \ \; \ \ \; \ \ \; \ \ \; \ \ \; \ \ \; \eta\; \ \ \; \ \ \; \ \ \; \ \ \\
\ \ \; \ \ \; \ \ \; \ \ \; \ \ \; \ \ \; \ \ \; \ \ \; 1\; \ \ \; \ \ \; \ \ \; \ \ \; \ \ \; \ \ \; \ \ \\
\ \ \; \ \ \; \ \ \; \ \ \; \ \ \; \ \ \; \ \ \; \ \ \; \ \ \; \ \ \; 1\; \ \ \; \ \ \; \ \ \; \ \ \; \ \ \\
\ \ \; \ \ \; \ \ \; \ \ \; \ \ \; \ \ \; \ \ \; \ \ \; \ \ \; \ \ \; \ \ \; \ \ \; \ \ \; 1\; \ \ \; \ \ \\
\ \ \; \ \ \; \ \ \; \ \ \; \ \ \; \ \ \; \ \ \; \ \ \; \ \ \; \ \ \; \ \ \; \ \ \; \ \ \; \ \ \; \ \ \; \eta\ \\
\ \ \; \ \ \; \ \ \; \ \ \; \ \ \; \ \ \; \ \ \; \ \ \; \ \ \; \ \ \; \ \ \; \ \ \; 1\; \ \ \; \ \ \; \ \ \\
\ \ \; \ \ \; \ \ \; \ \ \; \ \ \; \ \ \; \ \ \; \ \ \; \ \ \; \ \ \; \ \ \; \ \ \; \ \ \; \ \ \; 1\; \ \ 
\epm
\end{align}
Similarly, we can apply the $90^{\circ}$ rotations and obtain the
$16$ $\times$ $16$ $S$-matrix on the same basis:
\begin{align}
S=
\bpm
1 \; \ \ \; \ \ \; \ \ \; \ \ \; \ \ \; \ \ \; \ \ \; \ \ \; \ \ \; \ \ \; \ \ \; \ \ \; \ \ \; \ \ \; \ \ \\
\ \ \; \ \ \; \ \ \; \ \ \; 1 \; \ \ \; \ \ \; \ \ \; \ \ \; \ \ \; \ \ \; \ \ \; \ \ \; \ \ \; \ \ \; \ \ \\
\ \ \; \ \ \; \ \ \; \ \ \; \ \ \; \ \ \; \ \ \; \ \ \; \ \ \; \ \ \; \ \ \; \ \ \; 1 \; \ \ \; \ \ \; \ \ \\
\ \ \; \ \ \; \ \ \; \ \ \; \ \ \; \ \ \; \ \ \; \ \ \; 1 \; \ \ \; \ \ \; \ \ \; \ \ \; \ \ \; \ \ \; \ \ \\
\ \ \; 1 \; \ \ \; \ \ \; \ \ \; \ \ \; \ \ \; \ \ \; \ \ \; \ \ \; \ \ \; \ \ \; \ \ \; \ \ \; \ \ \; \ \ \\
\ \ \; \ \ \; \ \ \; \ \ \; \ \ \; \ \ \; \ \ \; \ \ \; \ \ \; \ \ \; \ \ \; \ \ \; \ \ \; \ \ \; \ \ \; \eta\ \\
\ \ \; \ \ \; \ \ \; \ \ \; \ \ \; \ \ \; \ \ \; \ \ \; \ \ \; \ \ \; \ \ \; \eta\; \ \ \; \ \ \; \ \ \; \ \ \\
\ \ \; \ \ \; \ \ \; \ \ \; \ \ \; \ \ \; \ \ \; 1 \; \ \ \; \ \ \; \ \ \; \ \ \; \ \ \; \ \ \; \ \ \; \ \ \\
\ \ \; \ \ \; 1 \; \ \ \; \ \ \; \ \ \; \ \ \; \ \ \; \ \ \; \ \ \; \ \ \; \ \ \; \ \ \; \ \ \; \ \ \; \ \ \\
\ \ \; \ \ \; \ \ \; \ \ \; \ \ \; \ \ \; \ \ \; \ \ \; \ \ \; \ \ \; 1 \; \ \ \; \ \ \; \ \ \; \ \ \; \ \ \\
\ \ \; \ \ \; \ \ \; \ \ \; \ \ \; \ \ \; \ \ \; \ \ \; \ \ \; \ \ \; \ \ \; \ \ \; \ \ \; 1 \; \ \ \; \ \ \\
\ \ \; \ \ \; \ \ \; \ \ \; \ \ \; \eta\; \ \ \; \ \ \; \ \ \; \ \ \; \ \ \; \ \ \; \ \ \; \ \ \; \ \ \; \ \ \\
\ \ \; \ \ \; \ \ \; 1 \; \ \ \; \ \ \; \ \ \; \ \ \; \ \ \; \ \ \; \ \ \; \ \ \; \ \ \; \ \ \; \ \ \; \ \ \ \\
\ \ \; \ \ \; \ \ \; \ \ \; \ \ \; \ \ \; \ \ \; \ \ \; \ \ \; \ \ \; \ \ \; \ \ \; \ \ \; \ \ \; 1\; \ \ \ \\
\ \ \; \ \ \; \ \ \; \ \ \; \ \ \; \ \ \; \ \ \; \ \ \; \ \ \; 1 \; \ \ \; \ \ \; \ \ \; \ \ \; \ \ \; \ \ \\
\ \ \; \ \ \; \ \ \; \ \ \; \ \ \; \ \ \; \eta\; \ \ \; \ \ \; \ \ \; \ \ \; \ \ \; \ \ \; \ \ \; \ \ \; \ \ 
\epm.
\end{align}

As before, we can choose a basis so that
$T$-matrix is diagonalized
and $S$-matrix satisfies eqn. \ref{Sconditions}.
For $\eta=1$,
we have the resulting $T$ and $S$-
matrices:
\begin{align}
T=
\bpm
1\; \ \ \; \ \ \; \ \ \; \ \ \; \ \ \; \ \ \; \ \ \; \ \ \; \ \ \; \ \ \; \ \ \; \ \ \; \ \ \; \ \ \; \ \ \\
\ \ \; 1\; \ \ \; \ \ \; \ \ \; \ \ \; \ \ \; \ \ \; \ \ \; \ \ \; \ \ \; \ \ \; \ \ \; \ \ \; \ \ \; \ \ \\
\ \ \; \ \ \; 1\; \ \ \; \ \ \; \ \ \; \ \ \; \ \ \; \ \ \; \ \ \; \ \ \; \ \ \; \ \ \; \ \ \; \ \ \; \ \ \\
\ \ \; \ \ \; \ \ \; 1\; \ \ \; \ \ \; \ \ \; \ \ \; \ \ \; \ \ \; \ \ \; \ \ \; \ \ \; \ \ \; \ \ \; \ \ \\
\ \ \; \ \ \; \ \ \; \ \ \; 1\; \ \ \; \ \ \; \ \ \; \ \ \; \ \ \; \ \ \; \ \ \; \ \ \; \ \ \; \ \ \; \ \ \\
\ \ \; \ \ \; \ \ \; \ \ \; \ \ \; -1\; \ \ \; \ \ \; \ \ \; \ \ \; \ \ \; \ \ \; \ \ \; \ \ \; \ \ \; \ \ \\
\ \ \; \ \ \; \ \ \; \ \ \; \ \ \; \ \ \; -1\; \ \ \; \ \ \; \ \ \; \ \ \; \ \ \; \ \ \; \ \ \; \ \ \; \ \ \\
\ \ \; \ \ \; \ \ \; \ \ \; \ \ \; \ \ \; \ \ \; 1\; \ \ \; \ \ \; \ \ \; \ \ \; \ \ \; \ \ \; \ \ \; \ \ \\
\ \ \; \ \ \; \ \ \; \ \ \; \ \ \; \ \ \; \ \ \; \ \ \; 1\; \ \ \; \ \ \; \ \ \; \ \ \; \ \ \; \ \ \; \ \ \\
\ \ \; \ \ \; \ \ \; \ \ \; \ \ \; \ \ \; \ \ \; \ \ \; \ \ \; -i\; \ \ \; \ \ \; \ \ \; \ \ \; \ \ \; \ \ \\
\ \ \; \ \ \; \ \ \; \ \ \; \ \ \; \ \ \; \ \ \; \ \ \; \ \ \; \ \ \; i\; \ \ \; \ \ \; \ \ \; \ \ \; \ \ \\
\ \ \; \ \ \; \ \ \; \ \ \; \ \ \; \ \ \; \ \ \; \ \ \; \ \ \; \ \ \; \ \ \; -1\; \ \ \; \ \ \; \ \ \; \ \ \\
\ \ \; \ \ \; \ \ \; \ \ \; \ \ \; \ \ \; \ \ \; \ \ \; \ \ \; \ \ \; \ \ \; \ \ \; 1\; \ \ \; \ \ \; \ \ \ \\
\ \ \; \ \ \; \ \ \; \ \ \; \ \ \; \ \ \; \ \ \; \ \ \; \ \ \; \ \ \; \ \ \; \ \ \; \ \ \; -i\; \ \ \; \ \ \\
\ \ \; \ \ \; \ \ \; \ \ \; \ \ \; \ \ \; \ \ \; \ \ \; \ \ \; \ \ \; \ \ \; \ \ \; \ \ \; \ \ \; i\; \ \ \\
\ \ \; \ \ \; \ \ \; \ \ \; \ \ \; \ \ \; \ \ \; \ \ \; \ \ \; \ \ \; \ \ \; \ \ \; \ \ \; \ \ \; \ \ \; -1 
\epm
\end{align}
\setlength{\arraycolsep}{1pt}
\footnotesize
\begin{align}
S=\frac{1}{4}
\bpm
1& 1& 1& 1& 1& 1& 1& 1& 1& 1& 1& 1& 1& 1& 1& 1\\
1& 1& 1& 1& 1& 1& 1& 1& -1& -1& -1& -1& -1& -1& -1& -1\\
1& 1& 1& 1& -1& -1& -1& -1& i& i& i& i& -i& -i& -i& -i\\
1& 1& 1& 1& -1& -1& -1& -1& -i& -i& -i& -i& i& i& i& i\\
1& 1& -1& -1& 1& -1& -1& 1& 1& -1& -1& 1& 1& -1& -1& 1\\
1& 1& -1& -1& -1& 1& 1& -1& -i& i& i& -i& i& -i& -i& i\\
1& 1& -1& -1& -1& 1& 1& -1& i& -i& -i& i& -i& i& i& -i\\
1& 1& -1& -1& 1& -1& -1& 1& -1& 1& 1& -1& -1& 1& 1& -1\\
1& -1& i&- i& 1& -i& i& -1& 1& i& -i& -1& 1& -i& i& -1\\
1& -1& i& -i& -1& i& -i& 1& i& -1& 1& -i& -i& -1& 1& i\\
1& -1& i& -i& -1& i& -i& 1& -i& 1& -1& i& i& 1& -1& -i\\
1& -1& i& -i& 1& -i& i& -1& -1& -i& i& 1& -1& i& -i& 1\\
1& -1& -i& i& 1& i& -i& -1& 1& -i& i& -1& 1& i& -i& -1\\
1& -1& -i& i& -1& -i& i& 1& -i& -1& 1& i& i& -1& 1& -i\\
1& -1& -i& i& -1& -i& i& 1& i& 1& -1& -i& -i& 1& -1& i\\
1& -1& -i& i& 1& i& -i& -1& -1& i& -i& 1& -1& -i& i& 1
\epm
\end{align}
\normalsize
\setlength{\arraycolsep}{2pt}
All information of the quasi-particles can be obtained
from the above $T$ and $S$ matrices.

Note that the eigenvalues of $T$-matrix are:
\begin{align*}
(1,1,1,1,1,1,1,1,-1,-1,-1,-1,i,i,-i,-i)
\end{align*}
which correspond exactly to the statistical angles of the
$\mathbb{Z}_4$ gauge theory,
or equivalently the $U(1)\times U(1)$ Chern-Simons gauge theory:
\begin{align}
 \cL=\frac{1}{4\pi} K^{IJ} a_{I\mu}\prt_\nu
a_{J\la}\eps^{\mu\nu\la} ,
\end{align}
with
$ K=\bpm  0 & 4 \\
         4 & 0 \\ \epm $.

On the other hand, when $\eta=-1$,
by properly choosing the basis,
we have the following $T$ and $S$-
matrices:
\begin{align}
T=
\bpm
1\; \ \ \; \ \ \; \ \ \; \ \ \; \ \ \; \ \ \; \ \ \; \ \ \; \ \ \; \ \ \; \ \ \; \ \ \; \ \ \; \ \ \; \ \ \\
\ \ \; -1\; \ \ \; \ \ \; \ \ \; \ \ \; \ \ \; \ \ \; \ \ \; \ \ \; \ \ \; \ \ \; \ \ \; \ \ \; \ \ \; \ \ \\
\ \ \; \ \ \; e^{{-i\frac{3\pi}{4}}}\; \ \ \; \ \ \; \ \ \; \ \ \; \ \ \; \ \ \; \ \ \; \ \ \; \ \ \; \ \ \; \ \ \; \ \ \; \ \ \\
\ \ \; \ \ \; \ \ \; e^{{-i\frac{3\pi}{4}}}\; \ \ \; \ \ \; \ \ \; \ \ \; \ \ \; \ \ \; \ \ \; \ \ \; \ \ \; \ \ \; \ \ \; \ \ \\
\ \ \; \ \ \; \ \ \; \ \ \; -1\; \ \ \; \ \ \; \ \ \; \ \ \; \ \ \; \ \ \; \ \ \; \ \ \; \ \ \; \ \ \; \ \ \\
\ \ \; \ \ \; \ \ \; \ \ \; \ \ \; 1\; \ \ \; \ \ \; \ \ \; \ \ \; \ \ \; \ \ \; \ \ \; \ \ \; \ \ \; \ \ \\
\ \ \; \ \ \; \ \ \; \ \ \; \ \ \; \ \ \; e^{{i\frac{\pi}{4}}}\; \ \ \; \ \ \; \ \ \; \ \ \; \ \ \; \ \ \; \ \ \; \ \ \; \ \ \\
\ \ \; \ \ \; \ \ \; \ \ \; \ \ \; \ \ \; \ \ \; e^{{i\frac{\pi}{4}}}\; \ \ \; \ \ \; \ \ \; \ \ \; \ \ \; \ \ \; \ \ \; \ \ \\
\ \ \; \ \ \; \ \ \; \ \ \; \ \ \; \ \ \; \ \ \; \ \ \; e^{{i\frac{3\pi}{4}}}\; \ \ \; \ \ \; \ \ \; \ \ \; \ \ \; \ \ \; \ \ \\
\ \ \; \ \ \; \ \ \; \ \ \; \ \ \; \ \ \; \ \ \; \ \ \; \ \ \; e^{{-i\frac{\pi}{4}}}\; \ \ \; \ \ \; \ \ \; \ \ \; \ \ \; \ \ \\
\ \ \; \ \ \; \ \ \; \ \ \; \ \ \; \ \ \; \ \ \; \ \ \; \ \ \; \ \ \; 1\; \ \ \; \ \ \; \ \ \; \ \ \; \ \ \\
\ \ \; \ \ \; \ \ \; \ \ \; \ \ \; \ \ \; \ \ \; \ \ \; \ \ \; \ \ \; \ \ \; 1\; \ \ \; \ \ \; \ \ \; \ \ \\
\ \ \; \ \ \; \ \ \; \ \ \; \ \ \; \ \ \; \ \ \; \ \ \; \ \ \; \ \ \; \ \ \; \ \ \;  e^{{i\frac{3\pi}{4}}}\; \ \ \; \ \ \; \ \ \ \\
\ \ \; \ \ \; \ \ \; \ \ \; \ \ \; \ \ \; \ \ \; \ \ \; \ \ \; \ \ \; \ \ \; \ \ \; \ \ \; e^{{-i\frac{\pi}{4}}}\; \ \ \; \ \ \\
\ \ \; \ \ \; \ \ \; \ \ \; \ \ \; \ \ \; \ \ \; \ \ \; \ \ \; \ \ \; \ \ \; \ \ \; \ \ \; \ \ \; 1\; \ \ \\
\ \ \; \ \ \; \ \ \; \ \ \; \ \ \; \ \ \; \ \ \; \ \ \; \ \ \; \ \ \; \ \ \; \ \ \; \ \ \; \ \ \; \ \ \; 1 
\epm
\end{align}

\setlength{\arraycolsep}{1pt}
\footnotesize
\begin{align}
S=\frac{1}{4}
\bpm
 1& 1& 1& 1& 1& 1& 1& 1& 1& 1& 1& 1& 1& 1& 1& 1&\\
 1& 1& -1& -1& 1& 1& -1& -1& 1& 1& -1& -1& 1& 1& -1& -1&\\
 1& -1& -i& i& 1& -1& -i& i& 1& -1& -i& i& 1& -1& -i& i&\\
 1& -1& i& -i& 1& -1& i& -i& 1& -1& i& -i& 1& -1& i& -i&\\
 1& 1& 1& 1& 1& 1& 1& 1& -1& -1& -1& -1& -1& -1& -1& -1&\\
 1& 1& -1& -1& 1& 1& -1& -1& -1& -1& 1& 1& -1& -1& 1& 1&\\
 1& -1& -i& i& 1& -1& -i& i& -1& 1& i& -i& -1& 1& i& -i&\\
 1& -1& i& -i& 1& -1& i& -i& -1& 1& -i& i& -1& 1& -i& i&\\
 1& 1& 1& 1& -1& -1& -1& -1& i& i& i& i& -i& -i& -i& -i&\\
 1& 1& -1& -1& -1& -1& 1& 1& i& i& -i& -i& -i& -i& i& i&\\
 1& -1& -i& i& -1& 1& i& -i& i& -i& 1& -1& -i& i& -1& 1&\\
 1& -1& i& -i& -1& 1& -i& i& i& -i& -1& 1& -i& i& 1& -1&\\
 1& 1& 1& 1& -1& -1& -1& -1& -i& -i& -i& -i& i& i& i& i&\\
 1& 1& -1& -1& -1& -1& 1& 1& -i& -i& i& i& i& i& -i& -i&\\
 1& -1& -i& i& -1& 1& i& -i& -i& i& -1& 1& i& -i& 1& -1&\\
 1& -1& i& -i& -1& 1& -i& i& -i& i& 1& -1& i& -i& -1& 1
\epm
\end{align}
\normalsize
\setlength{\arraycolsep}{2pt}
In this particular case, again, the above
$T$ and $S$ matrices can be further reduced
to the following forms:
\begin{align}
 T=
\bpm
1& 0& 0& 0&\\
0& -1& 0& 0&\\
0& 0& e^{{i\frac{3\pi}{4}}}& 0&\\
0& 0& 0& e^{{i\frac{3\pi}{4}}}&\\
\epm
\otimes
\bpm
1& 0& 0& 0&\\
0& -1& 0& 0&\\
0& 0& e^{{-i\frac{3\pi}{4}}}& 0&\\
0& 0& 0& e^{{-i\frac{3\pi}{4}}}&\\
\epm
\end{align}
\begin{align}
 S=
\frac{1}{2}
\bpm
1& 1& 1& 1&\\
1& 1& -1& -1&\\
1& -1& i& -i&\\
1& -1& -i& i&\\
\epm
\otimes
\frac{1}{2}
\bpm
1& 1& 1& 1&\\
1& 1& -1& -1&\\
1& -1& -i& i&\\
1& -1& i& -i&\\
\epm
\end{align}
which shows the ``doubled'' structure
of the quasi-particles.
As before, all information of the quasi-particles can be obtained
from the above $T$ and $S$ matrices.

Note that the eigenvalues of $T$ now
become:
\begin{align*}
(1,1,1,1,1,1,-1,-1,e^{\pm{i\frac{\pi}{4}}},e^{\pm{i\frac{3\pi}{4}}})
\end{align*}
which correspond exactly to the statistical angles of
the $U(1)\times U(1)$ Chern-Simons
gauge theory
\begin{align}
\cL=\frac {1}{4\pi}\Big(
4 a_{1\mu}\prt_{\nu}a_{2\la}\eps^{\mu\nu\la}
-4 a_{2\mu}\prt_{\nu}a_{1\la}\eps^{\mu\nu\la}
\Big) ,
\end{align}
or equivalently the $U(1)\times U(1)$ Chern-Simons gauge theory:
\begin{align}
 \cL=\frac{1}{4\pi} K^{IJ} a_{I\mu}\prt_\nu
a_{J\la}\eps^{\mu\nu\la} ,
\end{align}
with
$ K=\bpm  0 & 4 \\
         4 & 4 \\ \epm $.

\subsection{More $N=3$ string-net states -- the $\mathbb{Z}_2\times\mathbb{Z}_2$ states}

Now we will give an example of a non-orientable
string-net state with four edge states.
That is to set $N=3$, $0^*=0$, $1^*=1$,
$2^*=2$, $3^*=3$. We then have the following
solution for $N$:
\begin{align}
N_{000}&=
N_{011}=
N_{101}=
N_{110}=
N_{022}=
N_{202}=
N_{220}
\nonumber\\
&=
N_{033}=
N_{303}=
N_{330}=
N_{123}=
N_{231}=
N_{312}
\nonumber\\
&=
N_{132}=
N_{321}=
N_{213}=1.
\end{align}
The above $N_{ijk}$
satisfies \eqn{Phinonz}.

As before, we can recognize from the above that the
three edge labels of $N$ form a
$\mathbb{Z}_2\times\mathbb{Z}_2$ group if we flip the
positive direction of the third label. Since the state
is non-orientable, switching the direction has no real
effects. Note that any element square will give unity,
which is precisely the property for group
$\mathbb{Z}_2\times\mathbb{Z}_2$.

Similarly, due to relation \eqn{Phinonz},
different components of the tensor $F^{ijm}_{kln}$ are not
independent.  There are now twenty-two independent potentially 
non-zero components which are denoted as $f_0$,...,$f_{21}$:
\begin{align}
F^{000}_{000}
\bmm\begin{tikzpicture}[scale=0.26]
\FBox \iLnkaa \jLnkaa \kLnkaa \lLnkaa \mLnkaa \nLnkaa 
\end{tikzpicture}\emm
&=f_{0}
\nonumber\\
F^{000}_{111}
\bmm\begin{tikzpicture}[scale=0.26]
\FBox \iLnkaa \jLnkaa \kLnkbb \lLnkbb \mLnkaa \nLnkbb 
\end{tikzpicture}\emm
&=(F^{011}_{100}
\bmm\begin{tikzpicture}[scale=0.26]
\FBox \iLnkaa \jLnkbb \kLnkbb \lLnkaa \mLnkbb \nLnkaa 
\end{tikzpicture}\emm
)^*=(F^{101}_{010}
\bmm\begin{tikzpicture}[scale=0.26]
\FBox \iLnkbb \jLnkaa \kLnkaa \lLnkbb \mLnkbb \nLnkaa 
\end{tikzpicture}\emm
)^*
\nonumber\\
&=F^{110}_{001}
\bmm\begin{tikzpicture}[scale=0.26]
\FBox \iLnkbb \jLnkbb \kLnkaa \lLnkaa \mLnkaa \nLnkbb 
\end{tikzpicture}\emm
=f_{1}
\nonumber\\
F^{000}_{222}
\bmm\begin{tikzpicture}[scale=0.26]
\FBox \iLnkaa \jLnkaa \kLnkcc \lLnkcc \mLnkaa \nLnkcc 
\end{tikzpicture}\emm
&=(F^{022}_{200}
\bmm\begin{tikzpicture}[scale=0.26]
\FBox \iLnkaa \jLnkcc \kLnkcc \lLnkaa \mLnkcc \nLnkaa 
\end{tikzpicture}\emm
)^*=(F^{202}_{020}
\bmm\begin{tikzpicture}[scale=0.26]
\FBox \iLnkcc \jLnkaa \kLnkaa \lLnkcc \mLnkcc \nLnkaa 
\end{tikzpicture}\emm
)^*
\nonumber\\
&=F^{220}_{002}
\bmm\begin{tikzpicture}[scale=0.26]
\FBox \iLnkcc \jLnkcc \kLnkaa \lLnkaa \mLnkaa \nLnkcc 
\end{tikzpicture}\emm
=f_{2}
\nonumber\\
F^{000}_{333}
\bmm\begin{tikzpicture}[scale=0.26]
\FBox \iLnkaa \jLnkaa \kLnkdd \lLnkdd \mLnkaa \nLnkdd 
\end{tikzpicture}\emm
&=(F^{033}_{300}
\bmm\begin{tikzpicture}[scale=0.26]
\FBox \iLnkaa \jLnkdd \kLnkdd \lLnkaa \mLnkdd \nLnkaa 
\end{tikzpicture}\emm
)^*=(F^{303}_{030}
\bmm\begin{tikzpicture}[scale=0.26]
\FBox \iLnkdd \jLnkaa \kLnkaa \lLnkdd \mLnkdd \nLnkaa 
\end{tikzpicture}\emm
)^*
\nonumber\\
&=F^{330}_{003}
\bmm\begin{tikzpicture}[scale=0.26]
\FBox \iLnkdd \jLnkdd \kLnkaa \lLnkaa \mLnkaa \nLnkdd 
\end{tikzpicture}\emm
=f_{3}
\nonumber\\
F^{011}_{011}
\bmm\begin{tikzpicture}[scale=0.26]
\FBox \iLnkaa \jLnkbb \kLnkaa \lLnkbb \mLnkbb \nLnkbb 
\end{tikzpicture}\emm
&=(F^{101}_{101}
\bmm\begin{tikzpicture}[scale=0.26]
\FBox \iLnkbb \jLnkaa \kLnkbb \lLnkaa \mLnkbb \nLnkbb 
\end{tikzpicture}\emm
)^*=f_{4}
\nonumber
\end{align}

\begin{align}
F^{011}_{233}
\bmm\begin{tikzpicture}[scale=0.26]
\FBox \iLnkaa \jLnkbb \kLnkcc \lLnkdd \mLnkbb \nLnkdd 
\end{tikzpicture}\emm
&=(F^{123}_{301}
\bmm\begin{tikzpicture}[scale=0.26]
\FBox \iLnkbb \jLnkcc \kLnkdd \lLnkaa \mLnkdd \nLnkbb 
\end{tikzpicture}\emm
)^*=F^{231}_{013}
\bmm\begin{tikzpicture}[scale=0.26]
\FBox \iLnkcc \jLnkdd \kLnkaa \lLnkbb \mLnkbb \nLnkdd 
\end{tikzpicture}\emm
\nonumber\\
&=(F^{303}_{121}
\bmm\begin{tikzpicture}[scale=0.26]
\FBox \iLnkdd \jLnkaa \kLnkbb \lLnkcc \mLnkdd \nLnkbb 
\end{tikzpicture}\emm
)^*=f_{5}
\nonumber\\
F^{011}_{322}
\bmm\begin{tikzpicture}[scale=0.26]
\FBox \iLnkaa \jLnkbb \kLnkdd \lLnkcc \mLnkbb \nLnkcc 
\end{tikzpicture}\emm
&=(F^{132}_{201}
\bmm\begin{tikzpicture}[scale=0.26]
\FBox \iLnkbb \jLnkdd \kLnkcc \lLnkaa \mLnkcc \nLnkbb 
\end{tikzpicture}\emm
)^*=(F^{202}_{131}
\bmm\begin{tikzpicture}[scale=0.26]
\FBox \iLnkcc \jLnkaa \kLnkbb \lLnkdd \mLnkcc \nLnkbb 
\end{tikzpicture}\emm
)^*
\nonumber\\
&=F^{321}_{012}
\bmm\begin{tikzpicture}[scale=0.26]
\FBox \iLnkdd \jLnkcc \kLnkaa \lLnkbb \mLnkbb \nLnkcc 
\end{tikzpicture}\emm
=f_{6}
\nonumber\\
F^{022}_{022}
\bmm\begin{tikzpicture}[scale=0.26]
\FBox \iLnkaa \jLnkcc \kLnkaa \lLnkcc \mLnkcc \nLnkcc 
\end{tikzpicture}\emm
&=(F^{202}_{202}
\bmm\begin{tikzpicture}[scale=0.26]
\FBox \iLnkcc \jLnkaa \kLnkcc \lLnkaa \mLnkcc \nLnkcc 
\end{tikzpicture}\emm
)^*=f_{7}
\nonumber\\
F^{022}_{133}
\bmm\begin{tikzpicture}[scale=0.26]
\FBox \iLnkaa \jLnkcc \kLnkbb \lLnkdd \mLnkcc \nLnkdd 
\end{tikzpicture}\emm
&=F^{132}_{023}
\bmm\begin{tikzpicture}[scale=0.26]
\FBox \iLnkbb \jLnkdd \kLnkaa \lLnkcc \mLnkcc \nLnkdd 
\end{tikzpicture}\emm
=(F^{213}_{302}
\bmm\begin{tikzpicture}[scale=0.26]
\FBox \iLnkcc \jLnkbb \kLnkdd \lLnkaa \mLnkdd \nLnkcc 
\end{tikzpicture}\emm
)^*
\nonumber\\
&=(F^{303}_{212}
\bmm\begin{tikzpicture}[scale=0.26]
\FBox \iLnkdd \jLnkaa \kLnkcc \lLnkbb \mLnkdd \nLnkcc 
\end{tikzpicture}\emm
)^*=f_{8}
\nonumber\\
F^{022}_{311}
\bmm\begin{tikzpicture}[scale=0.26]
\FBox \iLnkaa \jLnkcc \kLnkdd \lLnkbb \mLnkcc \nLnkbb 
\end{tikzpicture}\emm
&=(F^{101}_{232}
\bmm\begin{tikzpicture}[scale=0.26]
\FBox \iLnkbb \jLnkaa \kLnkcc \lLnkdd \mLnkbb \nLnkcc 
\end{tikzpicture}\emm
)^*=(F^{231}_{102}
\bmm\begin{tikzpicture}[scale=0.26]
\FBox \iLnkcc \jLnkdd \kLnkbb \lLnkaa \mLnkbb \nLnkcc 
\end{tikzpicture}\emm
)^*
\nonumber\\
&=F^{312}_{021}
\bmm\begin{tikzpicture}[scale=0.26]
\FBox \iLnkdd \jLnkbb \kLnkaa \lLnkcc \mLnkcc \nLnkbb 
\end{tikzpicture}\emm
=f_{9}
\nonumber
\end{align}

\begin{align}
F^{033}_{033}
\bmm\begin{tikzpicture}[scale=0.26]
\FBox \iLnkaa \jLnkdd \kLnkaa \lLnkdd \mLnkdd \nLnkdd 
\end{tikzpicture}\emm
&=(F^{303}_{303}
\bmm\begin{tikzpicture}[scale=0.26]
\FBox \iLnkdd \jLnkaa \kLnkdd \lLnkaa \mLnkdd \nLnkdd 
\end{tikzpicture}\emm
)^*=f_{10}
\nonumber\\
F^{033}_{122}
\bmm\begin{tikzpicture}[scale=0.26]
\FBox \iLnkaa \jLnkdd \kLnkbb \lLnkcc \mLnkdd \nLnkcc 
\end{tikzpicture}\emm
&=F^{123}_{032}
\bmm\begin{tikzpicture}[scale=0.26]
\FBox \iLnkbb \jLnkcc \kLnkaa \lLnkdd \mLnkdd \nLnkcc 
\end{tikzpicture}\emm
=(F^{202}_{313}
\bmm\begin{tikzpicture}[scale=0.26]
\FBox \iLnkcc \jLnkaa \kLnkdd \lLnkbb \mLnkcc \nLnkdd 
\end{tikzpicture}\emm
)^*
\nonumber\\
&=(F^{312}_{203}
\bmm\begin{tikzpicture}[scale=0.26]
\FBox \iLnkdd \jLnkbb \kLnkcc \lLnkaa \mLnkcc \nLnkdd 
\end{tikzpicture}\emm
)^*=f_{11}
\nonumber\\
F^{033}_{211}
\bmm\begin{tikzpicture}[scale=0.26]
\FBox \iLnkaa \jLnkdd \kLnkcc \lLnkbb \mLnkdd \nLnkbb 
\end{tikzpicture}\emm
&=(F^{101}_{323}
\bmm\begin{tikzpicture}[scale=0.26]
\FBox \iLnkbb \jLnkaa \kLnkdd \lLnkcc \mLnkbb \nLnkdd 
\end{tikzpicture}\emm
)^*=F^{213}_{031}
\bmm\begin{tikzpicture}[scale=0.26]
\FBox \iLnkcc \jLnkbb \kLnkaa \lLnkdd \mLnkdd \nLnkbb 
\end{tikzpicture}\emm
\nonumber\\
&=(F^{321}_{103}
\bmm\begin{tikzpicture}[scale=0.26]
\FBox \iLnkdd \jLnkcc \kLnkbb \lLnkaa \mLnkbb \nLnkdd 
\end{tikzpicture}\emm
)^*=f_{12}
\nonumber\\
F^{110}_{110}
\bmm\begin{tikzpicture}[scale=0.26]
\FBox \iLnkbb \jLnkbb \kLnkbb \lLnkbb \mLnkaa \nLnkaa 
\end{tikzpicture}\emm
&=f_{13}
\nonumber\\
F^{110}_{223}
\bmm\begin{tikzpicture}[scale=0.26]
\FBox \iLnkbb \jLnkbb \kLnkcc \lLnkcc \mLnkaa \nLnkdd 
\end{tikzpicture}\emm
&=(F^{123}_{210}
\bmm\begin{tikzpicture}[scale=0.26]
\FBox \iLnkbb \jLnkcc \kLnkcc \lLnkbb \mLnkdd \nLnkaa 
\end{tikzpicture}\emm
)^*=(F^{213}_{120}
\bmm\begin{tikzpicture}[scale=0.26]
\FBox \iLnkcc \jLnkbb \kLnkbb \lLnkcc \mLnkdd \nLnkaa 
\end{tikzpicture}\emm
)^*
\nonumber\\
&=F^{220}_{113}
\bmm\begin{tikzpicture}[scale=0.26]
\FBox \iLnkcc \jLnkcc \kLnkbb \lLnkbb \mLnkaa \nLnkdd 
\end{tikzpicture}\emm
=f_{14}
\nonumber
\end{align}

\begin{align}
F^{110}_{332}
\bmm\begin{tikzpicture}[scale=0.26]
\FBox \iLnkbb \jLnkbb \kLnkdd \lLnkdd \mLnkaa \nLnkcc 
\end{tikzpicture}\emm
&=(F^{132}_{310}
\bmm\begin{tikzpicture}[scale=0.26]
\FBox \iLnkbb \jLnkdd \kLnkdd \lLnkbb \mLnkcc \nLnkaa 
\end{tikzpicture}\emm
)^*=(F^{312}_{130}
\bmm\begin{tikzpicture}[scale=0.26]
\FBox \iLnkdd \jLnkbb \kLnkbb \lLnkdd \mLnkcc \nLnkaa 
\end{tikzpicture}\emm
)^*
\nonumber\\
&=F^{330}_{112}
\bmm\begin{tikzpicture}[scale=0.26]
\FBox \iLnkdd \jLnkdd \kLnkbb \lLnkbb \mLnkaa \nLnkcc 
\end{tikzpicture}\emm
=f_{15}
\nonumber\\
F^{123}_{123}
\bmm\begin{tikzpicture}[scale=0.26]
\FBox \iLnkbb \jLnkcc \kLnkbb \lLnkcc \mLnkdd \nLnkdd 
\end{tikzpicture}\emm
&=(F^{213}_{213}
\bmm\begin{tikzpicture}[scale=0.26]
\FBox \iLnkcc \jLnkbb \kLnkcc \lLnkbb \mLnkdd \nLnkdd 
\end{tikzpicture}\emm
)^*=f_{16}
\nonumber\\
F^{132}_{132}
\bmm\begin{tikzpicture}[scale=0.26]
\FBox \iLnkbb \jLnkdd \kLnkbb \lLnkdd \mLnkcc \nLnkcc 
\end{tikzpicture}\emm
&=(F^{312}_{312}
\bmm\begin{tikzpicture}[scale=0.26]
\FBox \iLnkdd \jLnkbb \kLnkdd \lLnkbb \mLnkcc \nLnkcc 
\end{tikzpicture}\emm
)^*=f_{17}
\nonumber\\
F^{220}_{220}
\bmm\begin{tikzpicture}[scale=0.26]
\FBox \iLnkcc \jLnkcc \kLnkcc \lLnkcc \mLnkaa \nLnkaa 
\end{tikzpicture}\emm
&=f_{18}
\nonumber\\
F^{220}_{331}
\bmm\begin{tikzpicture}[scale=0.26]
\FBox \iLnkcc \jLnkcc \kLnkdd \lLnkdd \mLnkaa \nLnkbb 
\end{tikzpicture}\emm
&=(F^{231}_{320}
\bmm\begin{tikzpicture}[scale=0.26]
\FBox \iLnkcc \jLnkdd \kLnkdd \lLnkcc \mLnkbb \nLnkaa 
\end{tikzpicture}\emm
)^*=(F^{321}_{230}
\bmm\begin{tikzpicture}[scale=0.26]
\FBox \iLnkdd \jLnkcc \kLnkcc \lLnkdd \mLnkbb \nLnkaa 
\end{tikzpicture}\emm
)^*
\nonumber\\
&=F^{330}_{221}
\bmm\begin{tikzpicture}[scale=0.26]
\FBox \iLnkdd \jLnkdd \kLnkcc \lLnkcc \mLnkaa \nLnkbb 
\end{tikzpicture}\emm
=f_{19}
\nonumber\\
F^{231}_{231}
\bmm\begin{tikzpicture}[scale=0.26]
\FBox \iLnkcc \jLnkdd \kLnkcc \lLnkdd \mLnkbb \nLnkbb 
\end{tikzpicture}\emm
&=(F^{321}_{321}
\bmm\begin{tikzpicture}[scale=0.26]
\FBox \iLnkdd \jLnkcc \kLnkdd \lLnkcc \mLnkbb \nLnkbb 
\end{tikzpicture}\emm
)^*=f_{20}
\nonumber\\
F^{330}_{330}
\bmm\begin{tikzpicture}[scale=0.26]
\FBox \iLnkdd \jLnkdd \kLnkdd \lLnkdd \mLnkaa \nLnkaa 
\end{tikzpicture}\emm
&=f_{21}
\end{align}
There are now sixteen potentially non-zero components
in $P^{kj}_i$, which are denoted by
$p_0$,...,$p_{15}$:
\begin{align}
&
P^{00}_{0}=p_{0},\ \
P^{01}_{0}=p_{1},\ \
P^{02}_{0}=p_{2},\ \
P^{03}_{0}=p_{3},\ \
P^{00}_{1}=p_{4},
\nonumber\\
&
P^{01}_{1}=p_{5},\ \
P^{02}_{1}=p_{6},\ \
P^{03}_{1}=p_{7},\ \
P^{00}_{2}=p_{8},\ \
P^{01}_{2}=p_{9},
\nonumber\\
&
P^{02}_{2}=p_{10},\ \
P^{03}_{2}=p_{11},\ \
P^{00}_{3}=p_{12},\ \
P^{01}_{3}=p_{13},
\nonumber\\
&
P^{02}_{3}=p_{14},\ \
P^{03}_{3}=p_{15}.
\end{align}
Using the ``gauge fixing'' discussed in section
\Ref{CGW1038}, we can fix the phases of $f_1$, $f_2$, $f_3$,
$f_5$, $f_6$, $f_8$, $f_9$, $f_{11}$, $f_{12}$, $f_{14}$,
$f_{16}$, $p_0$ and $A^0$ to make them positive. As
before, these phases are not independent to each other,
so the gauge fixing has to be done in a self-consistent way.

The fixed-point conditions (\eqn{Phinonz}) form a set of non-linear
equation on the variables $f_i$, $p_i$, and $A^i$, which can
be solved exactly. After applying the ``gauge fixing''
discussed above, we find four sets of different
solutions, listing below:
\begin{align}
I.\ \ 
&
f_i=1,\ \ i=0,1,...,21,
\nonumber\\
&
p_i=\frac{1}{2},\ \ i=0,1,...,15,
\nonumber\\
&
A^0=A^1=A^2=A^3=\frac{1}{2}.
\nonumber\\
\nonumber\\
II.\ \ 
&
f_{13}=f_{15}=f_{17}=f_{18}=f_{19}=f_{20}= -1,
\nonumber\\
&
f_i=1,\ \ i=otherwise,
\nonumber\\
&
p_0=p_3=p_4=p_7=p_8=p_{11}=p_{12}=p_{15}=\frac{1}{2},
\nonumber\\
&
p_1=p_2=p_5=p_6=p_9=p_{10}=p_{13}=p_{14}=-\frac{1}{2},
\nonumber\\
&
A^0= -A^1= -A^2=A^3=\frac{1}{2}.
\nonumber
\end{align}

\begin{align}
III.\ \ 
&
f_{13}=f_{15}=f_{17}=f_{21}= -1,
\nonumber\\
&
f_i=1,\ \ i=otherwise,
\nonumber\\
&
p_0=p_2=p_4=...=p_{14}=\frac{1}{2},
\nonumber\\
&
p_1=p_3=p_5=...=p_{15}=-\frac{1}{2},
\nonumber\\
&
A^0= -A^1=A^2= -A^3=\frac{1}{2}.
\nonumber\\
\nonumber\\
IV.\ \ 
&
f_{18}=f_{19}=f_{20}=f_{21}= -1,
\nonumber\\
&
f_i=1,\ \ i=otherwise,
\nonumber\\
&
p_0=p_1=p_4=p_5=p_8=p_{9}=p_{12}=p_{13}=\frac{1}{2},
\nonumber\\
&
p_2=p_3=p_6=p_7=p_{10}=p_{11}=p_{14}=p_{15}=\frac{1}{2},
\nonumber\\
&
A^0=A^1= -A^2= -A^3=\frac{1}{2}.
\end{align}

We also find
\begin{align}
 \e^{\imth \th_F}=
 \e^{\imth \th_{P1}}=
 \e^{\imth \th_{P2}}=
 \e^{\imth \th_{A1}}=
 \e^{\imth \th_{A2}}=
1.
\end{align}

We can further
check if the results obtained above are consistent with the ones
in the $\mathbb{Z}_2$ case (c.f. \ref{Z2state}).
A quick check can be done as follows:
In \ref{Z2state}, we have
\begin{equation*}
f^\prime_0=1 \ \ \ and \ \ \  f^\prime_3=\eta.
\end{equation*}
(We use $f^\prime$ to differentiate it from the $f$ obtained
in this section.) Since
$\mathbb{Z}_2\times\mathbb{Z}_2$ can be seen as two copies
of $\mathbb{Z}_2$, we expect
\begin{align*}
&
f_0=f^{a\prime}_0 \times f^{b\prime}_0=1,\ \ 
f_{13}=f^{a\prime}_3 \times f^{b\prime}_0=\eta^a,
\\
&
f_{18}=f^{a\prime}_0 \times f^{b\prime}_3=\eta^b,\ \ 
f_{21}=f^{a\prime}_3 \times f^{b\prime}_3=\eta^a \times \eta^b.
\end{align*}
Thus $f_{13}$, $f_{18}$, $f_{21}$ would either all be one, or two
out of them would be $-1$. This precisely corresponds to
the four solutions obtained above. A more detailed check
also shows consistency between the two results.

\subsection{An $N=3$ string-net state -- the ``Chiral'' state}
\label{Chiral}

Finally, we present here a new state which was once thought
to be a non-string-net state (but eventually turned out not to be one,
fortunately or unfortunately).
In this state, interestingly enough, the 
chiral symmetry is spontaneously broken. As before, we choose 
$N=3$, $0^*=0$, $1^*=3$, $2^*=2$, $3^*=1$, and
\begin{align}
\label{ChiralN}
N_{000}&=
N_{013}=
N_{130}=
N_{301}=
N_{123}=
N_{231}=
N_{312}
\nonumber\\
&=
N_{222}=1.
\end{align}
The above $N_{ijk}$
satisfies \eqn{Phinonz}.

This state has several interesting features. Besides its chiral
symmetry being spontaneously broken, the three edge labels
of $N$ form a structure called a ``multi-fusion category''. This
mathematical structure has the following fusion rule between
its four elements:
\begin{equation*}
M_{ij} \times M_{kl} = \delta_{jk} M_{il}, \ \ where \  i, j, k, l = 1 \ or \ 2.
\end{equation*}
After making the following mapping, $M_{11} = state \ 0$,
$M_{12} = state \ 1$, $M_{22} = state \ 2$, $M_{21} = state \ 3$,
we can see that the
fusion rule above is exactly equivalent to \eq{ChiralN} 
(again, we need to flip the positive direction of the
third label of $N$). Although the fusion rule is fairly non-trivial,
we will see that at least in the no-symmetry case which we are
studying in this paper, the corresponding fixed-point state
is trivial. 

Due to relation \eqn{Phinonz},
different components of the tensor $F^{ijm}_{kln}$ are not
independent.  There are now six independent potentially non-zero
components which are denoted as $f_0$,...,$f_5$:
\begin{align}
F^{000}_{000}
\bmm\begin{tikzpicture}[scale=0.26]
\FBox \iLnkaa \jLnkaa \kLnkaa \lLnkaa \mLnkaa \nLnkaa 
\end{tikzpicture}\emm
&=f_{0}
\nonumber\\
F^{000}_{333}
\bmm\begin{tikzpicture}[scale=0.26]
\FBox \iLnkaa \jLnkaa \kLnkdb \lLnkdb \mLnkaa \nLnkdb 
\end{tikzpicture}\emm
&=(F^{033}_{100}
\bmm\begin{tikzpicture}[scale=0.26]
\FBox \iLnkaa \jLnkdb \kLnkbd \lLnkaa \mLnkdb \nLnkaa 
\end{tikzpicture}\emm
)^*=(F^{101}_{010}
\bmm\begin{tikzpicture}[scale=0.26]
\FBox \iLnkbd \jLnkaa \kLnkaa \lLnkbd \mLnkbd \nLnkaa 
\end{tikzpicture}\emm
)^*
\nonumber\\
&=F^{310}_{001}
\bmm\begin{tikzpicture}[scale=0.26]
\FBox \iLnkdb \jLnkbd \kLnkaa \lLnkaa \mLnkaa \nLnkbd 
\end{tikzpicture}\emm
=f_{1}
\nonumber\\
F^{033}_{233}
\bmm\begin{tikzpicture}[scale=0.26]
\FBox \iLnkaa \jLnkdb \kLnkcc \lLnkdb \mLnkdb \nLnkdb 
\end{tikzpicture}\emm
&=(F^{101}_{323}
\bmm\begin{tikzpicture}[scale=0.26]
\FBox \iLnkbd \jLnkaa \kLnkdb \lLnkcc \mLnkbd \nLnkdb 
\end{tikzpicture}\emm
)^*=F^{211}_{011}
\bmm\begin{tikzpicture}[scale=0.26]
\FBox \iLnkcc \jLnkbd \kLnkaa \lLnkbd \mLnkbd \nLnkbd 
\end{tikzpicture}\emm
\nonumber\\
&=(F^{323}_{101}
\bmm\begin{tikzpicture}[scale=0.26]
\FBox \iLnkdb \jLnkcc \kLnkbd \lLnkaa \mLnkdb \nLnkbd 
\end{tikzpicture}\emm
)^*=f_{2}
\nonumber\\
F^{132}_{110}
\bmm\begin{tikzpicture}[scale=0.26]
\FBox \iLnkbd \jLnkdb \kLnkbd \lLnkbd \mLnkcc \nLnkaa 
\end{tikzpicture}\emm
&=(F^{310}_{332}
\bmm\begin{tikzpicture}[scale=0.26]
\FBox \iLnkdb \jLnkbd \kLnkdb \lLnkdb \mLnkaa \nLnkcc 
\end{tikzpicture}\emm
)^*=f_{3}
\nonumber\\
F^{132}_{223}
\bmm\begin{tikzpicture}[scale=0.26]
\FBox \iLnkbd \jLnkdb \kLnkcc \lLnkcc \mLnkcc \nLnkdb 
\end{tikzpicture}\emm
&=(F^{211}_{322}
\bmm\begin{tikzpicture}[scale=0.26]
\FBox \iLnkcc \jLnkbd \kLnkdb \lLnkcc \mLnkbd \nLnkcc 
\end{tikzpicture}\emm
)^*=F^{222}_{111}
\bmm\begin{tikzpicture}[scale=0.26]
\FBox \iLnkcc \jLnkcc \kLnkbd \lLnkbd \mLnkcc \nLnkbd 
\end{tikzpicture}\emm
\nonumber\\
&=(F^{323}_{232}
\bmm\begin{tikzpicture}[scale=0.26]
\FBox \iLnkdb \jLnkcc \kLnkcc \lLnkdb \mLnkdb \nLnkcc 
\end{tikzpicture}\emm
)^*=f_{4}
\nonumber\\
F^{222}_{222}
\bmm\begin{tikzpicture}[scale=0.26]
\FBox \iLnkcc \jLnkcc \kLnkcc \lLnkcc \mLnkcc \nLnkcc 
\end{tikzpicture}\emm
&=f_{5}
\end{align}
There are eight potentially non-zero components
in $P^{kj}_i$, which are denoted by
$p_0$,...,$p_7$:
\begin{align}
&
P^{00}_{0}=p_{0},\ \
P^{03}_{0}=p_{1},\ \
P^{00}_{1}=p_{2},\ \
P^{03}_{1}=p_{3},\ \
P^{21}_{2}=p_{4},
\nonumber\\
&
P^{22}_{2}=p_{5},\ \
P^{21}_{3}=p_{6},\ \
P^{22}_{3}=p_{7}.
\end{align}
After a gauge fixing process,\cite{CGW1038} we can fix the phases of $f_1$, $f_3$,
$f_4$, $p_0$ and $A^0$ to make them positive.

The fixed-point conditions (\eqn{Phinonz}) form a set of non-linear
equation on the variables $f_i$, $p_i$, and $A^i$, which can
be solved exactly. After applying the ``gauge fixing''
discussed above, we find
the solution with an undetermined parameter $\eta$:
\begin{align}
&
f_i=1,\ \ i=0,1,...,5,
\nonumber\\
&
p_0=p_2=p_4=p_6=\eta,
\nonumber\\
&
p_1=p_3=p_5=p_7=\sqrt{1-{\eta}^2},
\nonumber\\
&
A^0=\eta^2, \ \ A^2=1-\eta^2,
\nonumber\\
&
A^1=A^3=\eta\sqrt{1-\eta^2}.
\end{align}
We also find
\begin{align}
 \e^{\imth \th_F}=
 \e^{\imth \th_{P1}}=
 \e^{\imth \th_{P2}}=
 \e^{\imth \th_{A1}}=
 \e^{\imth \th_{A2}}=
1.
\end{align}

\begin{figure} \begin{center}
\includegraphics[scale=0.4]{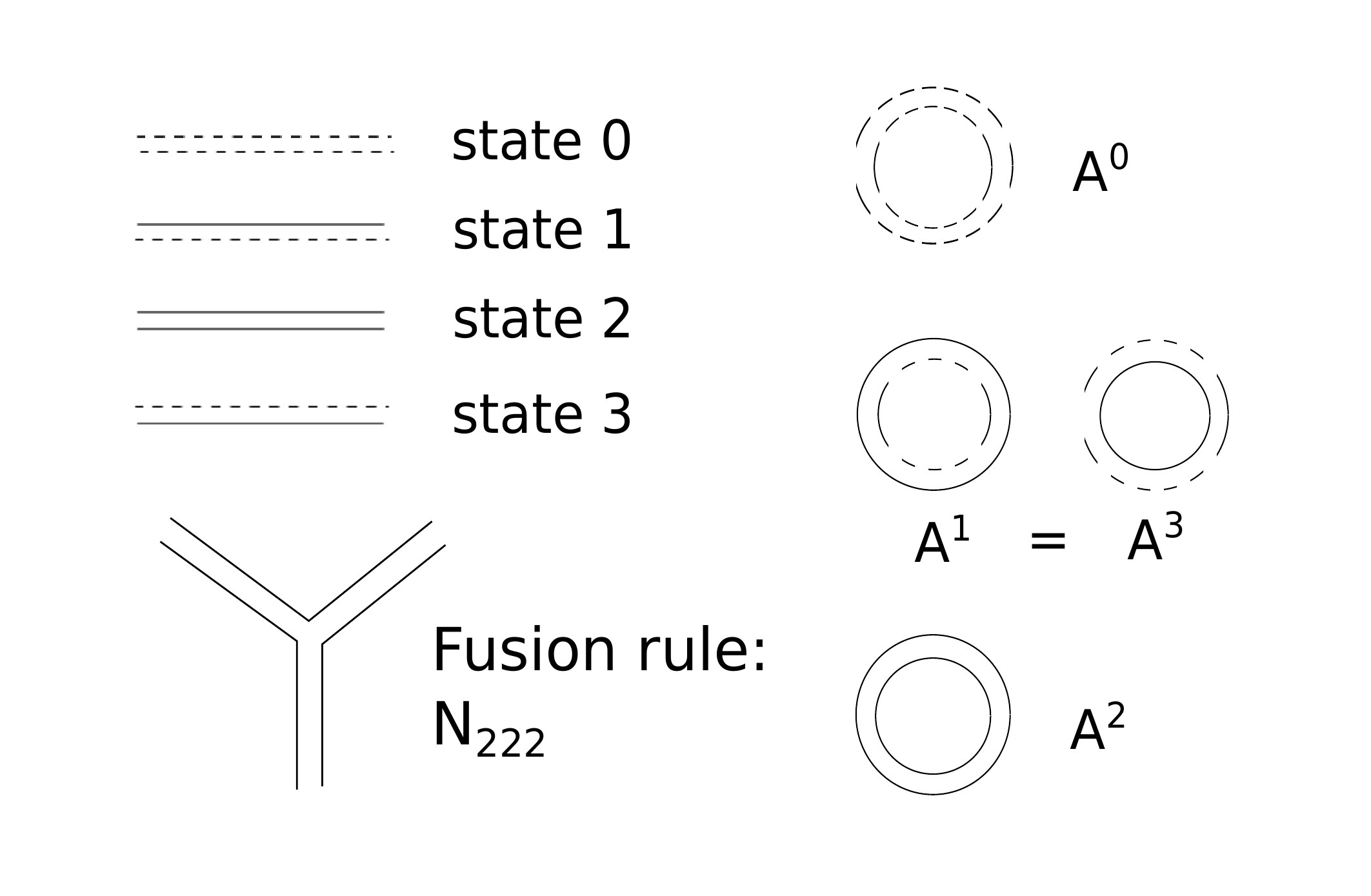}
\end{center}
%Fig 5
\caption{
The ``double-line rule''.
Replace the single-line trivalence graph by double lines.
If different link states are defined as shown above, then
the fusion rules are automatically satisfied. 
(The fusion rule $N_{222}=1$ was shown in the graph as an example.)
Suppose we associate a weight $\eta$ to a dashed-line loop and
a weight $\sqrt{1-\eta^2}$ to a solid-line loop respectively, then the
weight of any graph can be determined. (Note that all the loops are
independent, making it a trivial loop state.)
}
\label{doubleline}
\end{figure}

It can be shown that the above state corresponds to a trivial
loop state with adjustable loop weights. 
By introducing the ``double-line rule'' and associate different
weights to the ``dashed-line" loops $\&$ ``solid-line'' loops, the state
reduced to a state with independent loops.
This can be seen
very clearly in Fig. \ref{doubleline}.

\section{Summary}

In this paper, we compute the $T$ and $S$-matrices that was introduced to
define topological orders in 2D.\cite{Wrig,KW9327} We argue that this new label
can be used to fully describe topological orders.  We show explicitly how to
obtain these labels from their fixed-point states by applying the ``modular
transformations''.  The resulting $T$ and $S$-matrices are uniquely determined
after a proper choice of basis and can thus be used as labels for topological
orders.

The ``modular transformations'' can be defined
through two transformations, \ie the ``Dehn twist''
and ``$90^{\circ}$ rotation''. After applying both
transformations in the ground-state subspace,
we can get two matrices, $T$ and $S$,
describing ``modular transformations''.
We can then change the basis
to get the proper form of these two matrices.

Applying the two transformations in the ground-state subspace can be achieved
by projection.  We first define $T$ and $S$ for all non-contractable graphs on
a torus, then apply unitary transformation to diagonalize the Hamiltonian. We
then project $T$ and $S$ only onto the subspace with the smallest eigenvalue of
the Hamiltonian, thus obtaining $T$ and $S$ for the ground states.

To get the ``proper form'' of the two labeling matrices,
we first diagonalize $T$-matrix, and then search for
unitary transformation that doesn't change the diagonal
form of $T$, and at the same time, make $S$ satisfy
a few requirement including the ``Verlinde'' formula
(see \eqn{Sconditions}).
After this unitary transformation, the form
of $T$ and $S$ will be uniquely determined only up to
permutations of basis.

The resulting $T$ and $S$-matrices are unique labels for different topological
orders. We believe they contain full information of the phase and can fully
characterize different topological orders. Future work can be done to further
generalize this method to obtain $T$ and $S$ matrices for general
tensor-product states or other many-body states, so as to determine the
topological order of an arbitrary many-body state. Even further, using this $T$
and $S$-matrices description, we may eventually be able to describe phase
transitions between different topological orders.

We would like to thank Z.-X. Liu, L. Kong for some very helpful discussions.
This research is supported by NSF Grant No. DMR-1005541, NSFC 11074140, and
NSFC 11274192. It is also supported by the John Templeton Foundation. Research
at Perimeter Institute is supported by the Government of Canada through
Industry Canada and by the Province of Ontario through the Ministry of
Research.

\appendix

\section{Ideal Hamiltonian for fixed-point states on a torus}
\label{HamTorus}
The Hamiltonian construction is similar to the one
in appendix of \Ref{GWW1017}. The basic idea
is to construct a Hamiltonian which is a sum
of commuting projectors and is thus
exactly solvable.
On specific lattices for example
the honeycomb lattice,
it can be shown that all fixed-point
wave functions $(N_{ijk},
F^{ijm,\al\bt}_{kln,\chi\del},P_i^{kj,\al\bt},A^i)$
that we obtained from solving \eqn{Phinonz}
are exact gapped ground states
of such a local Hamiltonian.\cite{GWW1017}

\begin{figure}[tb]
\begin{center}
\includegraphics[scale=0.6]
{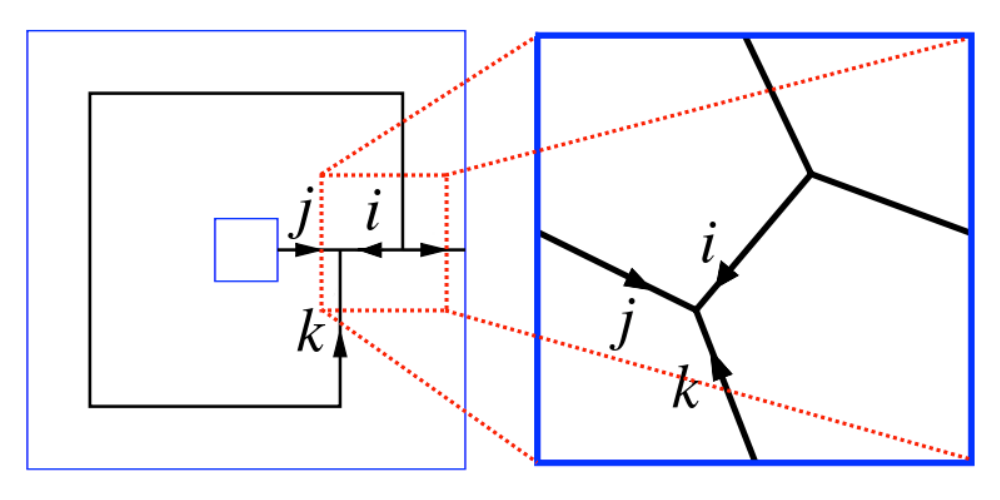}
\end{center}
\caption{
(Color online)
Both graphs represent non-contractable graphs
on a torus. The graph on the right is a zoom-in
version of the graph on the left. When
constructing the Hamiltonian, we
use the graph on the right to show the details
of $F$ and $P$-moves.
}
\label{AppendZoom}
\end{figure}

The Hamiltonian is of the following form:
\begin{align}
 \hat H=\sum_{\v v} (1-\hat Q_{\v v}) + \sum_{\v p} (1-\hat B_{\v p})
\end{align}
where $\sum_{\v v}$ sums over all vertices and
$\sum_{\v p}$ sums over all plaquettes. 
The Hamiltonian $\hat H$ acts on the Hilbert space
$V_G$ formed by all the graph states.
Operator $\hat Q_{\v v}$ in
$\hat H$ acts on the
states of the 3 links that connect to the vertex $\v v$:
\begin{align}
\label{LocalHamiltonian}
& \hat Q_{\v v} \left | \bmm \includegraphics[scale=.40]{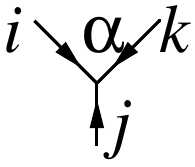} \emm  \right \>
=
\left | \bmm \includegraphics[scale=.40]{ijk} \emm  \right \>
\ \text{ if } N_{ijk}>0,
\nonumber\\
& \hat Q_{\v v} \left | \bmm \includegraphics[scale=.40]{ijk} \emm  \right \>
=0
\ \ \ \ \text{ otherwise} .
\end{align} 
Clearly, $\hat Q_{\v v}$ is a projector $\hat Q_{\v
v}^2=\hat Q_{\v v}$.  
Operator $\hat B_{\v p}$ in $\hat H$
acts on the states of all the links and vertices
belonging to the same plaquette $\v p$.
  
For our purpose in this paper,
we only need to consider the Hamiltonian
acting on the subspace of all
non-contractable graphs on a torus
(See Fig. \ref{AppendZoom}).
The $\hat B_{\v p}$ operator 
on such a torus can be constructed from a
combination of the $F$-moves and $P$-moves
as follows:
\begin{widetext}
\begin{align}
\Phi
\bpm \includegraphics[scale=.50]{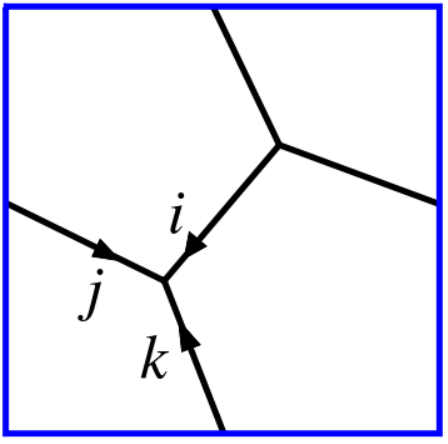} \epm
=
\sum_{t,s}
(P_{i}^{ts})^*
\Phi
\bpm \includegraphics[scale=.50]{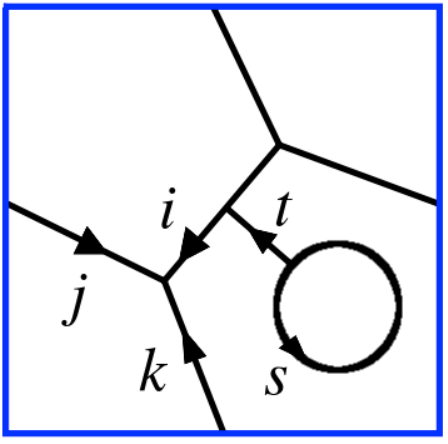} \epm
\nonumber
\end{align}
\begin{align}
=
\sum_{t,s}
\sum_{i'}
(P_{i}^{ts})^*
F^{i^*it^*}_{ssi'}
\Phi
\bpm \includegraphics[scale=.50]{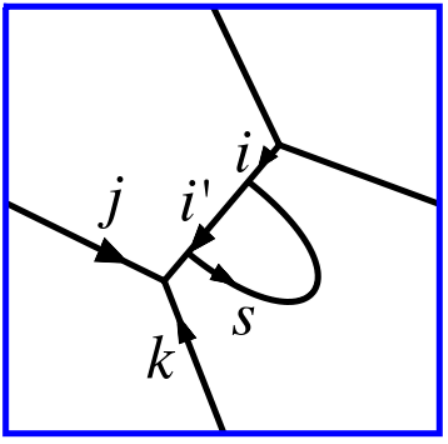} \epm
\nonumber
\end{align}
\begin{align}
=
\sum_{t,s}
\sum_{i'}
\sum_{k'}
(P_{i}^{ts})^*
F^{i^*it^*}_{ssi'}
F^{i's^*i}_{kj^*k'}
\Phi
\bpm \includegraphics[scale=.50]{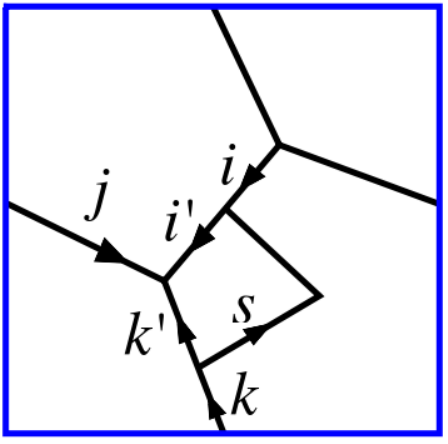} \epm
\nonumber
\end{align}
\begin{align}
=
\sum_{t,s}
\sum_{i'}
\sum_{k'}
\sum_{j'^*}
(P_{i}^{ts})^*
F^{i^*it^*}_{ssi'}
F^{i's^*i}_{kj^*k'}
F^{k'^*s^*k^*}_{j^*ij'^*}
\Phi
\bpm \includegraphics[scale=.50]{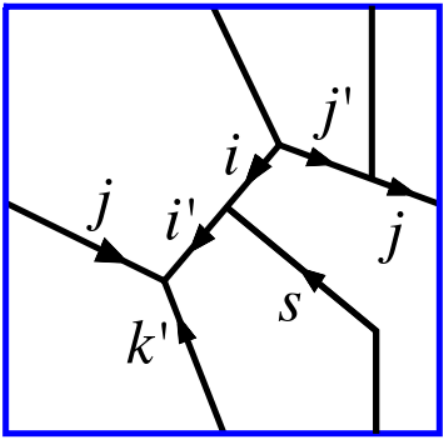} \epm
\nonumber
\end{align}
\begin{align}
=
\sum_{t,s}
\sum_{i'}
\sum_{k'}
\sum_{j'^*}
\sum_{i''}
(P_{i}^{ts})^*
F^{i^*it^*}_{ssi'}
F^{i's^*i}_{kj^*k'}
F^{k'^*s^*k^*}_{j^*ij'^*}
F^{j's^*j}_{i'k'^*i''}
\Phi
\bpm \includegraphics[scale=.50]{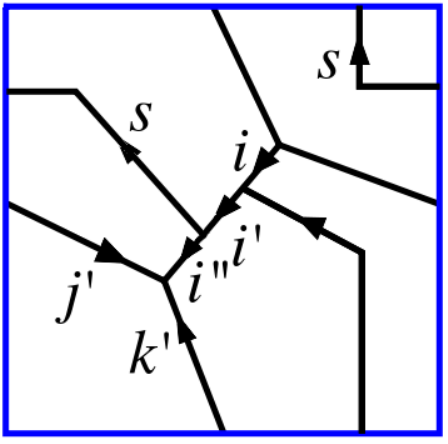} \epm
\nonumber
\end{align}
\begin{align}
=
\sum_{t,s}
\sum_{i'}
\sum_{k'}
\sum_{j'^*}
\sum_{i''}
\sum_{i'''}
(P_{i}^{ts})^*
F^{i^*it^*}_{ssi'}
F^{i's^*i}_{kj^*k'}
F^{k'^*s^*k^*}_{j^*ij'^*}
F^{j's^*j}_{i'k'^*i''}
F^{isi'}_{i''^*si'''^*}
\Phi
\bpm \includegraphics[scale=.50]{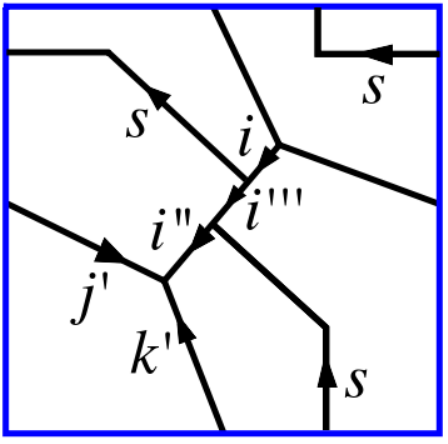} \epm
\nonumber
\end{align}
\begin{align}
=
\sum_{t,s}
\sum_{i'}
\sum_{k'}
\sum_{j'^*}
\sum_{i''}
\sum_{i'''}
\sum_{k''}
(P_{i}^{ts})^*
F^{i^*it^*}_{ssi'}
F^{i's^*i}_{kj^*k'}
F^{k'^*s^*k^*}_{j^*ij'^*}
F^{j's^*j}_{i'k'^*i''}
F^{isi'}_{i''^*si'''^*}
F^{i'''^*s^*i^*}_{k'^*j'k''^*}
\Phi
\bpm \includegraphics[scale=.50]{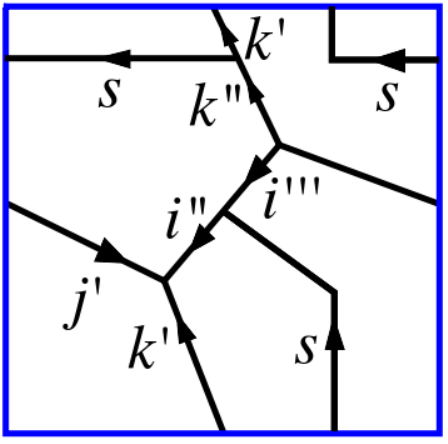} \epm
\nonumber
\end{align}
\begin{align}
=
\sum_{t,s}
\sum_{i'}
\sum_{k'}
\sum_{j'^*}
\sum_{i''}
\sum_{i'''}
\sum_{k''}
\sum_{j''}
(P_{i}^{ts})^*
F^{i^*it^*}_{ssi'}
F^{i's^*i}_{kj^*k'}
F^{k'^*s^*k^*}_{j^*ij'^*}
F^{j's^*j}_{i'k'^*i''}
F^{isi'}_{i''^*si'''^*}
F^{i'''^*s^*i^*}_{k'^*j'k''^*}
F^{k''s^*k'}_{j'i''*j''}
\Phi
\bpm \includegraphics[scale=.50]{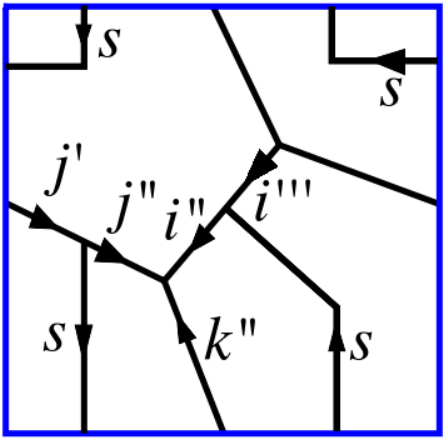} \epm
\nonumber
\end{align}
\begin{align}
=
\sum_{t,s}
\sum_{i'}
\sum_{k'}
\sum_{j'^*}
\sum_{i''}
\sum_{i'''}
\sum_{k''}
\sum_{j''}
(P_{i}^{ts})^*
F^{i^*it^*}_{ssi'}
F^{i's^*i}_{kj^*k'}
F^{k'^*s^*k^*}_{j^*ij'^*}
F^{j's^*j}_{i'k'^*i''}
F^{isi'}_{i''^*si'''^*}
F^{i'''^*s^*i^*}_{k'^*j'k''^*}
F^{k''s^*k'}_{j'i''*j''}\times
&
\nonumber\\
F^{j''^*s^*j'^*}_{i'''^*k''i''*}
\Phi
\bpm \includegraphics[scale=.50]{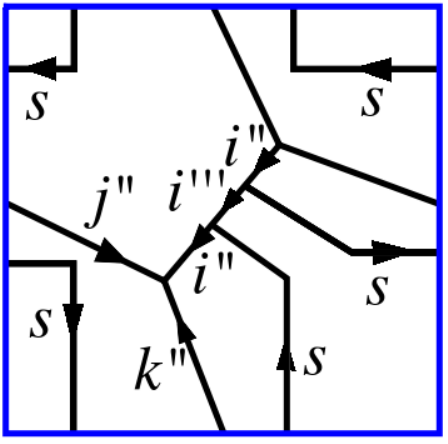} \epm
&
\nonumber
\end{align}
\begin{align}
=
\sum_{t,s}
\sum_{i'}
\sum_{k'}
\sum_{j'^*}
\sum_{i''}
\sum_{i'''}
\sum_{k''}
\sum_{j''}
(P_{i}^{ts})^*
F^{i^*it^*}_{ssi'}
F^{i's^*i}_{kj^*k'}
F^{k'^*s^*k^*}_{j^*ij'^*}
F^{j's^*j}_{i'k'^*i''}
F^{isi'}_{i''^*si'''^*}
F^{i'''^*s^*i^*}_{k'^*j'k''^*}
F^{k''s^*k'}_{j'i''*j''}\times
&
\nonumber\\
F^{j''^*s^*j'^*}_{i'''^*k''i''*}
\Phi
\bpm \includegraphics[scale=.50]{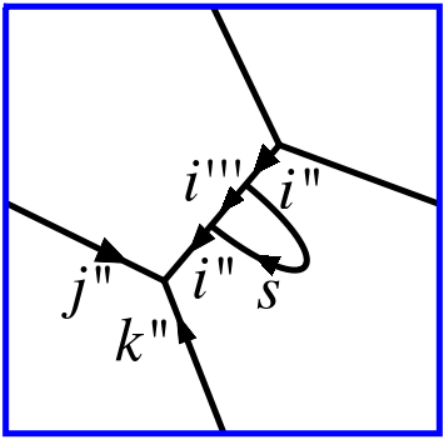} \epm
&
\nonumber
\end{align}
\begin{align}
=
\sum_{t,s}
\sum_{i'}
\sum_{k'}
\sum_{j'^*}
\sum_{i''}
\sum_{i'''}
\sum_{k''}
\sum_{j''}
\sum_{t'}
(P_{i}^{ts})^*
F^{i^*it^*}_{ssi'}
F^{i's^*i}_{kj^*k'}
F^{k'^*s^*k^*}_{j^*ij'^*}
F^{j's^*j}_{i'k'^*i''}
F^{isi'}_{i''^*si'''^*}
F^{i'''^*s^*i^*}_{k'^*j'k''^*}
F^{k''s^*k'}_{j'i''*j''}\times
&
\nonumber\\
F^{j''^*s^*j'^*}_{i'''^*k''i''*}
F^{i''s^*i'''}_{si''t'}
P^{t's*}_{i''}
\Phi
\bpm \includegraphics[scale=.50]{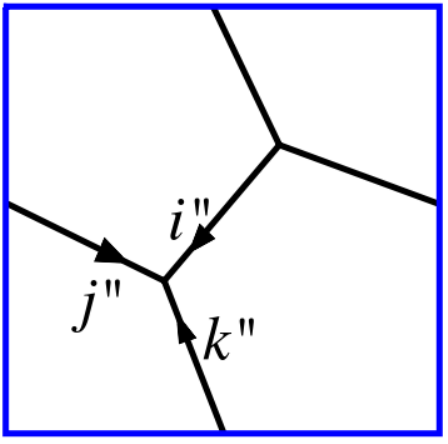} \epm
&
\end{align}
thus we have the matrix $\cB$ given by
\begin{align}
\cB
=
\sum_{t,s}
\sum_{i'}
\sum_{k'}
\sum_{j'^*}
\sum_{i''}
\sum_{i'''}
\sum_{k''}
\sum_{j''}
\sum_{t'}
(P_{i}^{ts})^*
F^{i^*it^*}_{ssi'}
F^{i's^*i}_{kj^*k'}
F^{k'^*s^*k^*}_{j^*ij'^*}
&
F^{j's^*j}_{i'k'^*i''}
F^{isi'}_{i''^*si'''^*}
F^{i'''^*s^*i^*}_{k'^*j'k''^*}
F^{k''s^*k'}_{j'i''*j''}\times
\nonumber\\
F^{j''^*s^*j'^*}_{i'''^*k''i''*}
F^{i''s^*i'''}_{si''t'}
P^{t's*}_{i''}
&
\end{align}
\end{widetext}
which is the matrix form of operator
$\hat B_{\v p}$.

Recall that all fixed-point states are
exact ground states of the Hamiltonian
$\hat H$ defined in \eqn{LocalHamiltonian}.
Thus by finding the exact ground states
of $\hat H$ among all the non-contractable
graph states,
we can find the subspace of all the
fixed-point states.

In sections \ref{Fibonacci} and \ref{Pfaffian}, we
are required to reduce the modular transformations
to only within the ground-state subspace of the
above Hamiltonian. This can now be easily achieved.
We can first define $T$ and $S$ on all the non-contractable
graphs on a torus, then find the ground-state
subspace by diagonalizing the Hamiltonian on
these non-contractable graphic states. We can then
project $T$ and $S$ only onto the ground-state subspace
of the Hamiltonian, thus obtain the required $T$
and $S$.

\bibliography{../../bib/wencross,../../bib/all,../../bib/publst,./local}

\end{document}